\documentclass[10pt,twocolumn]{article}
\usepackage[top=1.8cm, bottom=1.8cm, left=1.8cm, right=1.8cm]{geometry}
\usepackage{helvet}  %
\usepackage{courier}  %
\usepackage[hyphens]{url}  %
\usepackage{graphicx}  %
\usepackage{pifont}
\usepackage{stackengine}
\usepackage{amsmath}
\usepackage{amssymb} %
\renewcommand{\Bbb}{\mathbb}

\newif\ifcomment
 \commentfalse

\usepackage[hang,flushmargin]{footmisc}

\usepackage{xcolor}
\usepackage{booktabs}
\usepackage{graphicx}
\usepackage{paralist}
\usepackage[small,bf]{caption}
\usepackage{subfigure}
\usepackage{times}

\usepackage[compact]{titlesec}
\titlespacing*{\section}{0pt}{*3}{3pt}
\titlespacing{\subsection}{0pt}{*2}{2pt}
\titlespacing{\subsubsection}{0pt}{*2}{2pt}

\definecolor{linkcol}{rgb}{0,0,0.5}
\definecolor{citecol}{rgb}{0,0.5,0.3}
\definecolor{urlcol}{rgb}{0.3,0,0}

\usepackage{xspace}

\newcommand{\descr}[1]{\smallskip\noindent\textbf{#1}}
\newcommand{\descrit}[1]{\vspace{0.05cm}\noindent{#1}}

\newcommand{\dspol}{{{\fontsize{10}{10}\selectfont /pol/}}\xspace}
\newcommand{\dsb}{{{\fontsize{10}{10}\selectfont /b/}}\xspace}

\newcommand{\tdshort}{T\textunderscore D\xspace}
\newcommand{\td}{The\textunderscore Donald\xspace}

\renewenvironment{thebibliography}[1]{
  \begin{oldthebibliography}{#1}
    \setlength{\itemsep}{0.0em}
    \setlength{\parskip}{0.0em}
}
{
  \end{oldthebibliography}
}

\renewcommand{\footnoterule}{%
  \kern -3pt
  \hrule width 1in
  \kern 2pt
}

\makeatletter
\def\url@leostyle{%
  \@ifundefined{selectfont}{\def\UrlFont{}}%
  {\def\UrlFont{}}%
}
\makeatother
\usepackage[hyphenbreaks]{breakurl}

\definecolor{darkred}{RGB}{153,0,0}
\definecolor{darkblue}{RGB}{0,0,99}
\usepackage[colorlinks=true, linkcolor = darkred,   citecolor = darkred, urlcolor = darkblue]{hyperref}

\newif\ifwatermark
\watermarkfalse

\ifwatermark
    \usepackage{draftwatermark}
    \SetWatermarkScale{.7}
    \SetWatermarkText{\shortstack{In Submission:\\Do Not Distribute}}
    \SetWatermarkColor[rgb]{1,0.88,0.88}
\fi
\usepackage{etoolbox}
\makeatletter
\patchcmd\@combinedblfloats{\box\@outputbox}{\unvbox\@outputbox}{}{%
   \errmessage{\noexpand\@combinedblfloats could not be patched}%
}%
 \makeatother
\AtBeginShipout{%
  \ifnum\value{page}>1 %
    \typeout{* Additional boxing of page `\thepage'}%
    \setbox\AtBeginShipoutBox=\hbox{\copy\AtBeginShipoutBox}%
  \fi
}

\begin{document}
\title{\bf On the Origins of Memes by Means of Fringe Web Communities\thanks{A shorter version of this paper appears in the Proceedings of 18th ACM Internet Measurement Conference (IMC 2018). This is the full version.}}
\author{
  Savvas Zannettou$^{\star}$, Tristan Caulfield$^{\ddagger}$, Jeremy Blackburn$^\dagger$, Emiliano De Cristofaro$^\ddagger$,\\ Michael Sirivianos$^{\star}$, Gianluca Stringhini$^{\ddagger\diamond}$, and Guillermo Suarez-Tangil$^{\ddagger+}$\\[0.5ex]
\normalsize $^{\star}$Cyprus University of Technology, $^\ddagger$University College London\\[-0.5ex]
\normalsize $^\dagger$University of Alabama at Birmingham, ${^\diamond}$Boston University, ${^+}$King's College London \\
\normalsize sa.zannettou@edu.cut.ac.cy \{t.caulfield,e.decristofaro\}@ucl.ac.uk\\[-0.5ex]
 \normalsize blackburn@uab.edu, michael.sirivianos@cut.ac.cy, gian@bu.edu, guillermo.suarez-tangil@kcl.ac.uk}

\date{}

\maketitle

\begin{abstract}
Internet memes are increasingly used to sway and manipulate public opinion.
This prompts the need to study their propagation, evolution, and influence across the Web.
In this paper, we detect and measure the propagation of memes across multiple Web communities, using a processing pipeline based on perceptual hashing and clustering techniques, and a dataset of 160M images from 2.6B posts gathered from Twitter, Reddit, 4chan's Politically Incorrect board (\dspol), and Gab, over the course of 13 months.
We group the images posted on fringe Web communities (\dspol, Gab, and \td subreddit) into clusters, annotate them using meme metadata obtained from Know Your Meme, and also map images from mainstream communities (Twitter and Reddit) to the clusters.

Our analysis provides an assessment of the popularity and diversity of memes in the context of each community, showing, e.g., that racist memes are extremely common in fringe Web communities.
We also find a substantial number of politics-related memes on both mainstream and fringe Web communities, supporting media reports that memes might be used to enhance or harm politicians.
Finally, we use Hawkes processes to model the interplay between Web communities and quantify their reciprocal influence, finding that \dspol substantially influences the meme ecosystem with the number of memes it produces, while \td has a higher success rate in pushing them to other communities.
\end{abstract}

\section{Introduction}

The Web has become one of the most impactful vehicles for the propagation of ideas and culture.
Images, videos, and slogans are created and shared online at an unprecedented pace.
Some of these, commonly referred to as \emph{memes}, become viral, evolve, and eventually enter popular culture.
The term ``meme'' was first coined by Richard Dawkins \cite{dawkins1976selfish}, who framed them as cultural analogues to genes,
as they too self-replicate, mutate, and respond to selective pressures~\cite{graham2005genes}.
Numerous memes have become integral part of Internet culture, with well-known examples
including the Trollface~\cite{trollface_meme}, Bad Luck Brian~\cite{bad_luck_brian_meme}, and Rickroll~\cite{rickroll_meme}.

While most memes are generally ironic in nature, used with no bad intentions, others have assumed negative and/or hateful connotations, including outright racist and aggressive undertones~\cite{yoon2016not}.
These memes, often generated by fringe communities, are being ``weaponized'' and even becoming part of political and ideological propaganda~\cite{salon_weaponization}. 
For example, memes were adopted by candidates during the 2016 US Presidential Elections as part of their iconography~\cite{guardian_memes_election}; in October 2015, then-candidate Donald Trump retweeted an image depicting him as Pepe The Frog, 
a controversial character considered a hate symbol~\cite{adl_pepe_frog}.
In this context, polarized communities within 4chan and Reddit have been working hard to create new memes and make them go viral, aiming to increase the visibility of their ideas---a phenomenon known as ``attention hacking''~\cite{marwick2017media}.

\descr{Motivation.} Despite their increasingly relevant role, we have very little measurements and computational tools to understand the origins and the influence of memes.
The online information ecosystem is very complex; social networks do not operate in a vacuum but rather influence each other as to how information spreads~\cite{zannettou2017web}.
However, previous work (see Section~\ref{sec:related_work}) 
has mostly focused on social networks in an isolated manner.

In this paper, we aim to bridge these gaps by identifying and addressing a few research questions, which are oriented towards fringe Web communities:
1)~How can we characterize memes, and how do they evolve and propagate?
2)~Can we track meme propagation across multiple communities and measure their influence?
3)~How can we study variants of the same meme?
4)~Can we characterize Web communities through the lens of memes?

Our work focuses on four Web communities: Twitter, Reddit, Gab, and 4chan's Politically Incorrect board (\dspol), because of their impact on the information ecosystem~\cite{zannettou2017web} and anecdotal evidence of them disseminating weaponized memes~\cite{scott2018information}. %
We design a processing pipeline and use it over 160M images posted between July 2016 and July 2017. %
Our pipeline relies on perceptual hashing (pHash) and clustering techniques; the former extracts representative feature vectors from the images encapsulating their visual peculiarities, while the latter allow us to detect groups of images that are part of the same meme.
We design and implement a custom distance metric, based on both pHash and meme metadata, obtained from Know Your Meme (KYM), and use it to understand the interplay between the different memes.
Finally, using Hawkes processes, we quantify the reciprocal influence of each Web community with respect to the dissemination of image-based memes.

\descr{Findings.}
Some of our findings (among others) include: \smallskip
\begin{compactenum}
\item Our influence estimation analysis reveals that \dspol and \td are influential actors in the meme ecosystem, despite their modest size.
We find that \dspol substantially influences the meme ecosystem by posting a large number of memes, while \td is the most \emph{efficient} community in pushing memes to both fringe and mainstream Web communities.

\item Communities within 4chan, Reddit, and Gab use memes to share hateful and racist content. For instance, among the most popular cluster of memes, we find variants of the anti-semitic ``Happy Merchant'' meme~\cite{happy_merchant_meme} and the controversial Pepe the Frog~\cite{pepe_frog_meme}.

\item Our custom distance metric effectively reveals the phylogenetic relationships of clusters of images.
This is evident from the graph that shows the clusters obtained from \dspol, Reddit's \td subreddit, and Gab available for
exploration at~\cite{memes_graph_site}.

\end{compactenum}

\descr{Contributions.} 
First, we develop a robust processing pipeline for detecting and tracking memes across multiple Web communities.
Based on pHash and clustering algorithms, it supports large-scale measurements of meme ecosystems, while minimizing processing power and storage requirements.
Second, we introduce a custom distance metric,
geared to highlight hidden correlations between memes and better understand the interplay and overlap between them.
Third, we provide a characterization of multiple Web communities (Twitter, Reddit, Gab, and \dspol) with respect to the memes they share, and an analysis of their reciprocal influence using the Hawkes Processes statistical model.
Finally, we release our processing pipeline and datasets\footnote{\url{https://github.com/memespaper/memes_pipeline}}, in the hope to support further measurements in this space.

\section{Methodology}
\label{sec:methodology}
In this section, we present our methodology for measuring the propagation of memes across Web communities.

\subsection{Overview}
\label{sec:methodology:overview}

Memes are high-level concepts or ideas that spread within a culture~\cite{dawkins1976selfish}.
In Internet vernacular, a {\em meme} usually refers to variants of a particular image, video, clich\'e, etc.~that share a
common theme and are disseminated by a large number of users.
In this paper, we focus on their most common incarnation: \emph{static images}.

\begin{figure}[t]
\centering
\includegraphics[width=0.99\columnwidth]{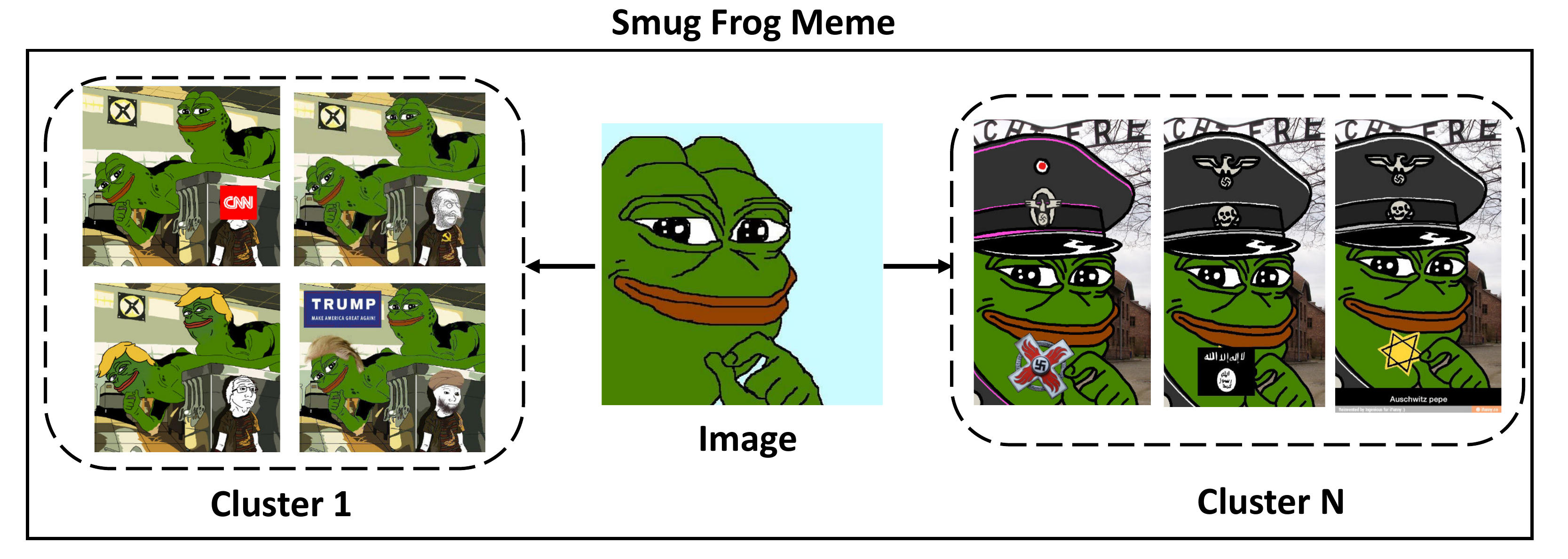}
\caption{An example of a meme (Smug Frog) that provides an intuition of what an image, a cluster, and a meme is.}
\label{fig:smug_frog_example}
\end{figure}

To gain an understanding of how memes propagate across the Web, with a particular focus on discovering the communities that are most
influential in spreading them, our intuition is to build \emph{clusters} of visually similar images, allowing us to track variants of a meme.
We then group clusters that belong to the same meme to study and track the meme itself.
In Figure~\ref{fig:smug_frog_example}, we provide a visual representation of the Smug Frog meme~\cite{smug_frog_meme}, which includes many variants of the same image (a ``smug'' Pepe the Frog) and several clusters of variants.
Cluster 1 has variants from a Jurassic Park scene, where one of the characters is hiding from two velociraptors behind a kitchen counter: the frogs are stylized to look similar to velociraptors, and the character hiding varies to express a particular message.
For example, in the image in the top right corner, the two frogs are searching for an anti-semitic caricature of a Jew (itself a meme known as the Happy Merchant~\cite{happy_merchant_meme}).
Cluster N shows variants of the smug frog wearing a Nazi officer military cap with a photograph of the infamous ``Arbeit macht frei'' slogan from the distinctive curved gates of Auschwitz in the background.
In particular, the two variants on the right display the death's head logo of the SS-Totenkopfverb{\"a}nde organization responsible for
running the concentration camps during World War II.
Overall, these clusters represent the branching nature of memes: as a new variant of a meme becomes prevalent, it often branches into its own sub-meme, potentially incorporating imagery from other memes.

\subsection{Processing Pipeline}\label{subsec:pipeline}

\begin{figure}[t]
\hspace*{-0.3cm}
\includegraphics[width=1.0\columnwidth]{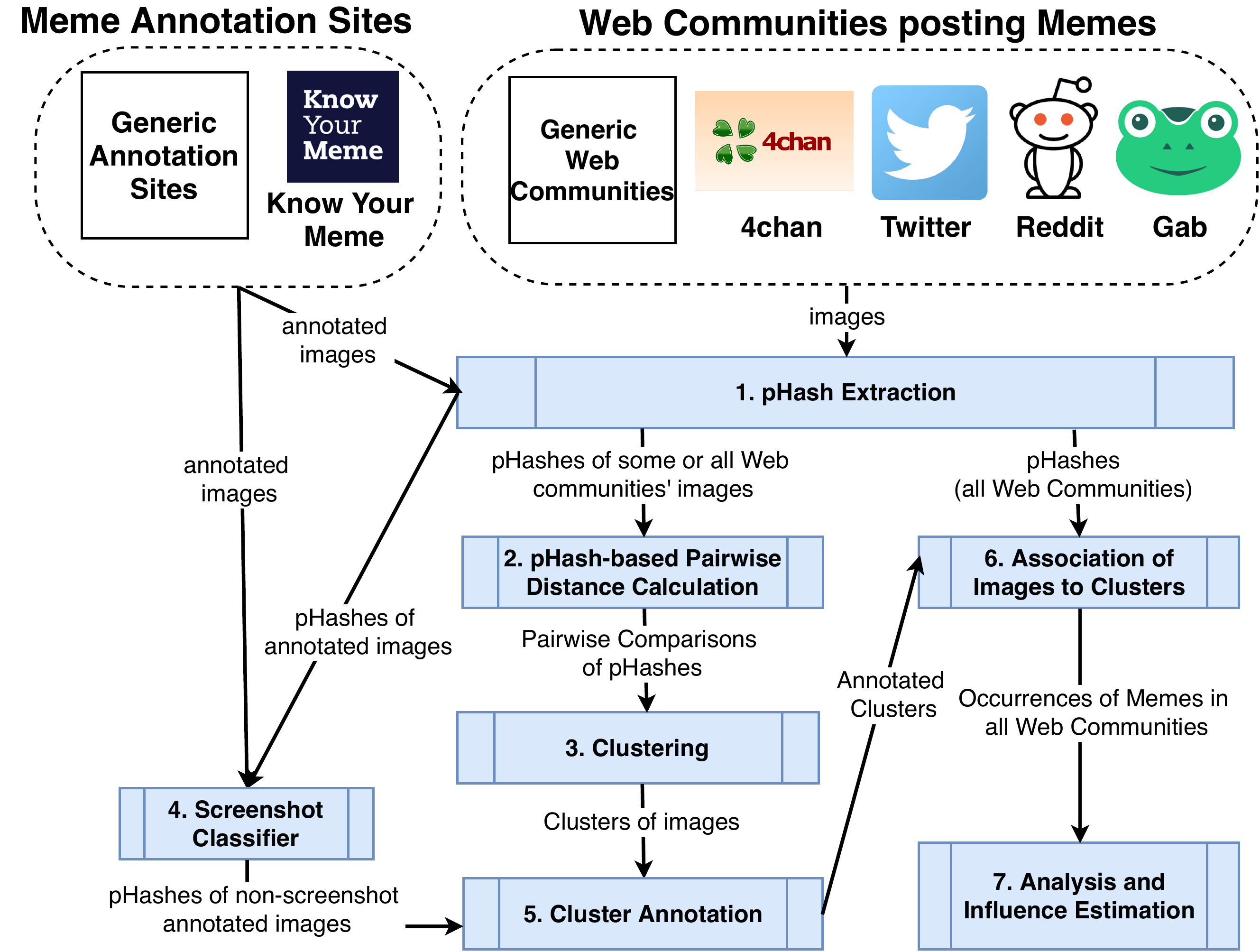}
\caption{High-level overview of our processing pipeline.}
\label{fig:pipeline}
\end{figure}

Our processing pipeline is depicted in Figure~\ref{fig:pipeline}.
As discussed above, our methodology aims at identifying clusters of similar images and assign them to higher level groups, which are the actual memes.
Note that the proposed pipeline is not limited to image macros and can be used to identify any image. 
We first discuss the types of data sources needed for our approach, i.e., meme annotation sites and Web communities that post memes
(dotted rounded rectangles in the figure).
Then, we describe each of the operations performed by our pipeline (Steps 1-7, see regular rectangles).

\descr{Data Sources.} Our pipeline uses two types of data sources: 1)~sites providing meme annotation and 2)~Web communities that disseminate memes.
In this paper, we use Know  Your Meme for the former, and Twitter, Reddit, \dspol, and Gab for the latter.
We provide more details about our datasets in Section~\ref{sec:dataset}.
Note that our methodology supports any annotation site and any Web community, and this is why we add the ``{\em Generic}'' sites/communities notation in Figure~\ref{fig:pipeline}.

\descr{pHash Extraction (Step 1).} We use the Perceptual Hashing (pHash) algorithm~\cite{monga2006perceptual} to calculate a fingerprint of each image in such a way that any two images that look similar to the human eye map to a ``similar'' hash value. pHash generates a feature vector of 64 elements that describe an image, computed from the Discrete Cosine Transform among the different frequency domains of the image.
Thus, visually similar images have minor differences in their vectors, hence allowing to search for and detect visually similar images.
For example, the string representation of the pHashes obtained from the images in cluster N (see Figure~\ref{fig:smug_frog_example}) are 55352b0b8d8b5b53, 55952b0bb58b5353, and 55952b2b9da58a53, respectively.
The algorithm is also robust against changes in the images, e.g., signal processing operations and direct
manipulation~\cite{zauner2011rihamark}, and effectively reduces the dimensionality of the raw images.

\descr{Clustering via pairwise distance calculation (Steps 2-3).}
Next, we cluster images from one or more Web Communities using the pHash values.
We perform a pairwise comparison of all the pHashes using Hamming distance (Step 2).
To support large numbers of images, we implement a highly parallelizable system on top of TensorFlow~\cite{abadi2016tensorflow}, which uses multiple GPUs to enhance performance. 
Images are clustered using a density-based algorithm (Step 3).
Our current implementation uses DBSCAN~\cite{ester1996density}, mainly because it can discover clusters of arbitrary shape
and performs well over large, noisy datasets.
Nonetheless, our architecture can be easily tweaked to support any clustering algorithm and distance metric.

We also perform an analysis of the clustering performance and the rationale for selecting the clustering threshold.
We refer to Appendix~\ref{sec:appendix_clustering} for more details.

\descr{Screenshots Removal (Step 4).} Meme annotation sites like KYM often include, in their image galleries, screenshots of social network posts that are not variants of a meme but just comments about it.
Hence, we discard social-network screenshots from the annotation sites data sources using a deep learning classifier. 
We refer to Appendix~\ref{sec:appendix_classifier} for details about the model and the training dataset.

\descr{Cluster Annotation (Steps 5).} Clustering annotation uses the \emph{medoid} of each cluster, i.e., the element with the minimum square average distance from all images in the cluster. 
In other words, the medoid is the image that best represents the cluster.
The clusters' medoids are compared with all images from meme annotation sites, by calculating the Hamming distance between each pair of pHash vectors. We consider that an image matches a cluster if the distance is less than or equal to a threshold $\theta$, which we set to $8$, as it allows us to capture the diversity of images that are part of the same meme while maintaining a low number of false positives.

As the annotation process considers all the images of a KYM entry's image gallery, it is likely we will get multiple annotations for a single cluster.
To find the representative KYM entry for each cluster, we select the one with the largest proportion of matches of KYM images with
the cluster medoid. In case of ties, we select the one with the minimum average Hamming distance.

As KYM is based on community contributions it is unclear how good our annotations are. To evaluate KYM entries and our cluster annotations, three authors of this paper assessed 200 annotated clusters and 162 KYM entries. We find that only 1.85\% of the assessed KYM entries were regarded as ``bad'' or not sufficient.
When it comes to the clustering annotation, we note that the three annotators had substantial agreement (Fleis agreement score equal to 0.67) and that the clustering accuracy, after majority agreement, of the assessed clusters is 89\% .
We refer to Appendix~\ref{sec:appendix_clustering_annotation_evaluation} for details about the annotation process and results.

\descr{Association of images to memes (Step 6).}
To associate images posted on Web communities (e.g., Twitter, Reddit, etc.) to memes, we compare them with the clusters' medoids, using the
same threshold $\theta$.
This is conceptually similar to Step 5, but uses images from Web communities instead of images from annotation sites.
This lets us identify memes posted in generic Web communities and collect relevant metadata from the posts (e.g., the timestamp of a tweet). 
Note that we track the propagation of memes in generic Web communities (e.g., Twitter) using a {\em seed} of memes obtained by clustering images from other (fringe) Web communities.
More specifically, our seeds will be memes generated on three fringe Web communities (\dspol, \td subreddit, Gab); nonetheless, our methodology can be applied to any community.

\descr{Analysis and Influence Estimation (Step 7).} We analyze all relevant clusters and the occurrences of memes, 
aiming to assess: 1)~their popularity and diversity in each community; 2)~their temporal evolution; and 3)~how communities influence each other with respect to meme dissemination. %

\subsection{Distance Metric}
\label{sec:methodology:distance}

To better understand the interplay and connections between the clusters, we introduce a custom distance metric, which relies on both the visual peculiarities of the images (via pHash) and  data available from annotation sites.
The distance metric supports one of two modes: 1)~one for when both clusters are annotated ({\it full-mode}), and 2)~another for when one or none of the clusters is annotated ({\it partial-mode}). 

\descr{Definition.}
Let $c$ be a cluster of images and $\mathsf{F}$ a set of features extracted from the clusters.
The custom distance metric between cluster $c_i$ and $c_j$ is defined as:

\begin{equation}\label{eq:distance}
\small
\mathsf{distance}(c_i, c_j) = 1 - \sum_\mathsf{f \in \mathsf{F}} w_\mathsf{f} \times \texttt{r}_\mathsf{f}(c_i, c_j) %
\end{equation}

\noindent where $\texttt{r}_\mathsf{f}(c_i, c_j)$ denotes the similarity between the features of type $\mathsf{f} \in \mathsf{F}$ of cluster $c_i$ and $c_j$, and $w_f$ is a weight that represents the relevance of each feature.
Note that $\sum_\mathsf{f} w_\mathsf{f} = 1$ and $\mathsf{r}_f(c_i, c_j) = \{x \in \Bbb{R} \mid 0 \leq x \leq 1\}$.
Thus, $\mathsf{distance}(c_i, c_j)$ is a number between 0 and 1. %

\descr{Features.} We consider four different features for $\texttt{r}_\mathsf{f \in \mathsf{F}}$, specifically, $\mathsf{F} = \{perceptual, meme, people, culture\}$; see below.

\descrit{\underline{${r_{perceptual}}$}:} this feature is the similarity between two clusters from a perceptual viewpoint.
Let $h$ be a pHash vector for an image $m$ in cluster $c$, where $m$ is the medoid of the cluster, and $d_{ij}$ the Hamming distance between vectors $h_i$ and $h_j$ (see Step 5 in Section~\ref{subsec:pipeline}).
We compute $d_{ij}$ from $c_i$ and $c_j$ as follows.
First, we obtain  obtain the medoid $m_i$ from cluster $c_i$.
Subsequently, we obtain $h_i{=}$\mbox{pHash}$(m_i)$.
Finally, we compute $d_{ij} {=} \mbox{Hamming}(h_i, h_j)$.
We simplify notation and use $d$ instead of $d_{ij}$ to denote the distance between two medoid images and refer to this distance as the Hamming {\it score}.

\begin{figure}[t]
\centering
\includegraphics[width=.65\columnwidth]{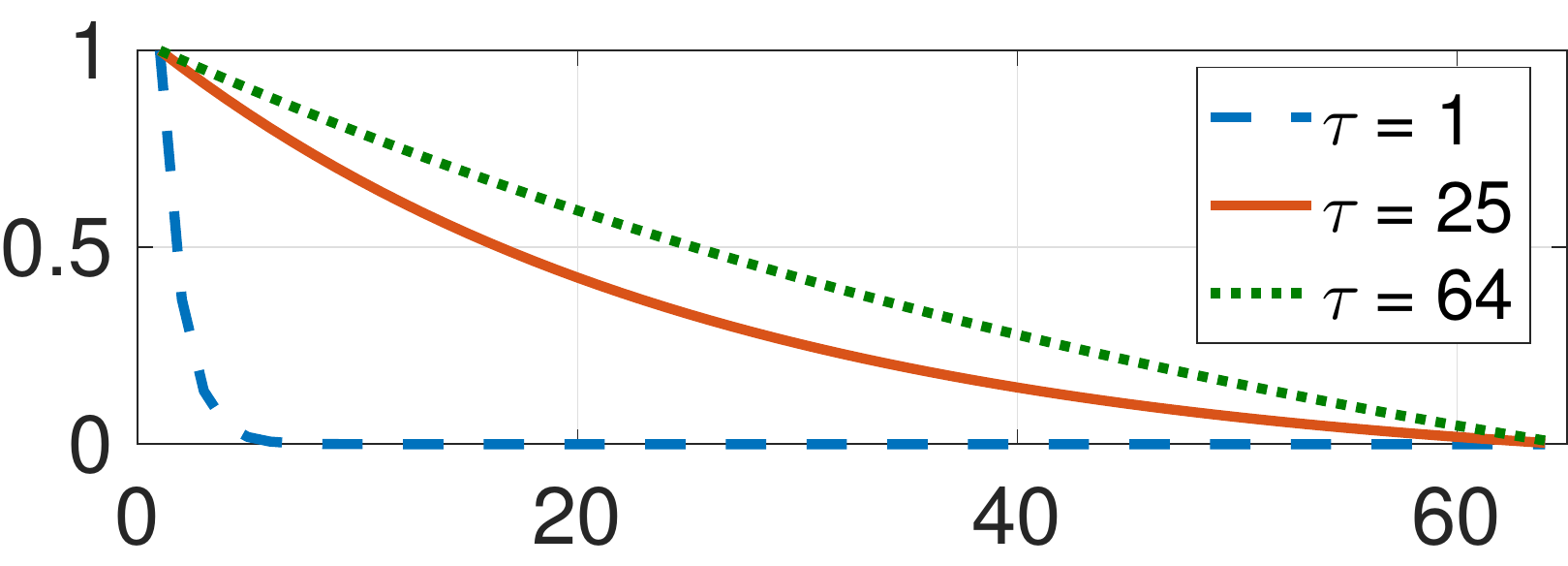}
\caption{Different values of $r_{perceptual}$ (y-axis) for all possible inputs of $d$ (x-axis) with respect to the smoother $\tau$.}
\label{fig:perceptual-relevance}
\end{figure}

We define the perceptual similarity between two clusters as an exponential decay function over the Hamming score $d$:
\begin{equation}\label{eq:score_perceptual}\small
r_{perceptual}(d) = 1 - \frac{d}{\tau \times e^{\texttt{max} / \tau}} 
\end{equation}

\noindent where $\texttt{max}$ represents the maximum pHash distance between two images and $\tau$ is a constant parameter, or {\it smoother}, that controls how fast the exponential function decays for all values of $d$ (recall that $\{d \in \Bbb{R} \mid 0 \leq d \leq \texttt{max}\}$).
Note that $\texttt{max}$ is bound to the precision given by the pHash algorithm.
Recall that each pHash has a size of $|d| {=} 64$, hence $\texttt{max} {=} 64$.
Intuitively, when $\tau << 64$, $r_{perceptual}$ is a high value only with perceptually indistinguishable images, e.g., for $\tau {=} 1$, two images with $d {=} 0$ have a similarity $r_{perceptual} {=} 1.0$.
With the same $\tau$, the similarity drops to $0.4$ when $d {=} 1$.
By contrast, when $\tau$ is close to $64$, $r_{perceptual}$ decays almost linearly.
For example, for $\tau {=} 64$, $r_{perceptual}(d{=}0) {=} 1.0$ and $r_{perceptual}(d{=}1) {=} 0.98$.
Figure~\ref{fig:perceptual-relevance} shows how $r_{perceptual}$ performs for different values of $\tau$.
As mentioned above, we observe that pairs of images with scores between $d{=}0$ and $d{=}8$ are usually part of the same variant (see Step 5 in Section~\ref{subsec:pipeline}).
In our implementation, we set $\tau {=} 25$ as $r_{perceptual}$ returns high values up to $d{=}8$, and rapidly decays thereafter.

\descrit{\underline{$r_{meme}$}, \underline{$r_{culture}$}, and \underline{$r_{people}$}:} the annotation process (Step 5) provides  contextualized information about the cluster medoid, including the name (i.e., the main identifier) given to a meme, the associated culture (i.e., high-level group of meme), and people that are included in a meme. Note that we use all the annotations for each category and not only the representative one (see Step 5).
Therefore, we model a different similarity for each of the these categories, by looking at the overlap of all the annotations among the medoids of both clusters ($m_i$, $m_j$, for $c_i$ and $c_j$, respectively).
Specifically, for each category, we calculate the Jaccard index between the annotations of both medoids, for memes, cultures, and people, thus acquiring $r_{meme}$, $r_{culture}$, $r_{people}$, respectively.

\descr{Modes.}
Our distance metric measures how similar two clusters are.
If both clusters are annotated, we operate in ``full-mode,'' and in ``partial-mode'' otherwise.
For each mode, we use different weights for the features in Eq.~\ref{eq:distance}, which we set empirically as we lack the ground-truth data needed to automate the computation of the optimal set of thresholds.

\descr{Full-mode.}  In full-mode, we set weights as follows.
1)~The features from the perceptual and meme categories should have higher relevance than people and culture, as they are intrinsically related to the definition of meme (see Section~\ref{sec:methodology:overview}).
The last two are non-discriminant features, yet are informative and should contribute to the metric.
Also, 2)~$r_{meme}$ should not outweigh $r_{perceptual}$ because of the relevance that visual similarities have on the different variants of a meme.
Likewise, $r_{perceptual}$ should not dominate over $r_{meme}$ because of the branching nature of the memes.
Thus, we want these two categories to play an equally important weight.
Therefore, we choose
$w_{perceptual} {=} 0.4$,
$w_{meme} {=} 0.4$,
$w_{people} {=} 0.1$,
$w_{culture} {=} 0.1.$

This means that when two clusters belong to the same meme and their medoids are perceptually similar, the distance between the clusters will be small.
In fact, it will be at most $0.2 = 1 - (0.4 + 0.4)$ if people and culture do not match, and $0.0$ if they also match.
Note that our metric also assigns small distance values for the following two cases:
1)~when two clusters are part of the same meme variant,
and 2)~when two clusters use the same image for different memes.

\descr{Partial-mode.} In this mode, we associate unannotated images with any of the known clusters.
This is a critical component of our analysis (Step 6), allowing us to study images from generic Web communities where annotations are unavailable.
In this case, we rely entirely on the perceptual features.
We once again use Eq.~\ref{eq:distance}, but simply set all weights to 0, except for $w_{perceptual}$ (which is set to 1).
That is, we compare the image we want to test with the medoid of the cluster and we apply Eq.~\ref{eq:score_perceptual} as described above.

\section{Datasets}
\label{sec:dataset}

We now present the datasets used in our measurements.

\subsection{Web Communities}\label{sec:communities}
As mentioned earlier, our data sources are Web communities that post memes and meme annotation sites.
For the former, we focus on four communities: Twitter, Reddit, Gab, and 4chan (more precisely, 4chan's Politically Incorrect board, \dspol).
This provides a mix of mainstream social networks (Twitter and Reddit) as well as fringe communities that are often associated with the alt-right and have an impact on the information ecosystem (Gab and \dspol)~\cite{zannettou2017web}.

There are several other platforms playing important roles in spreading memes, however, many are ``closed'' (e.g., Facebook) or do not involve memes based on static images (e.g., YouTube, Giphy).
In future work, we plan to extend our measurements to communities like Instagram and Tumblr, as well as to GIF and video memes.
Nonetheless, we believe our data sources already allow us to elicit comprehensive insights into the meme ecosystem.

\begin{table}[t]
\centering
\setlength{\tabcolsep}{3pt}
\small
\begin{tabular}{lrrrr}
\toprule
{\bf Platform} & {\bf \#Posts}    & {\bf \#Posts with} & {\bf \#Images} & {\bf \#Unique} \\
& & {\bf Images} &  & {\bf pHashes}\\
\midrule
\textbf{Twitter} &     1,469,582,378               & 242,723,732  & 114,459,736                               & 74,234,065                                                                                   \\
\textbf{Reddit}  & 1,081,701,536               & 62,321,628 & 40,523,275                                & 30,441,325                                                                                   \\
\textbf{/pol/}    & 48,725,043    & 13,190,390               & 4,325,648                                 & 3,626,184                                                                                    \\
\textbf{Gab}      &   12,395,575       &  955,440         & 235,222                                   & 193,783                                                                                      \\
\textbf{KYM}      & 15,584             & 15,584      & 706,940                                   & 597,060                                                                                      \\
 \bottomrule
\end{tabular}
\caption{Overview of our datasets.}
\label{tbl:datasets_summary}

\end{table}

Table~\ref{tbl:datasets_summary} reports the number of posts and images processed for each community.
Note that the number of images is lower than the number of posts with images because of duplicate image URLs and because some images get deleted.
Next, we discuss each dataset.

\descr{Twitter.} Twitter is a mainstream microblogging platform, allowing users to broadcast 280-character messages (tweets) to their followers.
Our Twitter dataset is based on tweets made available via the 1\% Streaming API,  between July 1, 2016 and July 31, 2017.
In total, we parse 1.4B tweets: 242M of them have at least one image.
We extract all the images, ultimately collecting 114M images yielding 74M unique pHashes.

\descr{Reddit.} Reddit is a news aggregator: users create submissions by posting a URL and others can reply in a structured way.
It is divided into multiple sub-communities called subreddits, each with its own topic and moderation policy.
Content popularity and ranking are determined via a voting system based on the up- and down-votes that users cast. 
We gather images from Reddit using publicly available data from Pushshift~\cite{reddit_data_pushshift}.
We parse all submissions and comments\footnote{See \cite{reddit_json_docs} for metadata associated with submissions and comments.} between July 1, 2016 and July, 31 2017, and extract 62M posts that contain at least one image.
We then download 40M images producing 30M unique pHashes.

\descr{4chan.} 4chan is an anonymous image board; users create new threads by posting an image with some text, which others can reply to.
It lacks many of the traditional social networking features like sharing or liking content, but has two characteristic features: anonymity and ephemerality.
By default, user identities are concealed and messages by the same users are not linkable across threads,
and all threads are deleted after one week. 
Overall, 4chan is known for its extremely lax moderation and the high degree of hate and racism, especially on boards like \dspol~\cite{hine2017kek}.
We obtain all threads posted on \dspol, between July 1, 2016 and July 31, 2017, using the same methodology of \cite{hine2017kek}.
Since all threads (and images) are removed after a week, we use a public archive service called 4plebs~\cite{4plebs_site} to collect 4.3M images, thus yielding 3.6M unique pHashes.

\descr{Gab.} Gab is a social network launched in August 2016 as a ``champion'' of free speech, providing ``shelter'' to users banned from other platforms.
It combines social networking features from Twitter (broadcast of 300-character messages) and Reddit (content is ranked according to up- and down-votes).
It also has extremely lax moderation as it allows everything except illegal pornography, terrorist propaganda, and doxing~\cite{snyder2017fifteen}. 
Overall, Gab attracts alt-right users, conspiracy theorists, and trolls, and high volumes of hate speech~\cite{zannettou2018what}.
We collect 12M posts, posted on Gab between August 10, 2016 and July 31, 2017, and 955K posts have at least one image, using the same methodology as in~\cite{zannettou2018what}.
Out of these, 235K images are unique, producing 193K unique pHashes.
Note that our Gab dataset starts one month later than the other ones, since Gab was launched in August 2016.

\descr{Ethics}. Although we only collect publicly available data, our study has been approved by the designated ethics officer at UCL.
Since 4chan content is typically posted with expectations of anonymity, we note that we have followed standard ethical guidelines~\cite{rivers2014ethical} and
encrypted data at rest, while making no attempt to de-anonymize users.

\subsection{Meme Annotation Site}
\label{sec:analysis:kym}

\descr{Know Your Meme (KYM).} We choose KYM as the source for meme annotation as it offers a comprehensive database of memes.
KYM is a sort of encyclopedia of Internet memes: for each meme, it provides information such as its origin (i.e., the platform on which it was first observed), the year it started, as well as descriptions and examples.
In addition, for each entry, KYM provides a set of keywords, called {\em tags}, that describe the entry.
Also, KYM provides a variety of higher-level categories that group meme entries; namely, cultures, subcultures, people, events, and sites.
``Cultures'' and ``subcultures'' entries refer to a wide variety of topics ranging from video games to various general categories.
For example, the Rage Comics {\em subculture}~\cite{kym_rage_comics_subculture} is a higher level category associated with memes related to comics like Rage Guy~\cite{rage_guy} or LOL Guy~\cite{lol_guy}, while the Alt-right {\em culture}~\cite{alt_right_culture} gathers entries from a loosely defined segment of the right-wing community.
The rest of the categories refer to specific individuals (e.g., Donald Trump~\cite{donald_trump_meme}), specific \em{events} (e.g.,\#CNNBlackmail~\cite{cnnblackmail_meme}), and \em{sites}  (e.g., \dspol~\cite{pol_kym}), respectively. It is also worth noting that KYM moderates all entries, hence entries that are wrong or incomplete are marked as so by the site.

As of May 2018, the site has 18.3K entries, specifically, 14K memes, 1.3K subcultures,  1.2K people, 1.3K events, and 427 websites~\cite{kym_memes_summary}.
We crawl KYM between October and December 2017, acquiring data for 15.6K entries; %
for each entry, we also download all the images related to it by crawling all the pages of the image gallery.
In total, we collect 707K images corresponding to 597K unique pHashes.
Note that we obtain 15.6K out of 18.3K entries, as we crawled the site several months before May 2018.

\begin{figure*}[t]
\centering
\subfigure[Categories]{\includegraphics[width=0.32\textwidth]{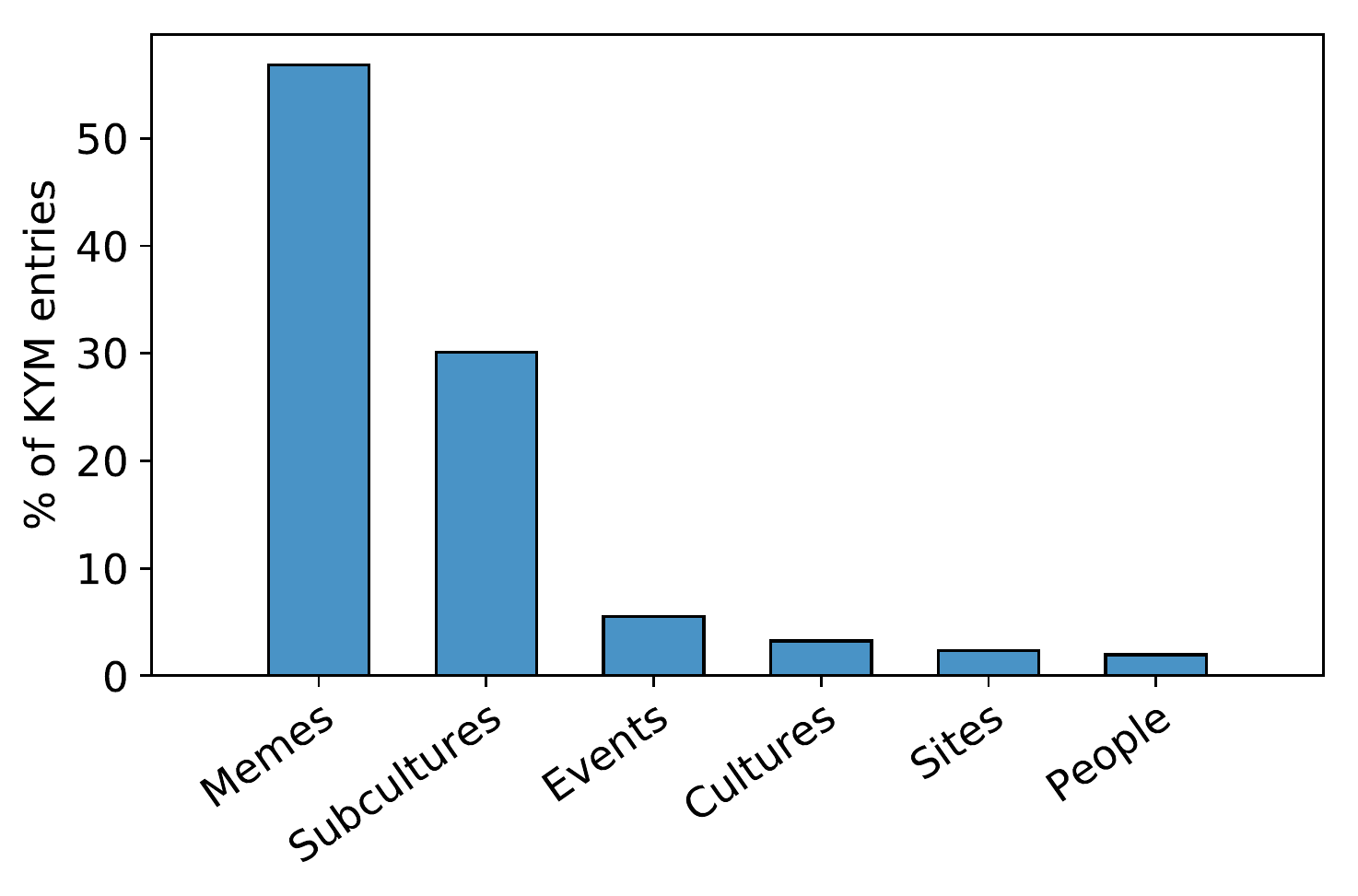}\label{fig:bc_kym_entry_categories}}~
\subfigure[Images]{\includegraphics[width=0.315\textwidth]{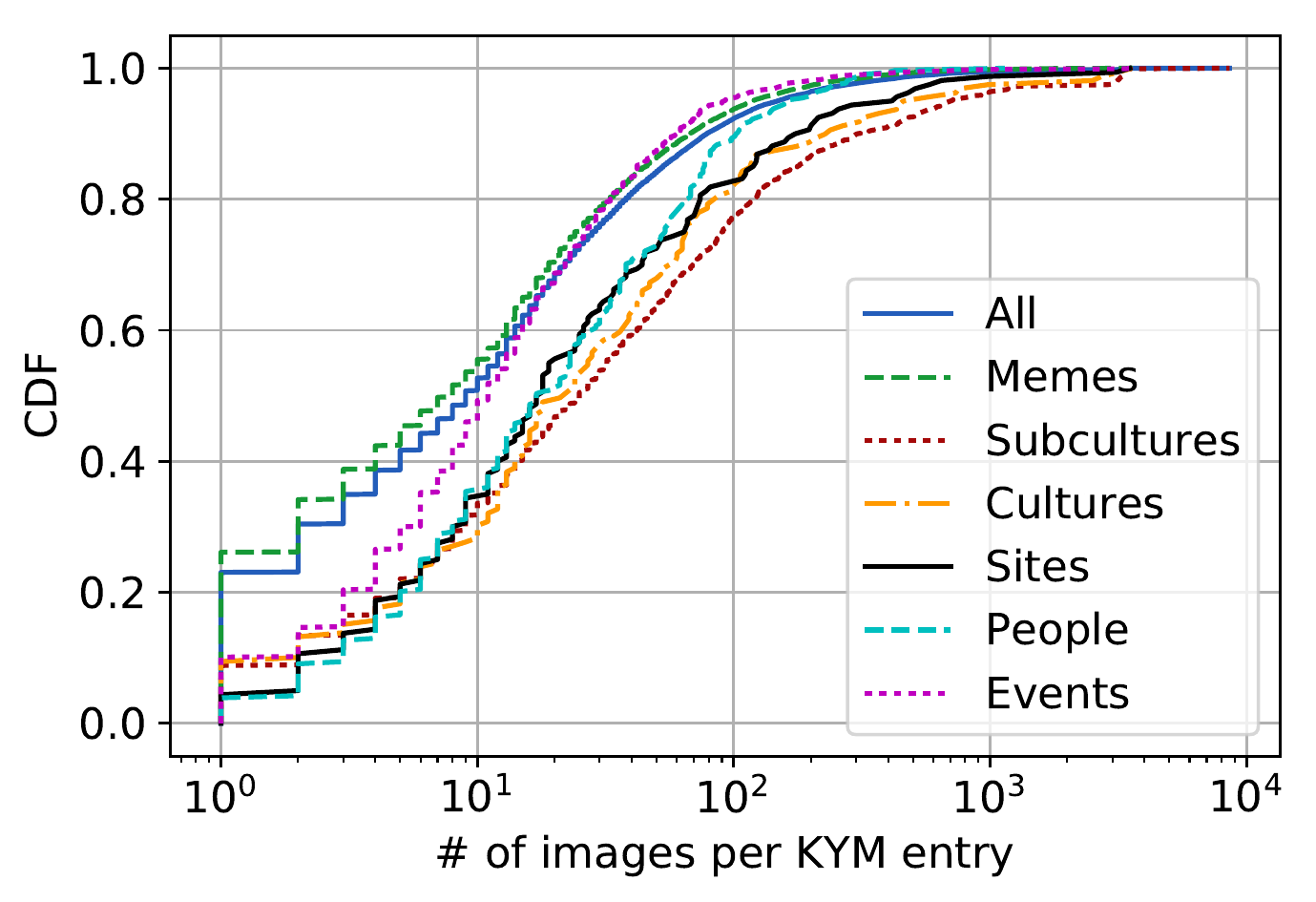}\label{fig:stats_kym_images}}~
\subfigure[Origins]{\includegraphics[width=0.32\textwidth]{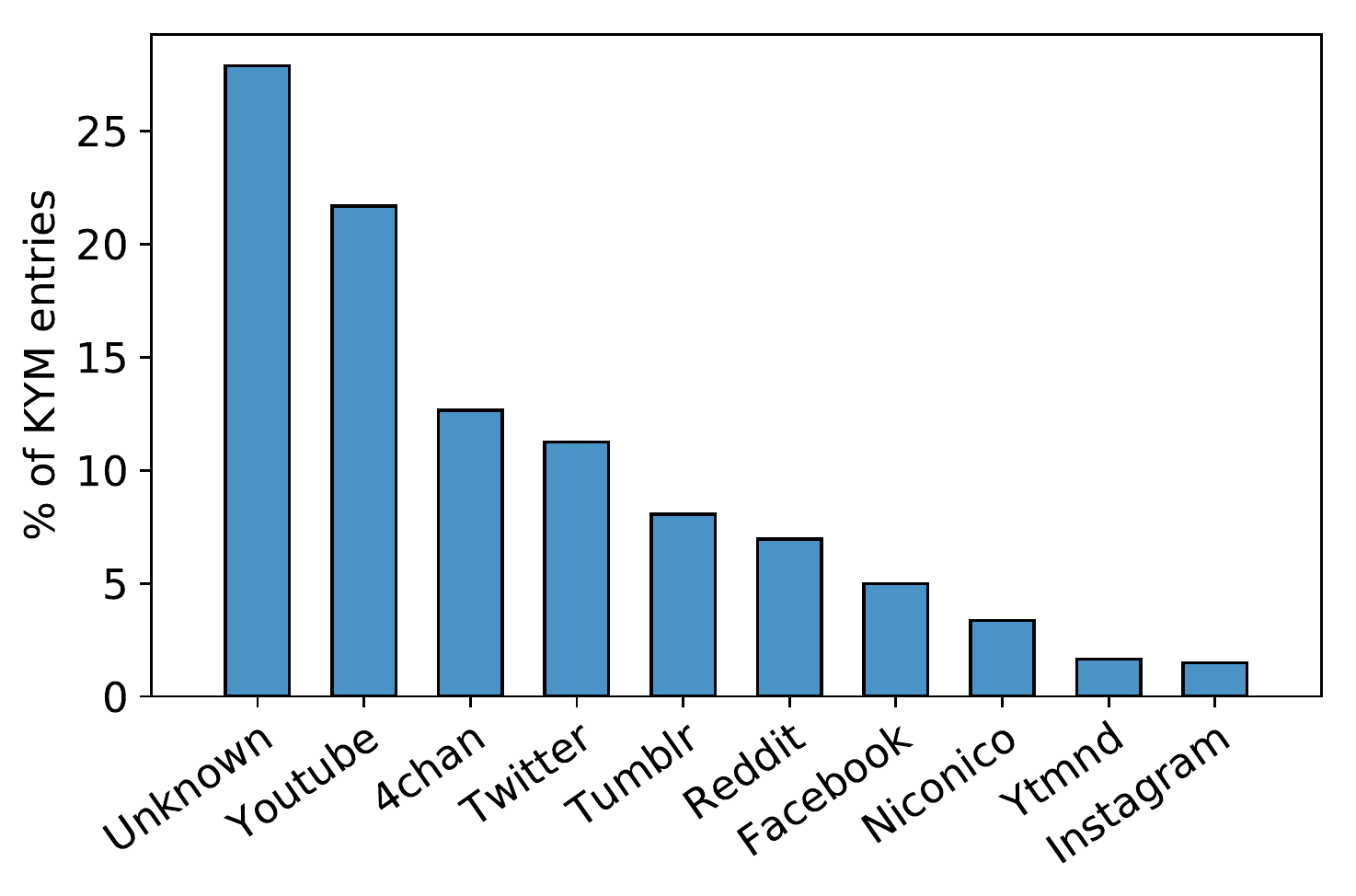}\label{fig:bc_origins_memes}}
\caption{Basic statistics of the KYM dataset.}
\label{fig:kym_overview}
\end{figure*}

\descr{Getting to know KYM.} We also perform a general characterization of KYM. %
First, we look at the distribution of entries across categories: as shown in Figure~\ref{fig:bc_kym_entry_categories}, %
as expected, the majority (57\%) are memes, followed by subcultures (30\%), cultures (3\%), websites (2\%), and people (2\%).

Next, we measure the number of images per entry: as shown in Figure~\ref{fig:stats_kym_images}, this varies considerably (note log-scale on x-axis).
KYM entries have as few as 1 and as many as 8K images, with an average of 45 and a median of 9 images.
Larger values may be related to the meme's popularity, but also to the ``diversity'' of image variants it generates.
Upon manual inspection, we find that the presence of a large number of images for the same meme happens either when images are visually very similar to each other (e.g., Smug Frog images {\em within} the two clusters in Figure~\ref{fig:smug_frog_example}), or if there are actually remarkably different variants of the same meme (e.g., images in `cluster 1' {\em vs.} images in `cluster N' in the same figure).
We also note that the distribution varies according to the category: e.g., higher-level concepts like cultures include more images than more specific entries like memes.

We then analyze the origin of each entry: see Figure~\ref{fig:bc_origins_memes}.
Note that a large portion of the memes (28\%) have an unknown origin, while YouTube, 4chan, and Twitter are the most popular platforms with, respectively, 21\%, 12\%, and 11\%, followed by Tumblr and Reddit with 8\% and 7\%.
This confirms our intuition that 4chan, Twitter, and Reddit, which are among our data sources, play an important role in the generation and dissemination of memes.
As mentioned, we do not currently study video memes originating from YouTube, due to the inherent complexity of video-processing tasks as well as scalability issues. However, a large portion of YouTube memes actually end up being morphed into image-based memes (see, e.g., the Overly Attached Girlfriend meme~\cite{overly_attached_girlfriend}).

\subsection{Running the pipeline on our datasets}
For all four Web communities (Twitter, Reddit, \dspol, and Gab), we perform Step 1 of the pipeline (Figure~\ref{fig:pipeline}), using the ImageHash library.\footnote{\url{https://github.com/JohannesBuchner/imagehash}}%
After computing the pHashes, we delete the images (i.e., we only keep the associated URL and pHash) due to space limitations of our infrastructure. 
We then perform Steps 2-3 (i.e., pairwise comparisons between all images and clustering), for all the images from \dspol, \td subreddit, and Gab, as we treat them as fringe Web communities.
Note that, we exclude mainstream communities like the rest of Reddit and Twitter as our main goal is to obtain clusters of memes from fringe Web communities and later characterize all communities by means of the clusters.
Next, we go through Steps 4-5 using all the images obtained from meme annotation websites (specifically, Know Your Meme, see Section~\ref{sec:analysis:kym}) and the medoid of each cluster from \dspol, \td, and Gab.
Finally, Steps 6-7 use all the pHashes obtained from Twitter, Reddit (all subreddits), \dspol, and Gab to find posts with images matching the annotated clusters.
This is an integral part of our process as it allows to characterize and study mainstream Web communities not used for clustering (i.e., Twitter and Reddit).

\section{Analysis} \label{sec:analysis}
In this section, we present a cluster-based measurement of memes and an analysis of a few Web communities from the ``perspective'' of memes.
We measure the prevalence of memes across the clusters obtained from fringe communities: \dspol, \td subreddit (\tdshort), and Gab.
We also use the distance metric introduced in Eq.~\ref{eq:distance} to perform a \emph{cross-community} analysis,
then, we group clusters into broad, but related, categories to gain a macro-perspective understanding of larger communities, including  Reddit and Twitter.

\subsection{Cluster-based Analysis}
\label{sec:analysis:clusters}
We start by analyzing the 12.6K annotated clusters consisting of 268K images from \dspol, \td, and Gab (Step 5 in Figure~\ref{fig:pipeline}).
We do so to understand the \emph{diversity} of memes in each Web community, as well as the interplay between \emph{variants} of memes. 
We then evaluate how clusters can be grouped into higher structures using hierarchical clustering and graph visualization techniques.

\begin{table}[t]
\centering
\small
\begin{tabular}{@{}lrrrr@{}}
\toprule
\textbf{Platform} & \textbf{\#Images} & \textbf{Noise} & \textbf{\#Clusters} & \hspace*{-0.2cm}\textbf{\#Clusters with}\\
& & & & \hspace*{-0.1cm}\textbf{KYM tags (\%)}\\
\midrule
\textbf{/pol/}    & 4,325,648                                                                            & 63\%                                                                               & 38,851                                                                                 & 9,265 (24\%)                                                                                              \\
\textbf{\tdshort}     & 1,234,940                                                                            & 64\%                                                                               & 21,917                                                                                 & 2,902 (13\%)                                                                                                \\
\textbf{Gab}      & 235,222                                                                              & 69\%                                                                               & 3,083                                                                                  & 447 (15\%)                                                                                                  \\ \bottomrule
\end{tabular}
\caption{Statistics obtained from clustering images from \dspol, \td, and Gab.}
\label{tbl:clustering-statistics}
\vspace{-0.2cm}
\end{table}

\begin{figure*}[t]
\center
\subfigure[]{\includegraphics[width=0.37\textwidth]{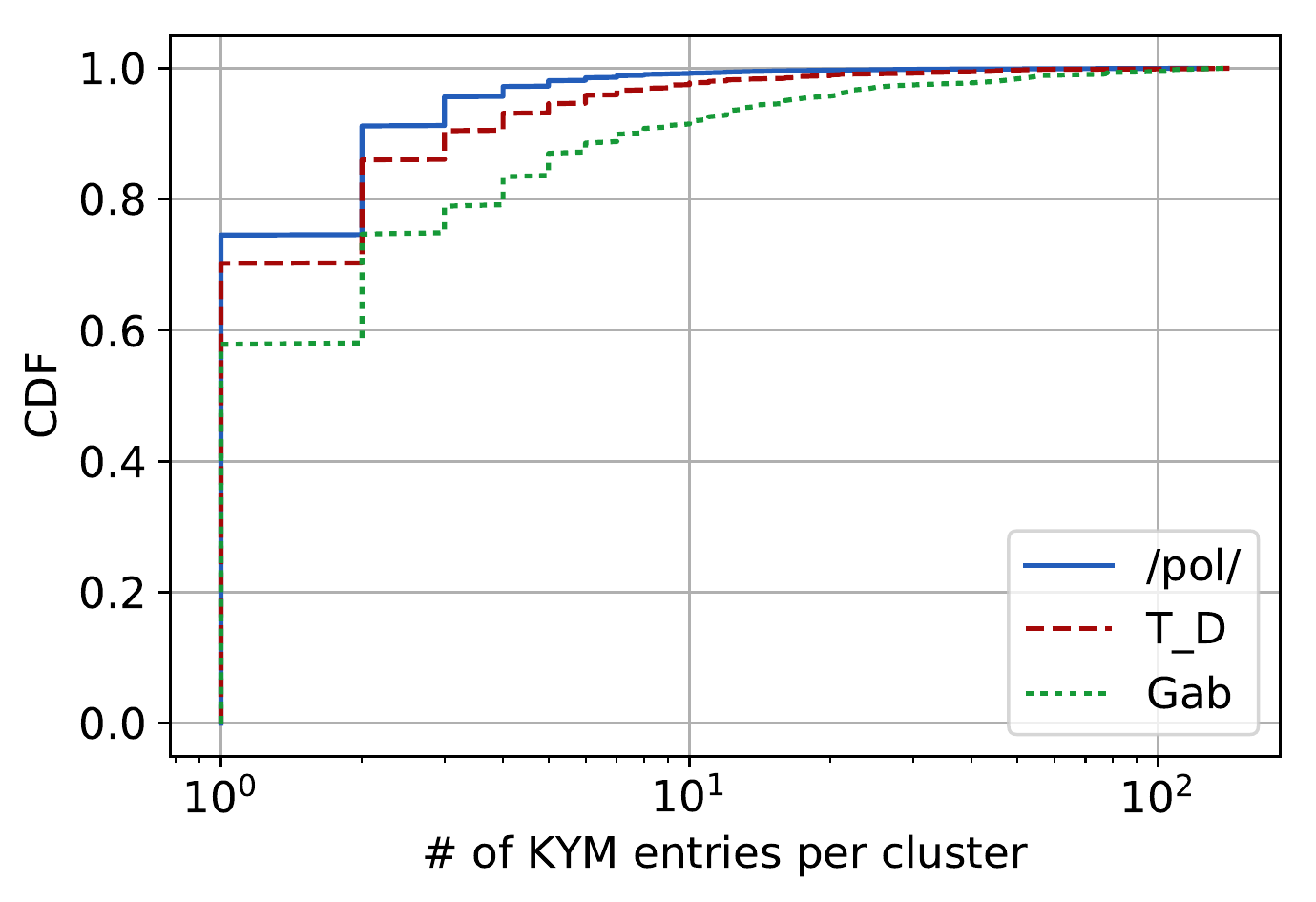}\label{subfig:cdf_kym_entries_per_cluster}} \hspace{0.5cm}
\subfigure[]{\includegraphics[width=0.37\textwidth]{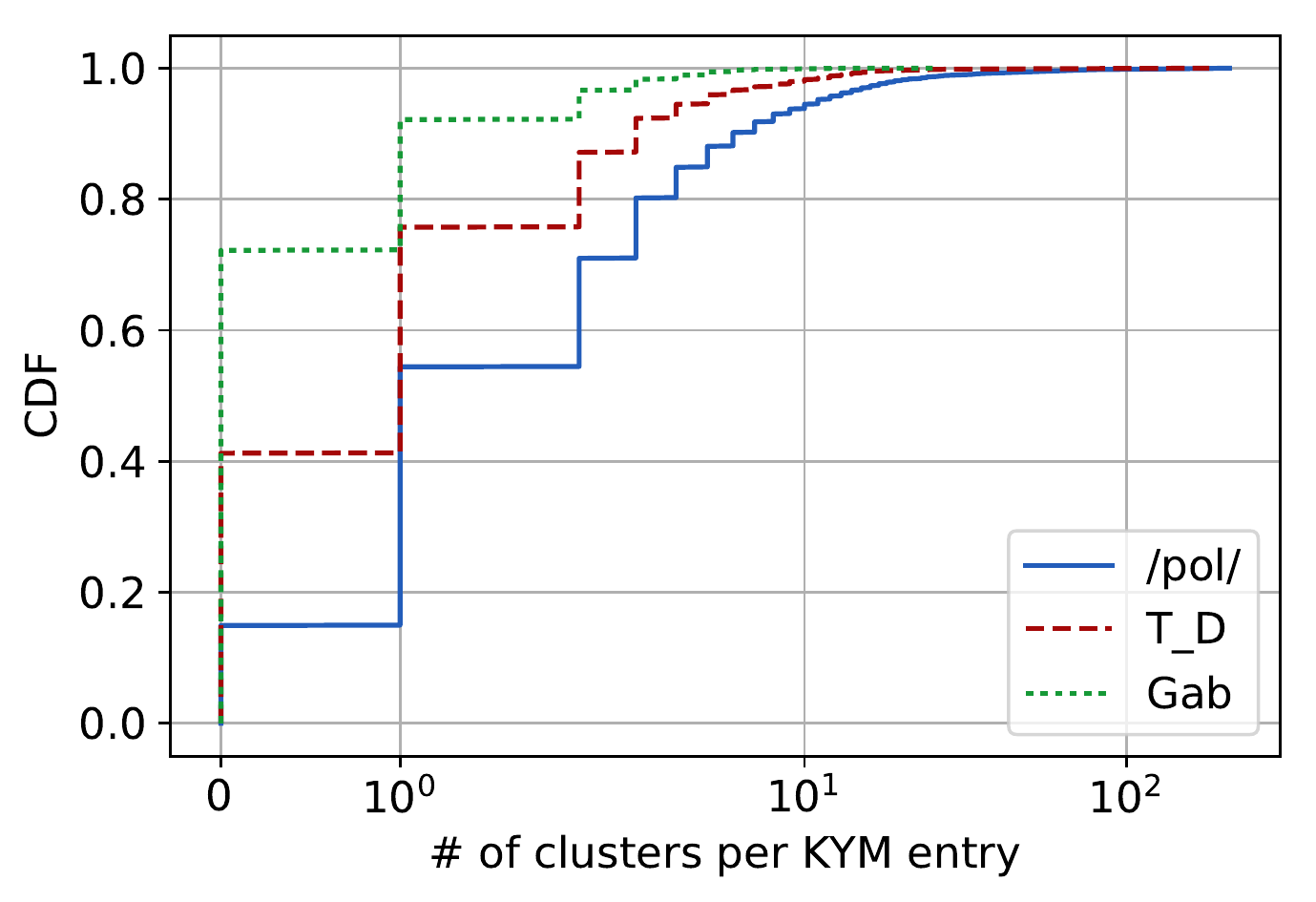}\label{subfig:cdf_clusters_per_kym_entry}}
\caption{CDF of KYM entries per cluster (a) and clusters per KYM entry (b). }
\label{fig:cdf_clusters_kym_entries}
\end{figure*}

\subsubsection{Clusters}

\noindent\textbf{Statistics.} In Table~\ref{tbl:clustering-statistics}, we report some basic statistics of the clusters obtained for each Web community.
A relatively high percentage of images (63\%--69\%) are not clustered, i.e., are labeled as noise.
While in DBSCAN ``noise'' is just an instance that does not fit in any cluster (more specifically, there are less than 5 images with perceptual distance $\leq 8$ from that particular instance), we note that this likely happens as these images are not memes, but rather ``one-off images.''
For example, on \dspol there is a large number of pictures of random people taken from various social media platforms.

Overall, we have 2.1M images in 63.9K clusters: 38K clusters for \dspol, 21K for \td, and 3K for Gab.
12.6K of these clusters are successfully annotated using the KYM data: 
9.2K from \dspol (142K images), 2.9K from \td (121K images), and 447 from Gab (4.5K images).
Examples of clusters are reported in Appendix~\ref{sec:appendix_clusters}. 
As for the un-annotated clusters, manual inspection confirms that many include miscellaneous images unrelated to memes, e.g., similar screenshots of social networks posts (recall that we only filter out screenshots from the KYM image galleries), images captured from video games, etc.

\begin{table*}[t!]
\centering
\setlength{\tabcolsep}{0.18em} %
\resizebox{0.999\textwidth}{!}{%
\begin{tabular}{@{}llrllrllr@{}}
\toprule
\multicolumn{3}{c|}{\textbf{/pol/}}                                                                                                                                                                    & \multicolumn{3}{c|}{\textbf{\tdshort}}                                                                                                                                                     & \multicolumn{3}{c}{\textbf{Gab}}                                                                                                                                          \\ \midrule
\textbf{Entry} & {\bf Category} & \multicolumn{1}{l|}{\textbf{Clusters (\%)}} &
\textbf{Entry} & {\bf Category} & \multicolumn{1}{l|}{\textbf{Clusters (\%)}} &
\textbf{Entry} & {\bf Category} & \multicolumn{1}{l|}{\textbf{Clusters (\%)}} \\ \midrule
\href{http://knowyourmeme.com/memes/people/donald-trump}{Donald Trump}                                                                     & People                                                             & \multicolumn{1}{r|}{207 (2.2\%)}            & \href{http://knowyourmeme.com/memes/people/donald-trump}{Donald Trump}                                                      & People                                                             & \multicolumn{1}{r|}{177 (6.1\%)}            & \href{http://knowyourmeme.com/memes/people/donald-trump}{Donald Trump}                                                                 & People                                                            & 25 (5.6\%)             \\
\href{http://knowyourmeme.com/memes/happy-merchant}{Happy Merchant}                                                                   & Memes                                                              & \multicolumn{1}{r|}{124 (1.3\%)}            & \href{http://knowyourmeme.com/memes/smug-frog}{Smug Frog}                                                           & Memes                                                              & \multicolumn{1}{r|}{78 (2.7\%)}             & \href{http://knowyourmeme.com/memes/happy-merchant}{Happy Merchant}                                                                & Memes                                                             & 10 (2.2\%)             \\
\href{http://knowyourmeme.com/memes/smug-frog}{Smug Frog}                                                                        & Memes                                                              & \multicolumn{1}{r|}{114 (1.2\%)}            & \href{http://knowyourmeme.com/memes/pepe-the-frog}{Pepe the Frog}                                                      & Memes                                                              & \multicolumn{1}{r|}{63 (2.1\%)}             & \href{http://knowyourmeme.com/memes/demotivational-posters}{Demotivational Posters}                                                        & Memes                                                             & 7 (1.5\%)              \\
\href{http://knowyourmeme.com/memes/computer-reaction-faces}{Computer Reaction Faces}                                                          & Memes                                                              & \multicolumn{1}{r|}{112 (1.2\%)}            & \href{http://knowyourmeme.com/memes/feels-bad-man-sad-frog}{Feels Bad Man/ Sad Frog} & Memes                                                              & \multicolumn{1}{r|}{61 (2.1\%)}             & \href{http://knowyourmeme.com/memes/pepe-the-frog}{Pepe the Frog}                                                                  & Memes                                                             & 6 (1.3\%)              \\
\href{http://knowyourmeme.com/memes/feels-bad-man-sad-frog}{Feels Bad Man/ Sad Frog}              & Memes                                                              & \multicolumn{1}{r|}{94 (1.0\%)}             & \href{http://knowyourmeme.com/memes/make-america-great-again}{Make America Great Again}  & Memes                                                              & \multicolumn{1}{r|}{50 (1.7\%)}             & \href{http://knowyourmeme.com/memes/events/cnnblackmail}{\#Cnnblackmail}                                                               & Events                                                            & 6 (1.3\%)              \\
\href{http://knowyourmeme.com/memes/i-know-that-feel-bro}{I Know that Feel Bro}                                                             & Memes                                                              & \multicolumn{1}{r|}{90 (1.0\%)}             & \href{http://knowyourmeme.com/memes/people/bernie-sanders}{Bernie Sanders}                                                      & People                                                             & \multicolumn{1}{r|}{31 (1.0\%)}             & \href{http://knowyourmeme.com/memes/events/2016-united-states-presidential-election}{2016 US election}                                                             & Events                                                            & 6 (1.3\%)              \\
\href{http://knowyourmeme.com/memes/tony-kornheiser-s-why}{Tony Kornheiser's Why}                                                            & Memes                                                              & \multicolumn{1}{r|}{89 (1.0\%)}             & \href{http://knowyourmeme.com/memes/events/2016-united-states-presidential-election}{2016 US Election}                                                   & Events                                                             & \multicolumn{1}{r|}{27 (0.9\%)}             & \href{http://knowyourmeme.com/memes/sites/knowyourmeme}{Know Your Meme}                                                               & Sites                                                             & 6 (1.3\%)              \\
\href{http://knowyourmeme.com/memes/bait-this-is-bait}{Bait/This is Bait}                                                                & Memes                                                              & \multicolumn{1}{r|}{84 (0.9\%)}             & \href{http://knowyourmeme.com/memes/counter-signal-memes}{Counter Signal Memes}   & Memes                                                              & \multicolumn{1}{r|}{24 (0.8\%)}             & \href{http://knowyourmeme.com/memes/sites/tumblr}{Tumblr}                                                                       & Sites                                                             & 6 (1.3\%)              \\
\href{http://knowyourmeme.com/memes/events/trumpanime-rick-wilson-controversy}{\#TrumpAnime/Rick Wilson} & Events                                                             & \multicolumn{1}{r|}{76 (0.8\%)}             & \href{http://knowyourmeme.com/memes/events/cnnblackmail}{\#Cnnblackmail}                                                     & Events                                                             & \multicolumn{1}{r|}{24 (0.8\%)}             & \href{http://knowyourmeme.com/memes/cultures/feminism}{Feminism}                                                                     & Cultures                                                          & 5 (1.1\%)              \\
\href{http://knowyourmeme.com/memes/reaction-images}{Reaction Images}                                                                  & Memes                                                              & \multicolumn{1}{r|}{73 (0.8\%)}             & \href{http://knowyourmeme.com/memes/sites/knowyourmeme}{Know Your Meme}                                                     & Sites                                                              & \multicolumn{1}{r|}{20 (0.7\%)}             & \href{http://knowyourmeme.com/memes/people/barack-obama}{Barack Obama}                                                                 & People                                                            & 5 (1.1\%)              \\
\href{http://knowyourmeme.com/memes/make-america-great-again}{Make America Great Again}               & Memes                                                              & \multicolumn{1}{r|}{72 (0.8\%)}             & \href{http://knowyourmeme.com/memes/angry-pepe}{Angry Pepe}                                                         & Memes                                                              & \multicolumn{1}{r|}{18 (0.6\%)}             & \href{http://knowyourmeme.com/memes/smug-frog}{Smug Frog}                                                                     & Memes                                                             & 5 (1.1\%)              \\
\href{http://knowyourmeme.com/memes/counter-signal-memes}{Counter Signal Memes}               & Memes                                                              & \multicolumn{1}{r|}{72 (0.8\%)}             & \href{http://knowyourmeme.com/memes/demotivational-posters}{Demotivational Posters}                                             & Memes                                                              & \multicolumn{1}{r|}{18 (0.6\%)}             & \href{http://knowyourmeme.com/memes/subcultures/rwby}{rwby}                                                                         & Subcultures                                                       & 5 (1.1\%)              \\
\href{http://knowyourmeme.com/memes/pepe-the-frog}{Pepe the Frog}                                                                      & Memes                                                              & \multicolumn{1}{r|}{65 (0.7\%)}             & \href{http://knowyourmeme.com/memes/sites/4chan}{4chan}                                                              & Sites                                                              & \multicolumn{1}{r|}{16 (0.5\%)}             & \href{http://knowyourmeme.com/memes/people/kim-jong-un}{Kim Jong Un}                                                                  & People                                                            & 5 (1.1\%)              \\
\href{http://knowyourmeme.com/memes/subcultures/spongebob-squarepants}{Spongebob Squarepants}                                                            & Subcultures                                                        & \multicolumn{1}{r|}{61 (0.7\%)}             & \href{http://knowyourmeme.com/memes/sites/tumblr}{Tumblr}                                                             & Sites                                                              & \multicolumn{1}{r|}{15 (0.5\%)}             & \href{http://knowyourmeme.com/memes/murica}{Murica}                                                                       & Memes                                                             & 5 (1.1\%)              \\
\href{http://knowyourmeme.com/memes/doom-paul-its-happening}{Doom Paul its Happening}                                                          & Memes                                                              & \multicolumn{1}{r|}{57 (0.6\%)}             & \href{http://knowyourmeme.com/memes/events/gamergate}{Gamergate}                                                          & Events                                                             & \multicolumn{1}{r|}{15 (0.5\%)}             & \href{http://knowyourmeme.com/memes/events/united-airlines-passenger-removal}{UA Passenger Removal} & Events                                                            & 5 (1.1\%)              \\
\href{http://knowyourmeme.com/memes/people/adolf-hitler}{Adolf Hitler}                                                                     & People                                                             & \multicolumn{1}{r|}{56 (0.6\%)}             & \href{http://knowyourmeme.com/memes/colbertposting}{Colbertposting}                                                     & Memes                                                              & \multicolumn{1}{r|}{15 (0.5\%)}              & \href{http://knowyourmeme.com/memes/make-america-great-again}{Make America Great Again}            & Memes                                                             & 4 (0.9\%)              \\
\href{http://knowyourmeme.com/memes/sites/pol}{pol}                                                                              & Sites                                                              & \multicolumn{1}{r|}{53 (0.6\%)}             & \href{http://knowyourmeme.com/memes/donald-trumps-wall}{Donald Trump's Wall}                                                & Memes                                                              & \multicolumn{1}{r|}{15 (0.5\%)}              & \href{http://knowyourmeme.com/memes/people/bill-nye}{Bill Nye}                                                                     & People                                                            & 4 (0.9\%)              \\
\href{http://knowyourmeme.com/memes/dubs-guy-check-em}{Dubs Guy/Check'em}                     & Memes                                                              & \multicolumn{1}{r|}{53 (0.6\%)}             & \href{http://knowyourmeme.com/memes/people/vladimir-putin}{Vladimir Putin}                                                     & People                                                             & \multicolumn{1}{r|}{15 (0.5\%)}              & \href{http://knowyourmeme.com/memes/cultures/trolling}{Trolling}                                                                     & Cultures                                                          & 4 (0.9\%)              \\
\href{http://knowyourmeme.com/memes/smug-anime-face}{Smug Anime Face}                                                                  & Memes                                                              & \multicolumn{1}{r|}{51 (0.6\%)}             & \href{http://knowyourmeme.com/memes/people/barack-obama}{Barack Obama}                                                        & People                                                             & \multicolumn{1}{r|}{15 (0.5\%)}              & \href{http://knowyourmeme.com/memes/sites/4chan}{4chan}                                                                        & Sites                                                             & 4 (0.9\%)              \\
\href{http://knowyourmeme.com/memes/subcultures/warhammer-40000}{Warhammer 40000}                                                                & Subcultures                                                        & \multicolumn{1}{r|}{51 (0.6\%)}             & \href{http://knowyourmeme.com/memes/people/hillary-clinton}{Hillary Clinton}                                                    & People                                                             & \multicolumn{1}{r|}{15 (0.5\%)}              & \href{http://knowyourmeme.com/memes/cultures/furries}{Furries}                                                                      & Cultures                                                          & 3 (0.7\%)              \\
\midrule
{\bf Total} && \multicolumn{1}{r|}{\bf 1,638 (17.7\%)} &&& \multicolumn{1}{r|}{\bf 695 (23.9\%)} &&& \multicolumn{1}{r|}{\bf 121 (27.1\%)} \\
\bottomrule
\end{tabular}%
}
\caption{Top 20 KYM entries appearing in the clusters of \dspol, \td, and Gab. We report the number of clusters and their respective percentage (per community).
Each item contains a hyperlink to the corresponding entry on the KYM website.}
\label{tbl:top_entries_cluster_numbers}
\end{table*}

\descr{KYM entries per cluster.} Each cluster may receive multiple annotations, depending on the KYM entries that have at least one image matching that cluster's medoid.
As shown in Figure~\ref{subfig:cdf_kym_entries_per_cluster}, the majority of the annotated clusters (74\% for \dspol, 70\% for \td, and 58\% for Gab) only have a single matching KYM entry.
However, a few clusters have a large number of matching entries, e.g., the one matching the Conspiracy Keanu meme~\cite{conspiracy_keanu} is annotated by 126 KYM entries (primarily, other memes that add text in an image associated with that meme).
This highlights that memes do overlap and that some are highly influenced by other ones.

\descr{Clusters per KYM entry.} We also look at the number of clusters annotated by the \emph{same} KYM entry.
Figure~\ref{subfig:cdf_clusters_per_kym_entry} plots the CDF of the number of clusters per entry.
About 40\% only annotate a single \dspol cluster, while 34\% and 20\% of the entries annotate a single \td and a single Gab cluster, respectively.
We also find that a small number of entries are associated to a large number of clusters: for example, the Happy Merchant meme~\cite{happy_merchant_meme} annotates 124 different clusters on \dspol.
This highlights the \emph{diverse} nature of memes, i.e., memes are mixed and matched, not unlike the way that genetic traits are combined in biological reproduction.

\begin{figure*}[t]
\vspace{0.2cm}
\centering
\stackunder[1pt]{\includegraphics[height=1.5cm]{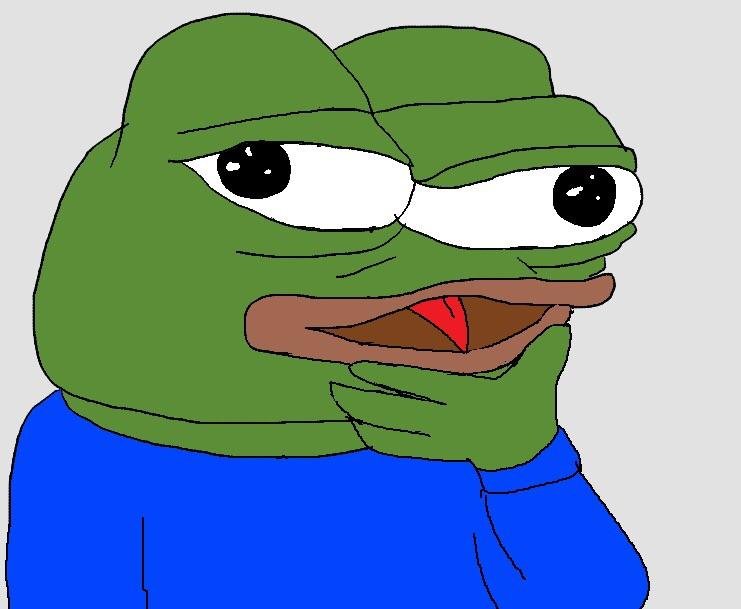}}{apustaja}~%
\stackunder[1pt]{\includegraphics[height=1.5cm]{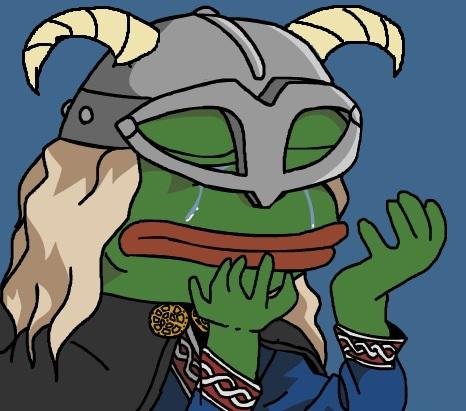}}{sad-frog}~~%
\stackunder[1pt]{\includegraphics[height=1.5cm]{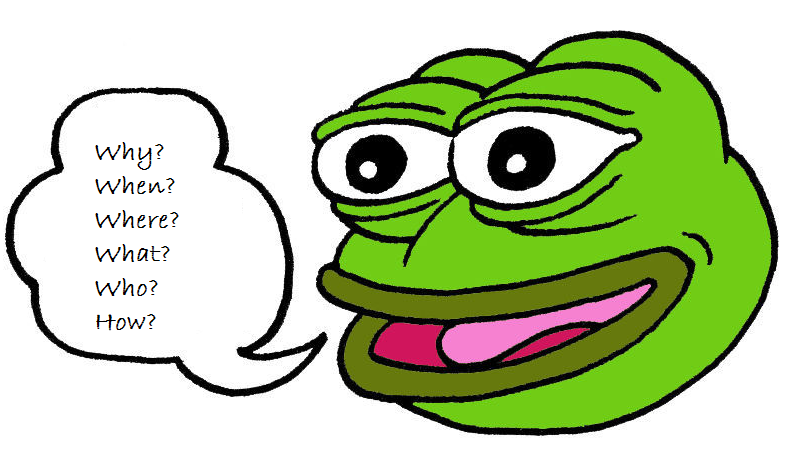}}{savepepe}~~%
\stackunder[1pt]{\includegraphics[height=1.5cm]{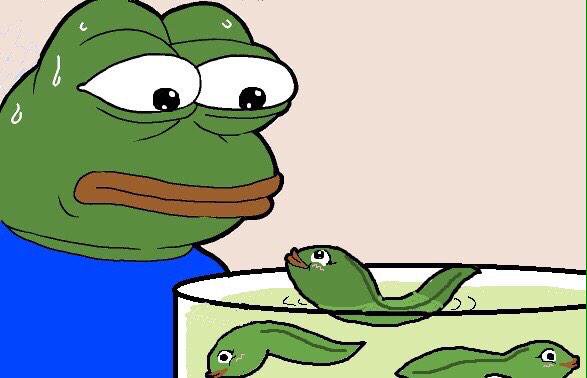}}{pepe}~~%
\stackunder[1pt]{\includegraphics[height=1.5cm]{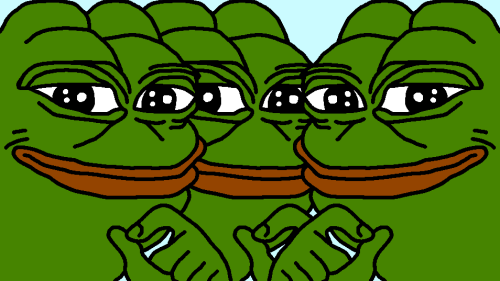}}{smug-frog-a}~~%
\stackunder[1pt]{\includegraphics[height=1.5cm]{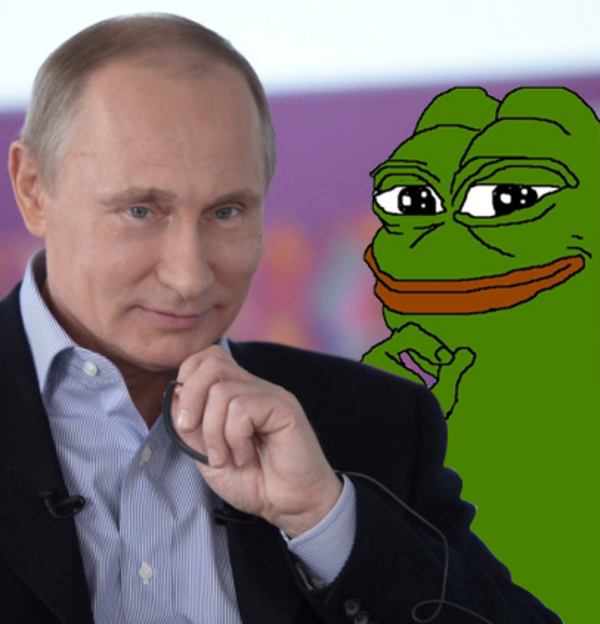}}{smug-frog-b}~~%
\stackunder[1pt]{\includegraphics[height=1.5cm]{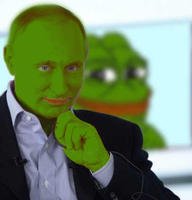}}{anti-meme}
\color{red}{\stackinset{l}{.4in}{b}{1.02in}{\rotatebox{0}{\rule{6.6in}{0.7pt}}}{
\subfigure{\includegraphics[width=1\textwidth, trim=255 200 255 85, clip]{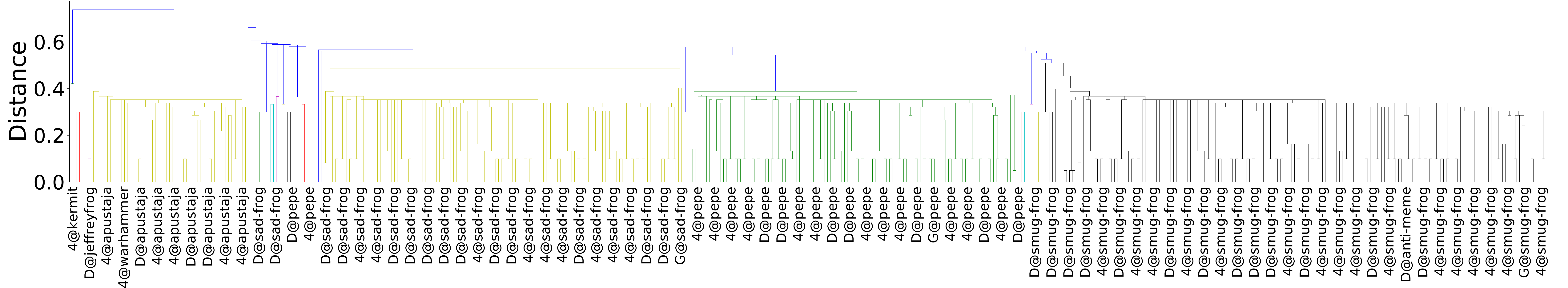}}}}
\vspace*{-.5cm}
  \caption{Inter-cluster distance between all clusters with frog memes. Clusters are labeled with the origin ({\em 4} for 4chan, {\em D} for \td, and {\em G} for Gab) and the meme name. To ease  readability, we do not display all labels, abbreviate meme names, and only show an excerpt of all relationships.}
\label{fig:casestudy_frogs_dendrogram}
\end{figure*}

\descr{Top KYM entries.}
Because the majority of clusters match only one or two KYM entries (Figure~\ref{subfig:cdf_kym_entries_per_cluster}), we simplify things by giving all clusters a \emph{representative annotation} based on the most prevalent annotation given to the medoid, and, in the case of ties the average distance between all matches (see Section~\ref{subsec:pipeline}).
\emph{Thus, in the rest of the paper, we report our findings based on the representative annotation for each cluster.}

In Table~\ref{tbl:top_entries_cluster_numbers}, we report the top 20 KYM entries with respect to the number of clusters they annotate.
These cover 17\%, 23\%, and 27\% of the clusters in \dspol, \td, and Gab, respectively, hence covering a relatively good sample of our datasets.
Donald Trump~\cite{donald_trump_meme}, Smug Frog~\cite{smug_frog_meme}, and Pepe the Frog~\cite{pepe_frog_meme} appear in the top 20 for all three communities, while the Happy Merchant~\cite{happy_merchant_meme} only in \dspol and Gab.
In particular, Donald Trump annotates the most clusters (207 in \dspol, 177 in \td, and 25 in Gab).
In fact, politics-related entries appear several times in the Table, 
e.g., Make America Great Again~\cite{maga_meme} as well as political personalities like Bernie Sanders, Barack Obama, Vladimir Putin, and Hillary Clinton.

When comparing the different communities, we observe the most prevalent categories are memes (6 to 14 entries in each community) and people (2-5).
Moreover, in \dspol, the 2nd most popular entry, related to people, is Adolf Hilter, which supports previous reports of the community's sympathetic views toward Nazi ideology~\cite{hine2017kek}.
Overall, there are several memes with hateful or disturbing content (e.g., holocaust).
This happens to a lesser extent in \td and Gab: the most popular people after Donald Trump are contemporary politicians, i.e., Bernie Sanders, Vladimir Putin, Barack Obama, and Hillary Clinton.

Finally, image posting behavior in fringe Web communities is greatly influenced by real-world events.
For instance, in \dspol, we find the \#TrumpAnime controversy event~\cite{trump_anime_meme}, where a political individual (Rick Wilson) offended the alt-right community, Donald Trump supporters, and anime fans (an oddly intersecting set of interests of \dspol users).
Similarly, on \td and Gab, we find the \#Cnnblackmail~\cite{cnnblackmail_meme} event, referring to the (alleged) blackmail of the Reddit user that created the infamous video of Donald Trump wrestling the CNN.

\subsubsection{Memes' Branching Nature}\label{subsection:hierarchy} 
Next, we study how memes \emph{evolve} by looking at {\em variants} across different clusters.
Intuitively, clusters that look alike and/or are part of the same meme are grouped together under the same branch of an evolutionary tree.
We use the custom distance metric introduced in Section~\ref{sec:methodology:distance}, aiming to infer the phylogenetic relationship between variants of memes.
Since there are 12.6K annotated clusters, we only report on a subset of variants. %
In particular, we focus on ``frog'' memes (e.g., Pepe the Frog~\cite{pepe_frog_meme}); as discussed later in Section~\ref{sec:analysis:annotations}, this is one of the most popular memes in our datasets.

The dendrogram in Figure~\ref{fig:casestudy_frogs_dendrogram} shows the hierarchical relationship between groups of clusters of memes related to frogs.
Overall, there are 525 clusters of frogs, belonging to 23 different memes.
These clusters can be grouped into four large categories, dominated by Apu Apustaja~\cite{apu_meme}, Feels Bad Man/Sad Frog~\cite{sad_frog_meme}, Pepe the Frog~\cite{pepe_frog_meme}, and Smug Frog~\cite{smug_frog_meme}.
The different memes express different ideas or messages: e.g., Apu Apustaja depicts a simple-minded non-native speaker using broken English, while the Feels Bad Man/Sad Frog (ironically) expresses dismay at a given situation, often accompanied with text like ``You will never do/be/have X.''
The dendrogram also shows a variant of Smug Frog ({\it smug-frog-b}) related to a variant of the Russian Anti Meme Law~\cite{russian_anti_meme} ({\it anti-meme}) as well as relationships between clusters from Pepe the Frog and Isis meme~\cite{isis_meme},
and between Smug Frog and Brexit-related clusters~\cite{brexit_meme}, as shown in Appendix~\ref{sec:appendix_interesting_images}.

The distance metric quantifies the similarity of any two variants of {\em different} memes; however, recall that two clusters can be close to each other even when the medoids are perceptually different (see Section~\ref{sec:methodology:distance}), as in the case of Smug Frog variants in the {\it smug-frog-a} and {\it smug-frog-b} clusters (top of Figure~\ref{fig:casestudy_frogs_dendrogram}).
Although, due to space constraints, this analysis is limited to a single ``family'' of memes, our distance metric can actually provide useful insights regarding the phylogenetic relationships of any clusters.
In fact, more extensive analysis of these relationships (through our pipeline) can facilitate the understanding of the diffusion of ideas and information across the Web, and provide a rigorous technique for large-scale analysis of Internet culture.

\begin{figure*}[t!]
\centering
\includegraphics[width=0.85\textwidth]{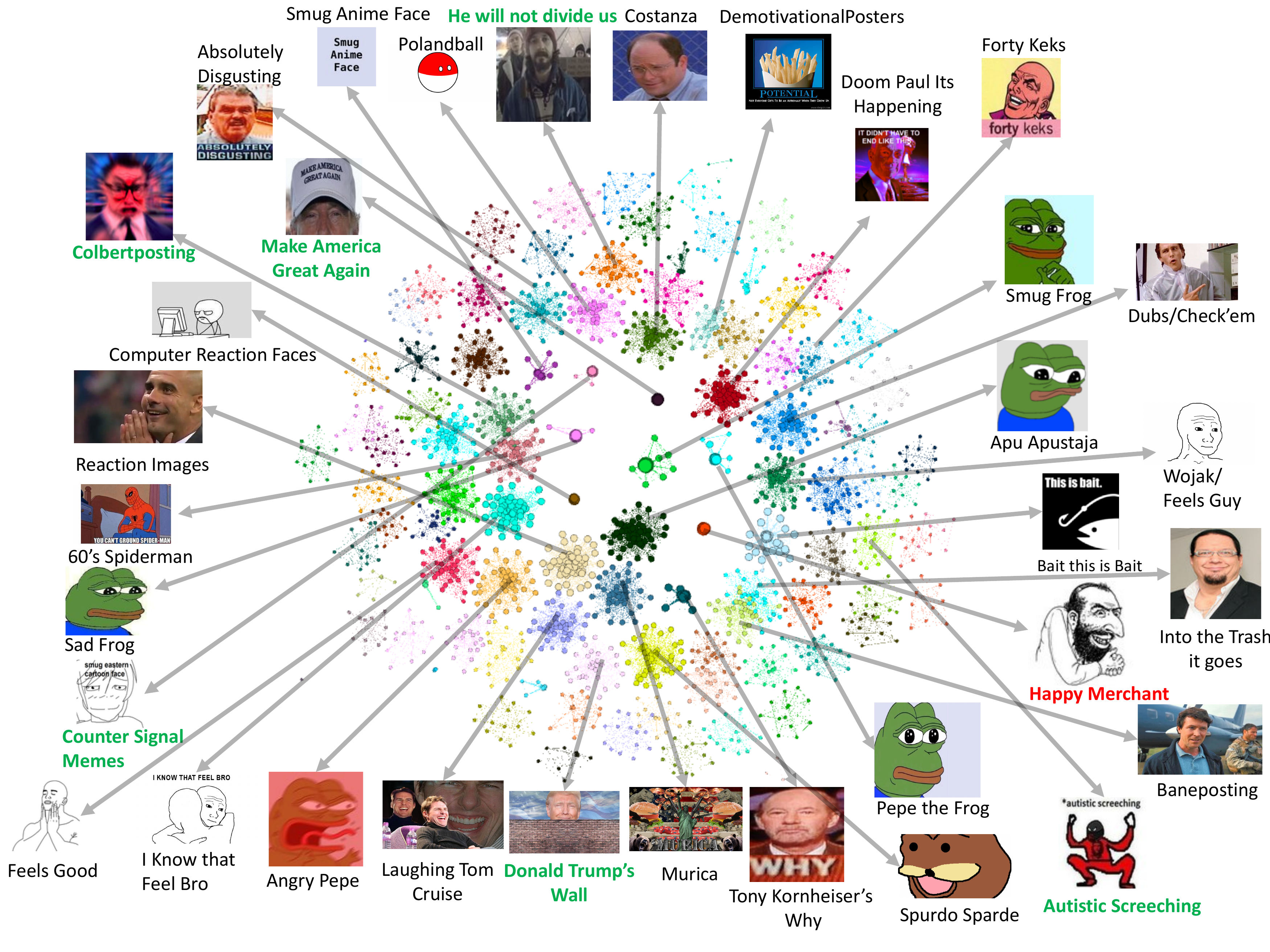}
\vspace{-0.2cm}
\caption{Visualization of the obtained clusters from \dspol, \td, and Gab. Note that memes with red labels are annotated as racist, while memes with green labels are annotated as politics (see Section~\ref{sec:meme_popularity} for the selection criteria).}
\label{fig:clusters_graph}
\end{figure*}

\subsubsection{Meme Visualization}
\label{sec:analysis:clusters:visualization}
We also use the custom distance metric (see Eq.~\ref{eq:distance}) to visualize the clusters with annotations.
We build a graph $\boldsymbol{G}=(\boldsymbol{V},\boldsymbol{E})$, where $\boldsymbol{V}$ are the medoids of annotated clusters and $\boldsymbol{E}$ the connections between medoids with distance under a threshold $\kappa$.
Figure~\ref{fig:clusters_graph} shows a snapshot of the graph for $\kappa=0.45$, chosen based on the frogs analysis above (see red horizontal line in Figure~\ref{fig:casestudy_frogs_dendrogram}).
In particular, we select this threshold as the majority of the clusters from the same meme (note coloration in Figure~\ref{fig:casestudy_frogs_dendrogram}) are hierarchically connected with a higher-level cluster at a distance close to 0.45.
To ease readability, we filter out nodes and edges that have a sum of in- and out-degree less than 10, which leaves 40\% of the nodes and 92\% of the edges.
Nodes are colored according to their KYM annotation.
NB: the graph is laid out using the OpenOrd algorithm~\cite{martin2011openord} and the distance between the components in it does not exactly match the actual distance metric.
We observe a large set of disconnected components, with each component containing nodes of primarily one color.
This indicates that our distance metric is indeed capturing the peculiarities of different memes. %
Finally, note that an interactive version of the full graph is publicly available from~\cite{memes_graph_site}. %

\subsection{Web Community-based Analysis}
\label{sec:analysis:annotations}
We now present a macro-perspective analysis of the Web communities through the lens of memes.
We assess the presence of different memes in each community, how popular they are, and how they evolve.
To this end, we examine the {\em posts} from all four communities (Twitter, Reddit, \dspol, and Gab) that contain {\em images} matching {\em memes} from fringe Web communities (\dspol, \td, and Gab).

\subsubsection{Meme Popularity}\label{sec:meme_popularity}

\descr{Memes.}
We start by analyzing clusters grouped by KYM `meme' entries, looking at the number of posts for each meme in \dspol, Reddit, Gab, and Twitter.

\begin{table*}[t]
\centering
\setlength{\tabcolsep}{2pt}
\resizebox{\textwidth}{!}{%
\begin{tabular}{lrlrlrlr}
\toprule
\multicolumn{2}{c}{\textbf{/pol/}}                                            & \multicolumn{2}{c}{\textbf{Reddit}}                                           & \multicolumn{2}{c}{\textbf{Gab}}                                                                                           & \multicolumn{2}{c}{\textbf{Twitter}}                                        \\ \midrule
\multicolumn{1}{c}{\textbf{Entry}} & \multicolumn{1}{c|}{\textbf{Posts (\%)}} & \multicolumn{1}{c}{\textbf{Entry}} & \multicolumn{1}{c|}{\textbf{Posts (\%)}} & \multicolumn{1}{c}{\textbf{Entry}}                                              & \multicolumn{1}{c|}{\textbf{Posts (\%)}} & \multicolumn{1}{c}{\textbf{Entry}} & \multicolumn{1}{c}{\textbf{Posts(\%)}} \\ \midrule
\href{http://knowyourmeme.com/memes/feels-bad-man-sad-frog}{Feels Bad Man/Sad Frog}           & \multicolumn{1}{r|}{64,367 (4.9\%)}      & \href{http://knowyourmeme.com/memes/manningface}{Manning Face}                       & \multicolumn{1}{r|}{12,540 (2.2\%)}       & \href{http://knowyourmeme.com/memes/jesusland}{Jesusland \textbf{(P)}}                                                                       & \multicolumn{1}{r|}{454 (1.6\%)}         & \href{http://knowyourmeme.com/memes/roll-safe}{Roll Safe}                                  & 55,010 (5.9\%)                                      \\
\href{http://knowyourmeme.com/memes/smug-frog}{Smug Frog}                          & \multicolumn{1}{r|}{63,290 (4.8\%)}      & \href{http://knowyourmeme.com/memes/thats-the-joke}{That's the Joke}                    & \multicolumn{1}{r|}{7,626 (1.3\%)}       & \href{http://knowyourmeme.com/memes/demotivational-posters}{Demotivational Posters}                                                           & \multicolumn{1}{r|}{414 (1.5\%)}         & \href{http://knowyourmeme.com/memes/evil-kermit}{Evil Kermit}                                    & 50,642 (5.4\%)                                      \\
\href{http://knowyourmeme.com/memes/happy-merchant}{Happy Merchant \textbf{(R)}}                     & \multicolumn{1}{r|}{49,608 (3.8\%)}      & \href{http://knowyourmeme.com/memes/feels-bad-man-sad-frog}{Feels Bad Man/ Sad Frog}           & \multicolumn{1}{r|}{7,240 (1.3\%)}       & \href{http://knowyourmeme.com/memes/smug-frog}{Smug Frog}                                                                        & \multicolumn{1}{r|}{392 (1.4\%)}         &                                 \href{http://knowyourmeme.com/memes/arthurs-fist}{Arthur's Fist} & 37,591 (4.0\%)                                      \\
\href{http://knowyourmeme.com/memes/apu-apustaja}{Apu Apustaja}                       & \multicolumn{1}{r|}{29,756 (2.2\%)}      & \href{http://knowyourmeme.com/memes/confession-bear}{Confession Bear}                    & \multicolumn{1}{r|}{7,147 (1.3\%)}       & \href{http://knowyourmeme.com/memes/based-stickman}{Based Stickman \textbf{(P)}}                                                                  & \multicolumn{1}{r|}{391 (1.4\%)}         &   \href{http://knowyourmeme.com/memes/nut-button}{Nut Button}                               &     13,598 (1,5\%)                                 \\
\href{http://knowyourmeme.com/memes/pepe-the-frog}{Pepe the Frog}                      & \multicolumn{1}{r|}{25,197 (1.9\%)}      & \href{http://knowyourmeme.com/memes/this-is-fine}{This is Fine}                           & \multicolumn{1}{r|}{5,032 (0.9\%)}       & \href{http://knowyourmeme.com/memes/pepe-the-frog}{Pepe the Frog}                                                                   & \multicolumn{1}{r|}{378 (1.3\%)}         &  \href{http://knowyourmeme.com/memes/spongebob-mock}{Spongebob Mock}                                  &  11,136 (1,2\%)                                     \\
\href{http://knowyourmeme.com/memes/make-america-great-again}{Make America Great Again \textbf{(P)}\hspace*{-0.4cm}}            & \multicolumn{1}{r|}{21,229 (1.6\%)}      & \href{http://knowyourmeme.com/memes/smug-frog}{Smug Frog}         & \multicolumn{1}{r|}{4,642 (0.8\%)}       & \href{http://knowyourmeme.com/memes/happy-merchant}{Happy Merchant \textbf{(R)}}                                                                  & \multicolumn{1}{r|}{297 (1.1\%)}         &   \href{http://knowyourmeme.com/memes/reaction-images}{Reaction Images}                             &  9,387 (1.0\%)                                      \\
\href{http://knowyourmeme.com/memes/angry-pepe}{Angry Pepe}                         & \multicolumn{1}{r|}{20,485 (1.5\%)}      & \href{http://knowyourmeme.com/memes/roll-safe}{Roll Safe}                     & \multicolumn{1}{r|}{4,523 (0.8\%)}       & \href{http://knowyourmeme.com/memes/murica}{Murica}                                                                          & \multicolumn{1}{r|}{274 (1.0\%)}         &      \href{http://knowyourmeme.com/memes/conceited-reaction}{Conceited Reaction}  &    9,106 (1.0\%)                                  \\
\href{http://knowyourmeme.com/memes/bait-this-is-bait}{Bait this is Bait}                 & \multicolumn{1}{r|}{16,686 (1.2\%)}      & \href{http://knowyourmeme.com/memes/rage-guy-fffffuuuuuuuu}{Rage Guy}                        & \multicolumn{1}{r|}{4,491 (0.8\%)}       & \href{http://knowyourmeme.com/memes/and-its-gone}{And Its Gone}                                                                    & \multicolumn{1}{r|}{235 (0.9\%)}         &  \href{http://knowyourmeme.com/memes/expanding-brain}{Expanding Brain}                                                               & 8,701 (0.9\%)                                       \\
\href{http://knowyourmeme.com/memes/i-know-that-feel-bro}{I Know that Feel Bro}               & \multicolumn{1}{r|}{14,490 (1.1\%)}      & \href{http://knowyourmeme.com/memes/make-america-great-again}{Make America Great Again \textbf{(P)\hspace*{-0.4cm}}}             & \multicolumn{1}{r|}{4,440 (0.8\%)}       & \href{http://knowyourmeme.com/memes/make-america-great-again}{Make America Great Again \textbf{(P)}}                                                         & \multicolumn{1}{r|}{207 (0.8\%)}         & \href{http://knowyourmeme.com/memes/demotivational-posters}{Demotivational Posters}                            &                                      7,781 (0.8\%)  \\
\href{http://knowyourmeme.com/memes/cult-of-kek}{Cult of Kek}                        & \multicolumn{1}{r|}{14,428 (1.1\%)}      & \href{http://knowyourmeme.com/memes/fake-ccg-cards}{Fake CCG Cards}         & \multicolumn{1}{r|}{4,438 (0.8\%)}       & \href{http://knowyourmeme.com/memes/feels-bad-man-sad-frog}{Feels Bad Man/ Sad Frog} & \multicolumn{1}{r|}{206 (0.8\%)}         &   \href{http://knowyourmeme.com/memes/cash-me-ousside-howbow-dah}{Cash Me Ousside/Howbow Dah}                                     &      5,972 (0.6\%)                              \\
\href{http://knowyourmeme.com/memes/laughing-tom-cruise}{Laughing Tom Cruise}                & \multicolumn{1}{r|}{14,312 (1.1\%)}      &  \href{http://knowyourmeme.com/memes/confused-nick-young}{Confused Nick Young}                & \multicolumn{1}{r|}{4,024 (0.7\%)}       & \href{http://knowyourmeme.com/memes/trump-s-first-order-of-business}{Trump's First Order of Business \textbf{(P)}\hspace*{-0.4cm}}      & \multicolumn{1}{r|}{192 (0.7\%)}         &  \href{http://knowyourmeme.com/memes/salt-bae}{Salt Bae}                                                          & 5,375 (0.6\%)    \\
\href{http://knowyourmeme.com/memes/awoo}{Awoo}                               & \multicolumn{1}{r|}{13,767 (1.0\%)}     & \href{http://knowyourmeme.com/memes/daily-struggle}{Daily Struggle}                 & \multicolumn{1}{r|}{4,015 (0.7\%)}       & \href{http://knowyourmeme.com/memes/kekistan}{Kekistan}                                                                        & \multicolumn{1}{r|}{186 (0.6\%)}         &   \href{http://knowyourmeme.com/memes/feels-bad-man-sad-frog}{Feels Bad Man/ Sad Frog}                                  & 4,991 (0.5\%)                                                                           \\
\href{http://knowyourmeme.com/memes/tony-kornheiser-s-why}{Tony Kornheiser's Why}              & \multicolumn{1}{r|}{13,577 (1.0\%)}      & \href{http://knowyourmeme.com/memes/expanding-brain}{Expanding Brain}                          & \multicolumn{1}{r|}{3,757 (0.7\%)}       & \href{http://knowyourmeme.com/memes/picardia}{Picardia \textbf{(P)}}                                                                        & \multicolumn{1}{r|}{183 (0.6\%)}         &   \href{http://knowyourmeme.com/memes/math-lady-confused-lady}{Math Lady/Confused Lady}                                &  4,722 (0.5\%)                                     \\
\href{http://knowyourmeme.com/memes/picardia}{Picardia \textbf{(P)}}                           & \multicolumn{1}{r|}{13,540 (1.0\%)}      &   \href{http://knowyourmeme.com/memes/demotivational-posters}{Demotivational Posters}        & \multicolumn{1}{r|}{3,419 (0.6\%)}       & \href{http://knowyourmeme.com/memes/things-with-faces-pareidolia}{Things with Faces (Pareidolia) }                                                 & \multicolumn{1}{r|}{156 (0.5\%)}         &  \href{http://knowyourmeme.com/memes/computer-reaction-faces}{Computer Reaction Faces}
                                 & 4,720 (0.5\%)                                      \\
\href{http://knowyourmeme.com/memes/big-grin-never-ever}{Big Grin / Never Ever}              & \multicolumn{1}{r|}{12,893 (1.0\%)}      &  \href{http://knowyourmeme.com/memes/actual-advice-mallard}{Actual Advice Mallard}                  & \multicolumn{1}{r|}{3,293 (0.6\%)}       & \href{http://knowyourmeme.com/memes/serbia-strong-remove-kebab}{Serbia Strong/Remove Kebab}                                                      & \multicolumn{1}{r|}{149 (0.5\%)}         &   \href{http://knowyourmeme.com/memes/clinton-trump-duet}{Clinton Trump Duet \textbf{(P)}}
 &    3,901 (0.4\%)                                  \\
\href{http://knowyourmeme.com/memes/reaction-images}{Reaction Images}                    & \multicolumn{1}{r|}{12,608 (0.9\%)}      & \href{http://knowyourmeme.com/memes/reaction-images}{Reaction Images}                      & \multicolumn{1}{r|}{2,959 (0.5\%)}       & \href{http://knowyourmeme.com/memes/riot-hipster}{Riot Hipster}                                                                    & \multicolumn{1}{r|}{148 (0.5\%)}        &     \href{http://knowyourmeme.com/memes/kendrick-lamar-damn-album-cover}{Kendrick Lamar Damn Album Cover\hspace*{-0.4cm}}
                            &   3,656 (0.4\%)                                    \\
\href{http://knowyourmeme.com/memes/computer-reaction-faces}{Computer Reaction Faces}             & \multicolumn{1}{r|}{12,247 (0.9\%)}      &  \href{http://knowyourmeme.com/memes/handsome-face}{Handsome Face}                   & \multicolumn{1}{r|}{2,675 (0.5\%)}       & \href{http://knowyourmeme.com/memes/colorized-history}{Colorized History                                                             } & \multicolumn{1}{r|}{144 (0.5\%)}         & \href{http://knowyourmeme.com/memes/what-in-tarnation}{What in tarnation}                                    & 3,363 (0.3\%)     \\
\href{http://knowyourmeme.com/memes/wojak-feels-guy}{Wojak / Feels Guy}                  & \multicolumn{1}{r|}{11,682 (0.9\%)}      & \href{http://knowyourmeme.com/memes/absolutely-disgusting}{Absolutely Disgusting}                         & \multicolumn{1}{r|}{2,674 (0.5\%)}       &  \href{http://knowyourmeme.com/memes/the-most-interesting-man-in-the-world}{Most Interesting Man in World} & \multicolumn{1}{r|}{140 (0.5\%)}         &   \href{http://knowyourmeme.com/memes/harambe-the-gorilla}{Harambe the Gorilla}                               & 3,164 (0.3\%)  \\
\href{http://knowyourmeme.com/memes/absolutely-disgusting}{Absolutely Disgusting}              & \multicolumn{1}{r|}{11,436 (0.8\%)}      & \href{http://knowyourmeme.com/memes/pepe-the-frog}{Pepe the Frog}                    & \multicolumn{1}{r|}{2,672 (0.5\%)}       & \href{http://knowyourmeme.com/memes/chuck-norris-facts}{Chuck Norris Facts}                                                              & \multicolumn{1}{r|}{131 (0.4\%)}         &  \href{http://knowyourmeme.com/memes/i-know-that-feel-bro}{I Know that Feel Bro}                              &   3,137 (0.3\%)                                                                                                              \\
\href{http://knowyourmeme.com/memes/spurdo-sparde}{Spurdo Sparde}                      & \multicolumn{1}{r|}{9,581 (0.7\%)}       & \href{http://knowyourmeme.com/memes/i-was-only-pretending-to-be-retarded}{Pretending to be Retarded}                     & \multicolumn{1}{r|}{2,462 (0.4\%)}       & \href{http://knowyourmeme.com/memes/roll-safe}{Roll Safe}                                                                        & \multicolumn{1}{r|}{131 (0.4\%)}         & \href{http://knowyourmeme.com/memes/this-is-fine}{This is Fine}                                    & 3,094 (0.3\%)                                     \\
\midrule
{\bf Total} & \multicolumn{1}{r|}{\bf 445,179 (33.4\%)} && \multicolumn{1}{r|}{\bf 94,069 (16.7\%)} && \multicolumn{1}{r|}{\bf 4,808 (17.0\%)} && \multicolumn{1}{r|}{\bf 249,047 (26.4\%)} \\
 \bottomrule
\end{tabular}%
}
\caption{Top 20 KYM entries for memes that we find our datasets. We report the number of posts for each meme as well as the percentage over all the posts (per community) that contain images that match one of the annotated clusters. The (R) and (P) markers indicate whether a meme is annotated as racist or politics-related, respectively (see Section~\ref{sec:meme_popularity} for the selection criteria). }
\label{tbl:top_memes}
\end{table*}

In Table~\ref{tbl:top_memes}, we report the top 20 memes for each Web community sorted by the number of posts.
We observe that Pepe the Frog~\cite{pepe_frog_meme} and its variants are among the most popular memes for every platform.
While this might be an artifact of using fringe communities as a ``seed'' for the clustering, recall that the goal of this work is in fact to gain an understanding of how fringe communities disseminate memes and influence mainstream ones.
Thus, we leave to future work a broader analysis of the wider meme ecosystem.

Sad Frog~\cite{sad_frog_meme} is the most popular meme on \dspol (4.9\%), the 3rd on Reddit (1.3\%), the 10th on Gab (0.8\%), and the 12th on Twitter (0.5\%).
We also find variations like Smug Frog~\cite{smug_frog_meme}, Apu Apustaja~\cite{apu_meme}, Pepe the Frog~\cite{pepe_frog_meme}, and Angry Pepe~\cite{angry_pepe_meme}.
Considering that Pepe is treated as a hate symbol by the Anti-Defamation League~\cite{adl_pepe_frog} and that is often used in hateful or racist, this likely indicates that polarized communities like \dspol and Gab do use memes to incite hateful conversation.
This is also evident from the popularity of the anti-semitic Happy Merchant meme~\cite{happy_merchant_meme}, which depicts a ``greedy'' and ``manipulative'' stereotypical caricature of a Jew (3.8\% on \dspol and 1.1\% on Gab). 

By contrast, mainstream communities like Reddit and Twitter primarily share harmless/neutral memes, which are rarely used in hateful contexts.
Specifically, on Reddit the top memes are Manning Face~\cite{manning_face_meme} (2.2\%) and That's the Joke~\cite{thats_the_joke_meme} (1.3\%), while on Twitter the top ones are Roll Safe~\cite{roll_safe_meme} (5.9\%) and Evil Kermit~\cite{evil_kermit_meme} (5.4\%).

Once again, we find that users (in all communities) post memes to share politics-related information, possibly aiming to enhance or penalize the public image of politicians (see Appendix~\ref{sec:appendix_interesting_images} for an example of such memes).
For instance, we find Make America Great Again~\cite{maga_meme}, a meme dedicated to Donald Trump's US presidential campaign, among the top memes in \dspol (1.6\%), in Reddit (0.8\%), and Gab (0.8\%).
Similarly, in Twitter, we find the Clinton Trump Duet meme~\cite{clinton_trump_duet_meme} (0.4\%), a meme inspired by the 2nd US presidential debate.

\begin{table*}[t]
\centering
\resizebox{0.8\textwidth}{!}{%
\begin{tabular}{@{}lrlrlrlr@{}}
\toprule
\multicolumn{2}{c}{\textbf{/pol/}}                                            & \multicolumn{2}{c}{\textbf{Reddit}}                                           & \multicolumn{2}{c}{\textbf{Gab}}                                              & \multicolumn{2}{c}{\textbf{Twitter}}                                        \\ \midrule
\multicolumn{1}{c}{\textbf{Entry}} & \multicolumn{1}{c|}{\textbf{Posts (\%)}} & \multicolumn{1}{c}{\textbf{Entry}} & \multicolumn{1}{c|}{\textbf{Posts (\%)}} & \multicolumn{1}{c}{\textbf{Entry}} & \multicolumn{1}{c|}{\textbf{Posts (\%)}} & \multicolumn{1}{c}{\textbf{Entry}} & \multicolumn{1}{c}{\textbf{Posts(\%)}} \\ \midrule
\href{http://knowyourmeme.com/memes/people/donald-trump}{Donald Trump}                       & \multicolumn{1}{r|}{60,611 (4.6\%)}      & \href{http://knowyourmeme.com/memes/people/donald-trump}{Donald Trump}                      & \multicolumn{1}{r|}{34,533 (6.1\%)}      & \href{http://knowyourmeme.com/memes/people/donald-trump}{Donald Trump}                       & \multicolumn{1}{r|}{1,665 (6.1\%)}       & \href{http://knowyourmeme.com/memes/people/donald-trump}{Donald Trump}                                 & 10,208 (1.3\%)                                      \\
\href{http://knowyourmeme.com/memes/people/adolf-hitler}{Adolf Hitler}                       & \multicolumn{1}{r|}{8,759  (0.6\%)}      & \href{http://knowyourmeme.com/memes/people/steve-bannon}{Steve Bannon}                       & \multicolumn{1}{r|}{3,733 (0.6\%)}       & \href{http://knowyourmeme.com/memes/people/mitt-romney}{Mitt Romney}                        & \multicolumn{1}{r|}{455 (1.7\%)}         & \href{http://knowyourmeme.com/memes/people/barack-obama}{Barack Obama}                                   & 5,187 (0.6\%)                                      \\
\href{http://knowyourmeme.com/memes/people/mike-pence}{Mike Pence}                         & \multicolumn{1}{r|}{4,738 (0.3\%)}       & \href{http://knowyourmeme.com/memes/people/stephen-colbert}{Stephen Colbert}                     & \multicolumn{1}{r|}{3,121 (0.6\%)}       & \href{http://knowyourmeme.com/memes/people/bill-nye}{Bill Nye}                           & \multicolumn{1}{r|}{370 (1.3\%)}         & \href{http://knowyourmeme.com/memes/people/chelsea-manning}{Chelsea Manning}                                   & 4,173 (0.5\%)                                      \\
\href{http://knowyourmeme.com/memes/people/jeb-bush}{Jeb Bush}                           & \multicolumn{1}{r|}{4,217 (0.3\%)}       & \href{http://knowyourmeme.com/memes/people/chelsea-manning}{Chelsea Manning}                        & \multicolumn{1}{r|}{2,261 (0.4\%)}       & \href{http://knowyourmeme.com/memes/people/adolf-hitler}{Adolf Hitler}                       & \multicolumn{1}{r|}{106 (0.4\%)}         & \href{http://knowyourmeme.com/memes/people/kim-jong-un}{Kim Jong Un}                                 & 3,271 (0.4\%)                                      \\
\href{http://knowyourmeme.com/memes/people/vladimir-putin}{Vladimir Putin}                      & \multicolumn{1}{r|}{3,218 (0.2\%)}       & \href{http://knowyourmeme.com/memes/people/ben-carson}{Ben Carson}                      & \multicolumn{1}{r|}{2,148 (0.4\%)}       & \href{http://knowyourmeme.com/memes/people/barack-obama}{Barack Obama}                       & \multicolumn{1}{r|}{104 (0.4\%)}         & \href{http://knowyourmeme.com/memes/people/anita-sarkeesian}{Anita Sarkeesian}                                  & 2,764 (0.3\%)                                     \\
\href{http://knowyourmeme.com/memes/people/alex-jones}{Alex Jones}                         & \multicolumn{1}{r|}{3,206 (0.2\%)}       & \href{http://knowyourmeme.com/memes/people/bernie-sanders}{Bernie Sanders}                     & \multicolumn{1}{r|}{1,757 (0.3\%)}       & \href{http://knowyourmeme.com/memes/people/isis-daesh}{Isis Daesh}                         & \multicolumn{1}{r|}{92 (0.3\%)}          & \href{http://knowyourmeme.com/memes/people/bernie-sanders}{Bernie Sanders}                                  & 2,277 (0.3\%)                                      \\
\href{http://knowyourmeme.com/memes/people/ron-paul}{Ron Paul}                           & \multicolumn{1}{r|}{3,116 (0.2\%)}       & \href{http://knowyourmeme.com/memes/people/ajit-pai}{Ajit Pai}                           & \multicolumn{1}{r|}{1,658 (0.3\%)}       & \href{http://knowyourmeme.com/memes/people/death-grips}{Death Grips}                        & \multicolumn{1}{r|}{91 (0.3\%)}          & \href{http://knowyourmeme.com/memes/people/vladimir-putin}{Vladimir Putin}                                   &  1,733 (0.2\%)                                      \\
\href{http://knowyourmeme.com/memes/people/bernie-sanders}{Bernie Sanders}                      & \multicolumn{1}{r|}{3,022 (0.2\%)}       & \href{http://knowyourmeme.com/memes/people/barack-obama}{Barack Obama}                          & \multicolumn{1}{r|}{1,628 (0.3\%)}       & \href{http://knowyourmeme.com/memes/people/eminem}{Eminem}                             & \multicolumn{1}{r|}{89 (0.3\%)}          & \href{http://knowyourmeme.com/memes/people/billy-mays}{Billy Mays}                                  & 1,454 (0.2\%)                                      \\
\href{http://knowyourmeme.com/memes/people/massimo-dalema}{Massimo D'alema}                     & \multicolumn{1}{r|}{2,725 (0.2\%)}       & \href{http://knowyourmeme.com/memes/people/gabe-newell}{Gabe Newell}                        & \multicolumn{1}{r|}{1,518 (0.3\%)}       & \href{http://knowyourmeme.com/memes/people/kim-jong-un}{Kim Jong Un}                       & \multicolumn{1}{r|}{87 (0.3\%)}          & \href{http://knowyourmeme.com/memes/people/adolf-hitler}{Adolf Hitler}                                  & 1,304 (0.2\%)                                      \\
\href{http://knowyourmeme.com/memes/people/mitt-romney}{Mitt Romney}                        & \multicolumn{1}{r|}{2,468 (0.2\%)}       & \href{http://knowyourmeme.com/memes/people/bill-nye}{Bill Nye}                         & \multicolumn{1}{r|}{1,478 (0.3\%)}       & \href{http://knowyourmeme.com/memes/people/ajit-pai}{Ajit Pai}                            & \multicolumn{1}{r|}{76 (0.3\%)}          & \href{http://knowyourmeme.com/memes/people/kanye-west}{Kanye West}                                  & 1,261 (0.2\%)                                      \\
\href{http://knowyourmeme.com/memes/people/chelsea-manning}{Chelsea Manning}                        & \multicolumn{1}{r|}{2,403 (0.2\%)}       & \href{http://knowyourmeme.com/memes/people/hillary-clinton}{Hillary Clinton}                        & \multicolumn{1}{r|}{1,468 (0.3\%)}       & \href{http://knowyourmeme.com/memes/people/pewdiepie}{Pewdiepie}                           & \multicolumn{1}{r|}{73 (0.3\%)}          & \href{http://knowyourmeme.com/memes/people/bill-nye}{Bill Nye}                               & 968 (0.2\%)                                      \\
\href{http://knowyourmeme.com/memes/people/hillary-clinton}{Hillary Clinton}                       & \multicolumn{1}{r|}{2,378 (0.2\%)}       & \href{http://knowyourmeme.com/memes/people/death-grips}{Death Grips}                         & \multicolumn{1}{r|}{1,463 (0.3\%)}       & \href{http://knowyourmeme.com/memes/people/bernie-sanders}{Bernie Sanders}                            & \multicolumn{1}{r|}{71 (0.3\%)}          & \href{http://knowyourmeme.com/memes/people/mitt-romney}{Mitt Romney}                                  & 923 (0.1\%)                                      \\
\href{http://knowyourmeme.com/memes/people/a-wyatt-mann}{A. Wyatt Mann}                        & \multicolumn{1}{r|}{2,110 (0.2\%)}       & \href{http://knowyourmeme.com/memes/people/adolf-hitler}{Adolf Hitler}                        & \multicolumn{1}{r|}{1,449 (0.3\%)}       & \href{http://knowyourmeme.com/memes/people/alex-jones}{Alex Jones}                           & \multicolumn{1}{r|}{70 (0.3\%)}          & \href{http://knowyourmeme.com/memes/people/filthy-frank}{Filthy Frank}                                  & 777 (0.1\%)                                      \\
\href{http://knowyourmeme.com/memes/people/ben-carson}{Ben Carson}                       & \multicolumn{1}{r|}{1,780 (0.1\%)}       & \href{http://knowyourmeme.com/memes/people/mitt-romney}{Mitt Romney}                       & \multicolumn{1}{r|}{1,294 (0.2\%)}       & \href{http://knowyourmeme.com/memes/people/hillary-clinton}{Hillary Clinton}                           & \multicolumn{1}{r|}{59 (0.2\%)}          & \href{http://knowyourmeme.com/memes/people/hillary-clinton}{Hillary Clinton}                                   & 758 (0.1\%)                                      \\
\href{http://knowyourmeme.com/memes/people/filthy-frank}{Filthy Frank}                       & \multicolumn{1}{r|}{1,598 (0.1\%)}       & \href{http://knowyourmeme.com/memes/people/eminem}{Eminem}                        & \multicolumn{1}{r|}{1,274 (0.2\%)}       & \href{http://knowyourmeme.com/memes/people/anita-sarkeesian}{Anita Sarkeesian}                           & \multicolumn{1}{r|}{54 (0.2\%)}          & \href{http://knowyourmeme.com/memes/people/ajit-pai}{Ajit Pai}                                   & 715 (0.1\%)                                      \\ \bottomrule
\end{tabular}%
}
  \caption{Top 15 KYM entries about people that we find in each of our dataset. We report the number of posts and the percentage over all the posts (per community) that match a cluster with KYM annotations.}
\label{tbl:top_people}
\vspace{-0.2cm}
\end{table*}

 \begin{figure*}[t]
   \begin{minipage}[t]{1\textwidth}
 \centering
\subfigure[all memes]{\includegraphics[width=0.322\textwidth]{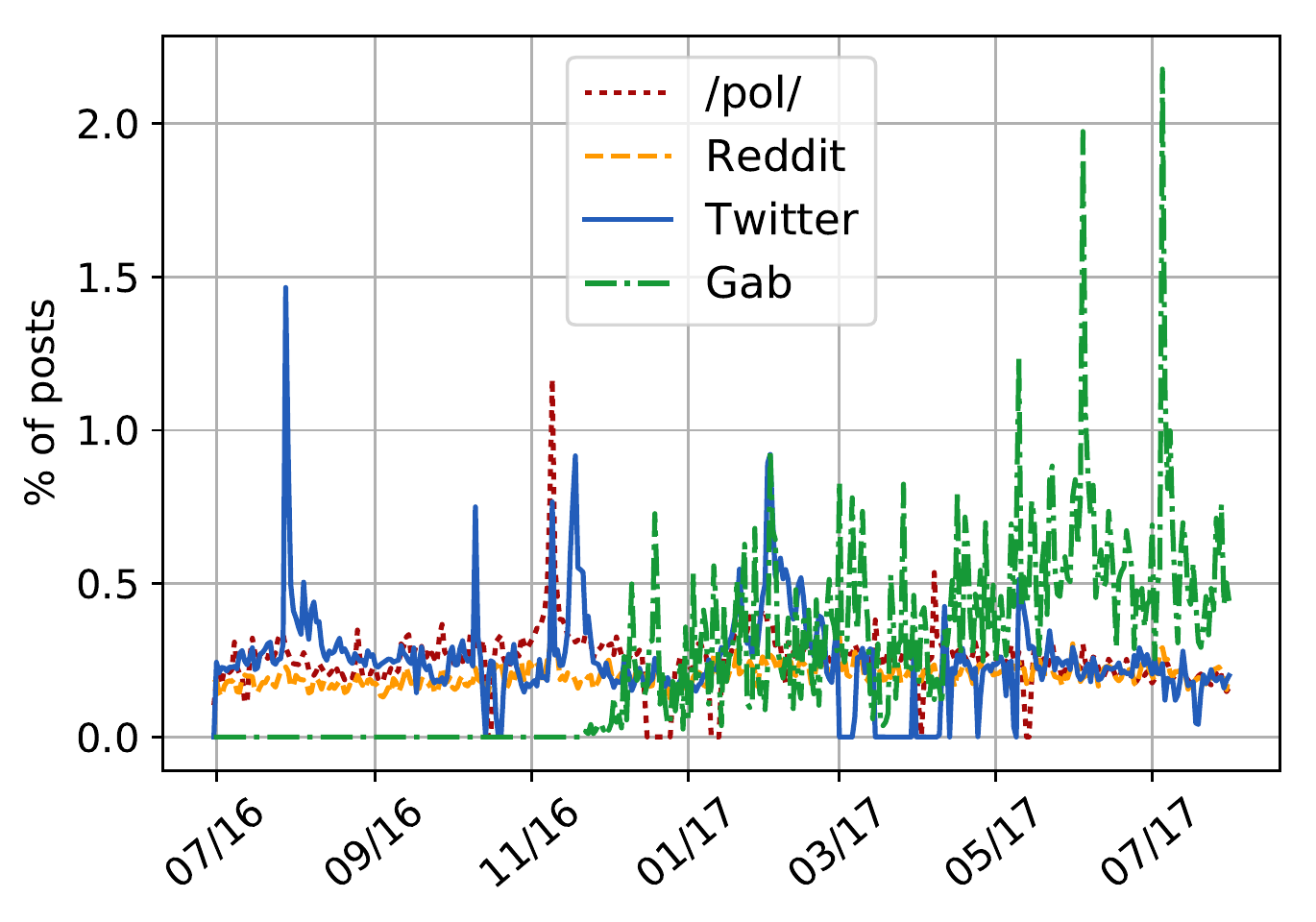}\label{temporal_all}}
\subfigure[racist]{\includegraphics[width=0.335\textwidth]{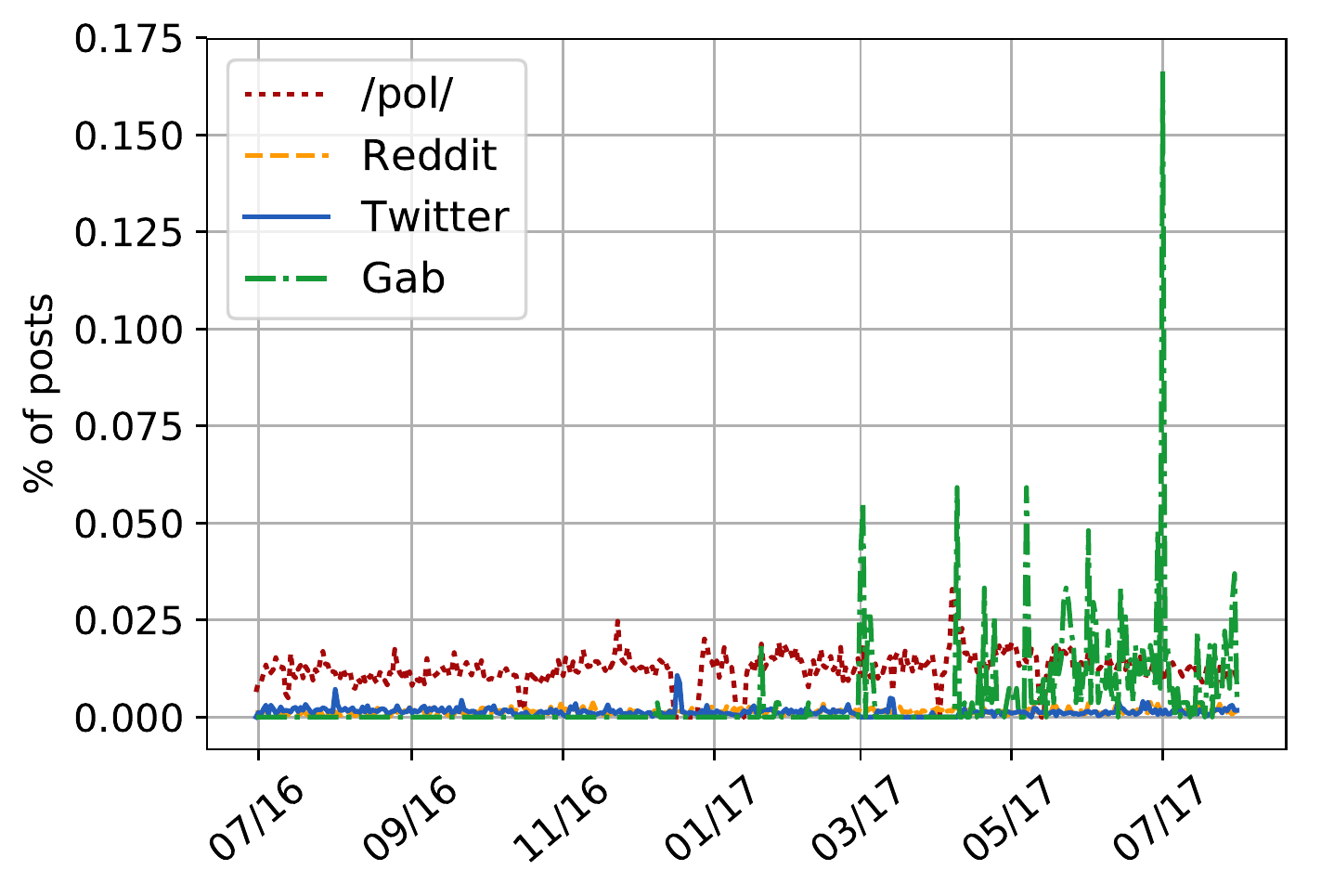}\label{temporal_racism}}
\subfigure[politics]{\includegraphics[width=0.322\textwidth]{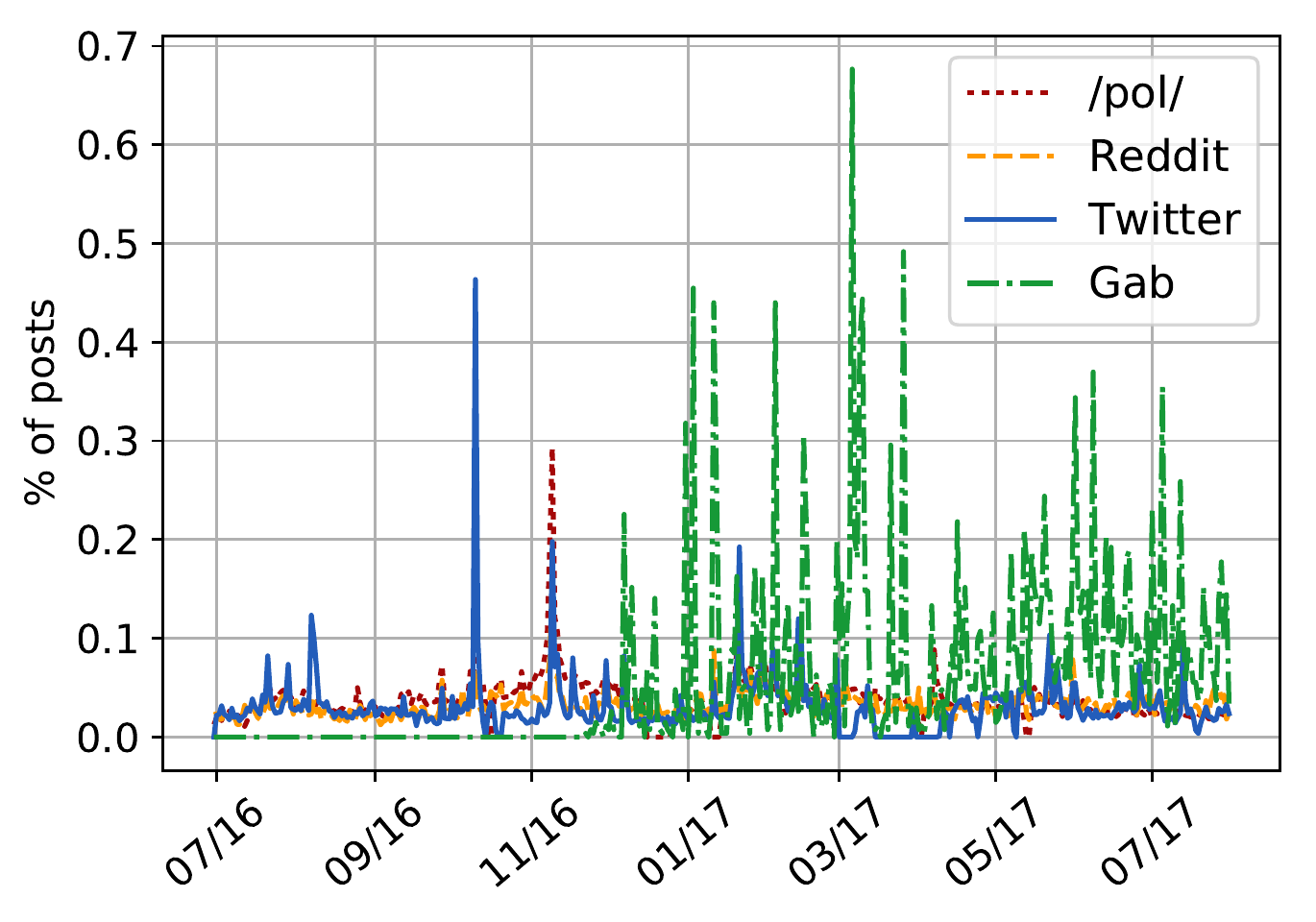}\label{temporal_politics}}
\caption{Percentage of posts per day in our dataset for all, racist, and politics-related memes.}
\label{fig:temporal_selected_memes_group}
\vspace{0.2cm}
\end{minipage}
 \centering
    \begin{minipage}[t]{0.85\textwidth}
    \centering
\subfigure[Reddit]{\includegraphics[width=0.37\textwidth]{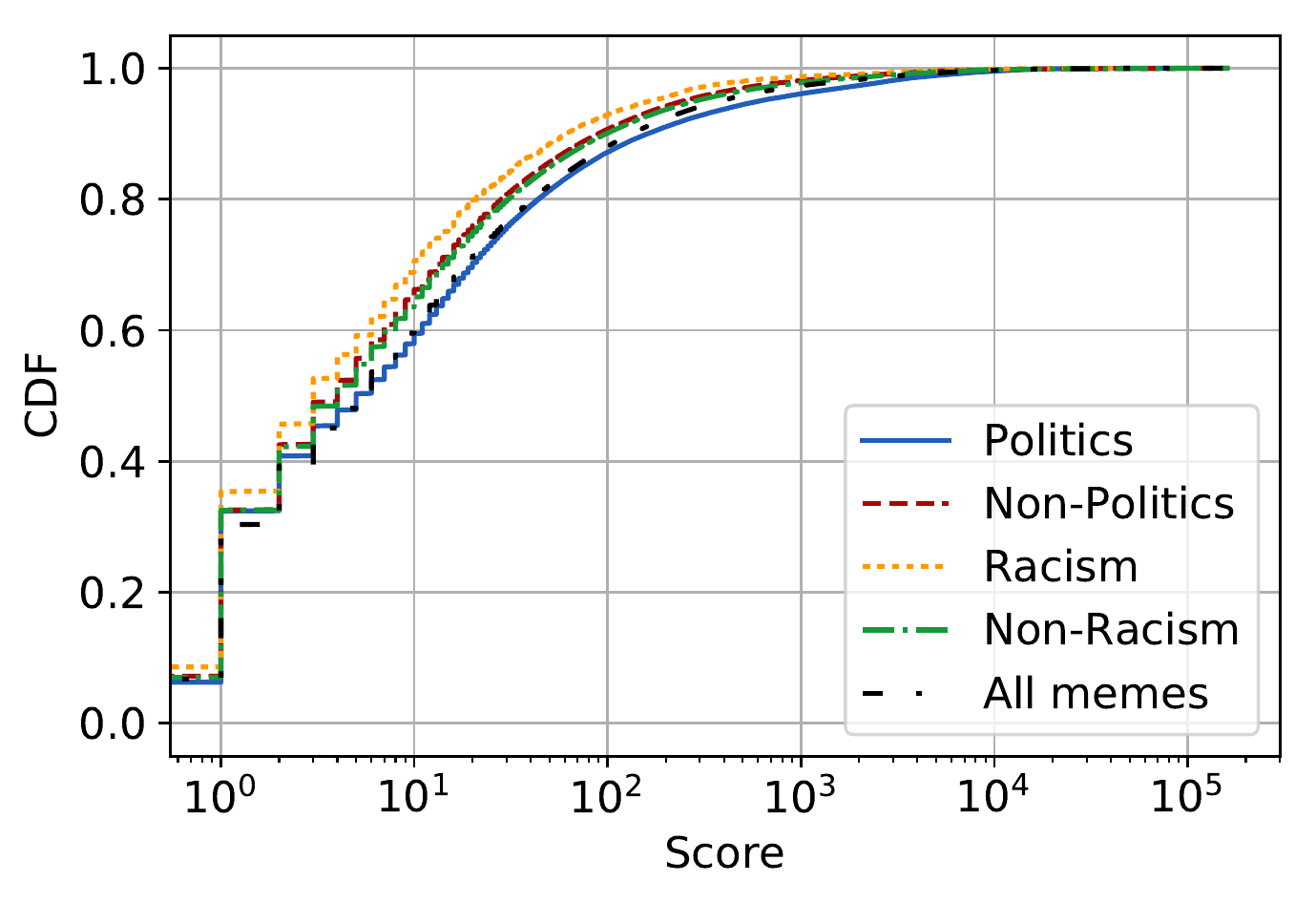}\label{scores_reddit_memes}}\hspace{0.5cm}
\subfigure[Gab]{\includegraphics[width=0.37\textwidth]{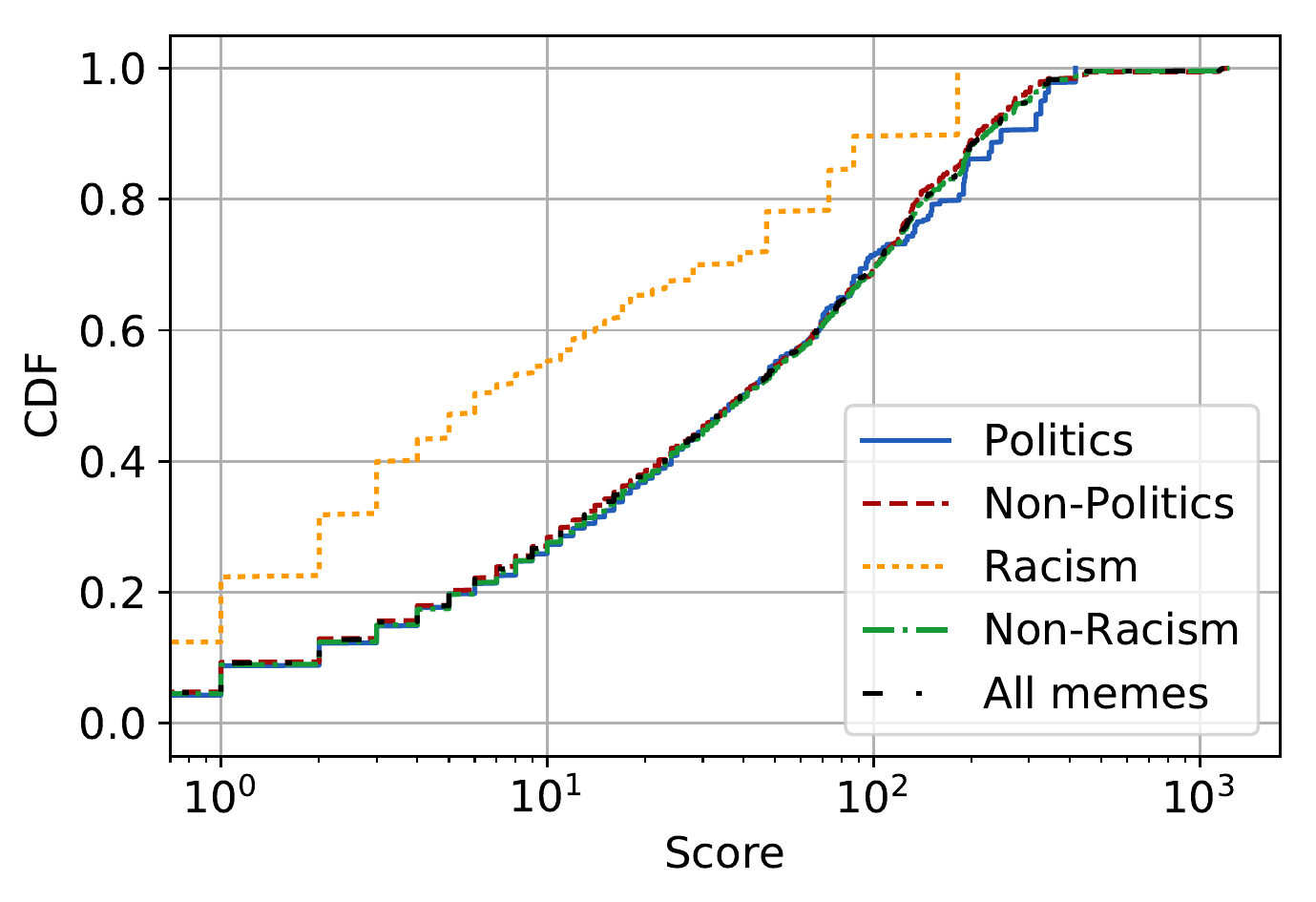}\label{scores_gab_memes}}
\caption{CDF of scores of posts that contain memes on Reddit and Gab.}
\label{fig:scores_groups}
\vspace{0.2cm}
\end{minipage}
\end{figure*}

\descr{People.}
We also analyze memes related to people (i.e., KYM entries with the people category). %
Table~\ref{tbl:top_people} reports the top 15 KYM entries in this category.
We observe that, in all Web Communities, the most popular person portrayed in memes is Donald Trump: he is depicted in 4.6\% of \dspol posts that contain annotated images, while for Reddit, Gab, and Twitter the percentages are 6.1\%, 6.1\%, and 1.3\%, respectively.
Other popular personalities, in all platforms, include several politicians.
For instance, in \dspol, we find Mike Pence (0.3\%), Jeb Bush (0.3\%), Vladimir Putin (0.2\%), while, in Reddit, we find Steve Bannon (0.6\%),  Chelsea Manning (0.6\%), and Bernie Sanders (0.3\%), in Gab, Mitt Romney (1.7\%) and Barack Obama (0.4\%), and, in Twitter, Barack Obama (0.6\%), Kim Jong Un (0.5\%), and Chelsea Manning (0.4\%).
This highlights the fact that users on these communities utilize memes to share information and opinions about politicians, and possibly try to either enhance or harm public opinion about them.
Finally, we note the presence of Adolf Hitler memes on all Web Communities, i.e., \dspol (0.6\%), Reddit (0.3\%),  Gab (0.4\%), and Twitter (0.2\%).

We further group memes into two high-level groups, racist and politics-related.
We use the {\em tags} that are available in our KYM dataset, i.e., we assign a meme to the politics-related group if it has the ``politics,'' ``2016 us presidential election,'' ``presidential election,'' ``trump,'' or ``clinton'' tags, and to the racism-related one if the tags include ``racism,'' ``racist,'' or ``antisemitism,''
obtaining 117 racist memes (4.4\% of all memes that appear in our dataset) and 556 politics-related memes (21.2\% of all memes that appear on our dataset).
In the rest of this section, we use these groups to further study the memes, and later in Section~\ref{sec:influence_estimation} to estimate influence.

\subsubsection{Temporal Analysis}
Next, we study the temporal aspects of posts that contain memes from \dspol, Reddit, Twitter, and Gab.
In Figure~\ref{fig:temporal_selected_memes_group}, we plot the percentage of posts per day that include memes.
For all memes (Figure~\ref{temporal_all}), we observe that \dspol and Reddit follow 
a steady posting behavior, with a peak in activity around the 2016 US elections.
We also find that memes are increasingly more used on Gab (see, e.g., 2016 vs 2017).

As shown in Figure~\ref{temporal_racism}, both \dspol and Gab include a substantially higher number of posts with racist memes, used over time with a difference in behavior: while \dspol users share them in a very steady and constant way, Gab exhibits a bursty behavior.
A possible explanation is that the former is inherently more racist, with the latter primarily reacting to particular world events.
As for political memes (Figure~\ref{temporal_politics}), we find a lot of activity overall on Twitter, Reddit, and \dspol, but with different spikes in time.
On Reddit and \dspol, the peaks coincide with the 2016 US elections.
On Twitter, we note a peak that coincides with the 2nd US Presidential Debate on October 2016.
For Gab, there is again an increase in posts with political memes after January 2017.

\begin{table*}[]
\centering
\resizebox{0.77\textwidth}{!}{%
\begin{tabular}{lrlrlr}
\hline
\multicolumn{2}{c}{\textbf{All Memes}} & \multicolumn{2}{c}{\textbf{Racism-Related Memes}} & \multicolumn{2}{c}{\textbf{Politics-Related Memes}} \\ \hline
\textbf{Subreddit} & \multicolumn{1}{r|}{\textbf{Posts (\%)}} & \textbf{Subreddit} & \multicolumn{1}{r|}{\textbf{Posts (\%)}} & \textbf{Subreddit} & \textbf{Posts (\%)} \\
\td & \multicolumn{1}{r|}{82,698 (12.5\%)} & \td & \multicolumn{1}{r|}{359 (9.3\%)} & \td & 24,343 (26.4\%) \\
AdviceAnimals & \multicolumn{1}{r|}{35,475 (5.3\%)} & AdviceAnimals & \multicolumn{1}{r|}{87 (2.2\%)} & politics & 2,751 (3.0\%) \\
me\_irl & \multicolumn{1}{r|}{15,366 (2.3\%)} & conspiracy & \multicolumn{1}{r|}{76 (2.0\%)} & EnoughTrumpSpam & 2,679 (2.9\%) \\
politics & \multicolumn{1}{r|}{8,875 (1,3\%)} & me\_irl & \multicolumn{1}{r|}{70 (1.8\%)} & TrumpsTweets & 2,363 (2.5\%) \\
funny & \multicolumn{1}{r|}{8,508 (1.3\%)} & funny & \multicolumn{1}{r|}{56 (1.4\%)} & AdviceAnimals & 1,740 (1.9\%) \\
dankmemes & \multicolumn{1}{r|}{7,744 (1,1\%)} & CringeAnarchy & \multicolumn{1}{r|}{43 (1.1\%)} & USE2016 & 1,653 (1.8\%) \\
EnoughTrumpSpam & \multicolumn{1}{r|}{6,973 (1.1\%)} & EDH & \multicolumn{1}{r|}{43 (1.1\%)} & PoliticsAll & 1,401(1.5\%) \\
pics & \multicolumn{1}{r|}{5,945 (0.9\%)} & magicTCG & \multicolumn{1}{r|}{42 (1.1\%)} & dankmemes & 881 (0.9\%) \\
AskReddit & \multicolumn{1}{r|}{5,482 (0.8\%)} & dankmemes & \multicolumn{1}{r|}{40 (1.0\%)} & pics & 877 (0.9\%) \\
HOTandTrending & \multicolumn{1}{r|}{4,674 (0.7\%)} & ImGoingToHellForThis & \multicolumn{1}{r|}{39 (1.0\%)} & me\_irl & 873 (0.9\%) \\ \hline
\end{tabular}
}
\caption{Top ten subreddits for all memes, racism-related memes, and politics-related memes.}
\label{tbl:top_subreddits_groups}
\end{table*}

\subsubsection{Score Analysis}
As discussed in Section~\ref{sec:communities}, Reddit and Gab incorporate a voting system that determines the popularity of content within the Web community and essentially captures the appreciation of other users towards the shared content.
To study how users react to racist and politics-related memes, we plot the CDF of the posts' scores that contain such memes in Figure~\ref{fig:scores_groups}.

For Reddit (Figure~\ref{scores_reddit_memes}), we find that posts that contain politics-related memes are rated highly (mean score of 224.7 and a median of 5) than posts that contain non-politics memes (mean 124.9, median 4).
On the contrary, posts that contain racist memes are rated lower (average score of 94.8 and a median of 3) than other non-racist memes (average 141.6 and median 4).
On Gab (Figure~\ref{scores_gab_memes}), posts that contain politics-related memes have a similar score as non-political memes (mean 87.3 vs 82.4).
However, this does not apply for racist and non-racist memes, as non-racist memes have over 2 times higher scores than racist memes (means 84.7 vs 35.5).

Overall, this suggests that posts that contain politics-related memes receive high scores by Reddit and Gab users, while for racist memes this applies only on Reddit.

\subsubsection{Sub-Communities}
Among all the Web communities that we study, only Reddit is divided into multiple sub-communities.
We now study which sub-communities share memes with a focus on racist and politics-related content.
In Table~\ref{tbl:top_subreddits_groups}, we report the top ten subreddits in terms of the percentage over all posts that contain memes in Reddit for: 1)~all memes; 2)~racist ones; and 3)~politics-related memes.

For all three groups, the most popular subreddit is \td with 12.5\%, 9.3\%, and 26.4\%, %
respectively.
Interestingly, AdviceAnimals, a general-purpose meme subreddit, is among the top-ten sub-communities also for racist and political memes, highlighting their infiltration in otherwise non-hateful communities.

Other popular subreddits for racist memes include conspiracy (2.0\%), me\_irl (1.8\%), and funny (1.4\%) subreddits.
For politics-related memes, the majority of the subreddits are related to Donald Trump, while there also are general subreddits that talk about politics, e.g., the politics (3.0\%) and the PoliticsAll subreddit (1.5\%).

\subsection{Take-Aways}
\noindent In summary, the main take-aways of our analysis include:\smallskip
\begin{compactenum}
\item Fringe Web communities use many variants of memes related to politics and world events, possibly aiming to share weaponized information about them (Appendix~\ref{sec:appendix_interesting_images} include some examples of such memes).
For instance, Donald Trump is the KYM entry with the largest number of clusters in \dspol (2.2\%), \td (6.1\%), and Gab (2.2\%).
\item \dspol and Gab share hateful and racist memes at a higher rate than mainstream communities, as we find a considerable number of anti-semitic and pro-Nazi clusters (e.g., The Happy Merchant meme~\cite{happy_merchant_meme} appears in 1.3\% of all \dspol annotated clusters and 2.2\% of Gab's, while Adolf Hitler in 0.6\% of \dspol's).
This trend is steady over time for \dspol but ramping up for Gab.
\item Seemingly ``neutral'' memes, like Pepe the Frog (or one of its variants), are used in conjunction with other memes to incite hate or influence public opinion on world events, e.g., with images related to terrorist organizations like ISIS or world events such as Brexit.
\item Our custom distance metric successfully allows us to study the interplay and the overlap of memes, as showcased by the visualizations of the clusters and the dendrogram (see Figs.~\ref{fig:casestudy_frogs_dendrogram} and \ref{fig:clusters_graph}).
\item Reddit users are more interested in politics-related memes than other type of memes.
That said, when looking at individual subreddits, we find that \td is the most active one when it comes to posting memes in general.
It is also the subreddit where most racism and politics-related memes are posted.\smallskip
\end{compactenum}

\section{Influence Estimation} \label{sec:influence_estimation}
So far we have studied the dissemination of memes by looking at Web communities in isolation.
However, in reality, these influence each other: e.g., memes posted on one community are often re-posted to another.
Aiming to capture the relationship between them, we use a statistical model known as Hawkes Processes~\cite{linderman2014,lindermanArxiv}, which describes how events occur over time on a collection of processes.
This maps well to the posting of memes on different platforms: each community can be seen as a process, and an event occurs each time a meme image is posted on one of the communities.
Events on one process can cause impulses that can increase the likelihood of subsequent events, including other processes, e.g., a person might see a meme on one community and re-post it, or share it to a different one.
This approach allows us to assess the causality of events, hence it is a far better approach when compared to simple approaches like looking at the timeline of specific memes or pHashes.

\subsection{Hawkes Processes}
To model the spread of memes on Web communities, we use a similar approach as in our previous work~\cite{zannettou2017web}, which looked at the spread of mainstream and alternative news URLs. Next, we provide a brief description, and present an improved method for estimating influence.

We use five processes, one for each of our seed Web communities (\dspol, Gab, and \td), as well as Twitter and Reddit, fitting a separate model for each meme cluster.
Fitting the model to the data yields a number of values: background rates for each process, weights from each process to each other, and the shape of the impulses an event on one process causes on the rates of the others.
The background rate is the expected rate at which events will occur on a process {\em without} influence from the communities modeled or previous events; this captures memes posted for the first time, or those seen on a community we do not model and then reposted on a community we do.
The weights from community-to-community indicate the effect an event on one has on the others; for example, a weight from Twitter to Reddit of 1.2 means that each event on Twitter will cause an expected $1.2$ additional events on Reddit.
The shape of the impulse from Twitter to Reddit determines how the probability of these events occurring is distributed over time; typically the probability of another event occurring is highest soon after the original event and decreases over time.

\begin{figure}[]
\centering
\includegraphics[width=0.8\columnwidth]{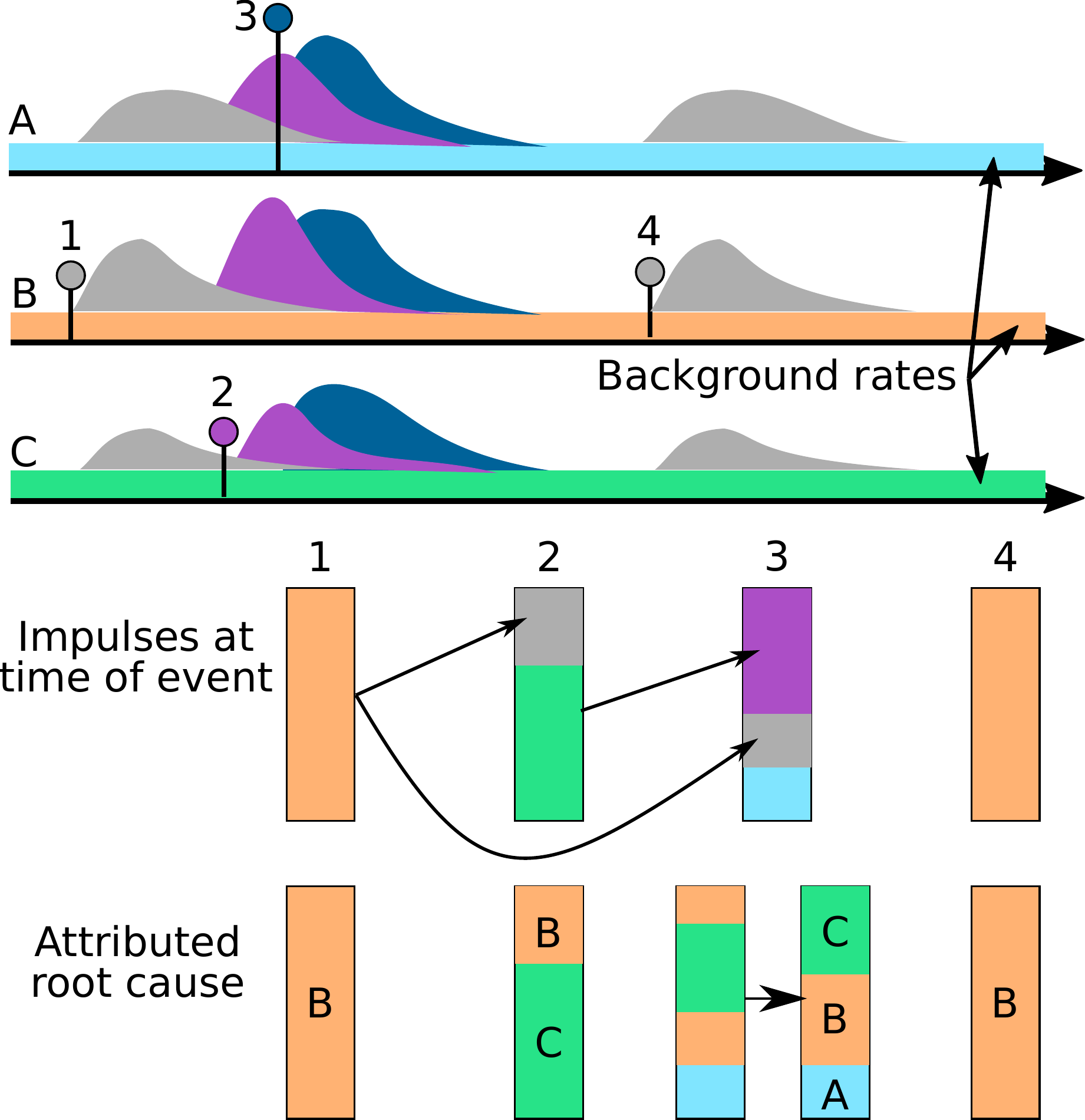}
\caption{A Hawkes model with three processes.  Events cause impulses that increase the rate of subsequent events in the same or other processes.  By looking at the impulses present when events occur, the probability of a process being the root cause of an event can be determined. Note that on the second part of the figure, colors represent events while arrows represent impulses between the events. }
\label{fig:hawkes_explanation}
\end{figure}

Figure~\ref{fig:hawkes_explanation} illustrates a Hawkes model with three processes.
The first event occurs on process B, which causes an increase in the rate of events on all three processes.
The second event then occurs on process C, again increasing the rate of events on the processes.
The third event occurs soon after, on process A.
The fourth event occurs later, again caused by the background arrival rate on process B, after the increases in arrival rate from the other events have disappeared.

To understand the influence different communities have on the spread of memes, we want to be able to attribute the cause of a meme being posted back to a specific community.
For example, if a meme is posted on \dspol and then someone sees it there and posts it on Twitter where it is shared several times, we would like to be able to say that \dspol was the \emph{root cause} of those events.
Obviously, we do not actually know where someone saw something and decided to share it, but we can, using the Hawkes models, determine the \emph{probability} of each community being the root cause of an event.

Looking again at Figure~\ref{fig:hawkes_explanation}, we see that events 1 and 4 are caused directly by the background rate of process B.
This is because, in the case of event 1, there are no previous events on other processes, and in the case of event 4, the impulses from previous events have already stopped.
Events 2 and 3, however, occur when there are multiple possible causes: the background rate for the community and the impulses from previous events.
In these cases, we assign the probability of being the root cause in proportion to the magnitudes of the impulses (including the background rate) present at the time of the event.
For event 2, the impulse from event 1 is smaller than the background rate of community C, so the background rate has a higher probability of being the cause of event 2 than event 1.
Thus, most of the cause for event 2 is attributed to community C, with a lesser amount to B (through event 1).
Event 3 is more complicated: impulses from both previous events are present, thus the probability of being the cause is split three ways, between the background rate and the two previous events.
The impulse from event 2 is the largest, with the background rate and event 1 impulse smaller.
Because event 2 is attributed both to communities B and C, event 3 is partly attributed to community B through both event 1 and event 2.

In the rest of our analysis, we use this new measure. 
This is a substantial improvement over the influence estimation in~\cite{zannettou2017web}, which used the weights from source to destination community, multiplied by the number of events on the source to estimate influence.
However, this only looks at influence across a single ``hop'' and would not allow us to understand the source community's influence as memes spread onwards from the destination community.
The new method allows us to gain an understanding of where memes that appear on a community originally come from, and how they are likely to spread from community to community from the original source.

\subsection{Influence}

\begin{table}[]
\centering
\small
\begin{tabular}{@{}llllll@{}}
\toprule
\dspol & Twitter & Reddit & \tdshort  & Gab   &  \\ \midrule
1,574,045               & 865,885  & 581,803 & 81,924 & 44,918 &  \\ \bottomrule
\end{tabular}
\caption{Events per community from the 12.6K clusters.} %
\label{tbl:hawkes_input}
\end{table}

We fit Hawkes models using Gibbs sampling as described in~\cite{lindermanArxiv} for the 12.6K annotated clusters; in Table~\ref{tbl:hawkes_input}, we report the total number of meme images posted to each community in these clusters.
As seen in Table~\ref{tbl:hawkes_input}, \dspol has the greatest number of memes posted, followed by Twitter and then Reddit.
In terms of total images collected (see Table~\ref{tbl:datasets_summary}), Twitter and Reddit have many more than \dspol.
However, many of the images on these communities might not be memes; additionally, because our clusters are created from the memes present on only \dspol, \td, and Gab (as these are the communities primarily of interest in this paper), it is possible that there are memes on Twitter and Reddit that are not included in the clusters.
This yields an additional interesting question: how \emph{efficient} are different communities at disseminating memes?

First, we report the source of events in terms of the percent of events on the destination community.
This describes the results in terms of the data as we have collected it, e.g., it tells us the percentage of memes posted on Twitter that were caused by \dspol.
The second way we report influence is by normalizing the values by the total number of events in the source community, which lets us see how much influence each community has, relative to the number of memes they post---in other words, their efficiency.

\begin{figure}[t!]
\centering
\includegraphics[width=\columnwidth]{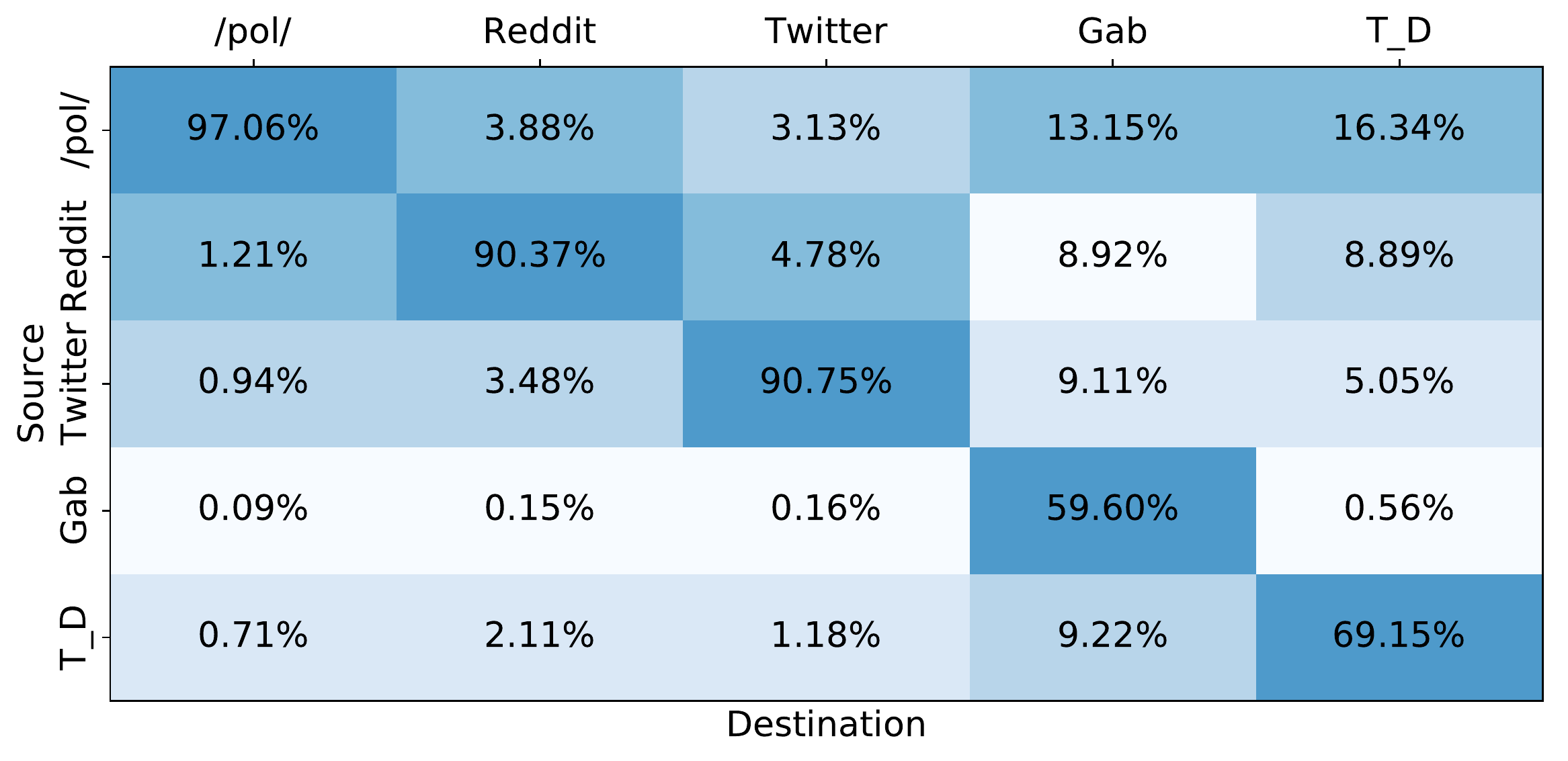}
\caption{Percent of \textit{destination} events caused by the source community on the destination community.  Colors indicate the largest-to-smallest influences per destination.}
\label{fig:hawkes_cause}
\end{figure}

We first look at the influence of all clusters together.
Figure~\ref{fig:hawkes_cause} shows the percent of events on each \emph{destination} community caused by each \emph{source} community.
The values from one community to the same community (for example, from \dspol to \dspol) include both events caused by the background rate of that community and events caused by previous events within that community; these values are the largest influence for each community.
After this, \dspol is the strongest source of influence for Reddit, \td, and Gab, but not for Twitter, which is most influenced by Reddit.
Interestingly, although Twitter has a greater number of memes posted than Reddit, it causes less influence.
Perhaps there is less original content posted directly to Twitter.

\begin{figure}[t!]
\centering
\includegraphics[width=\columnwidth]{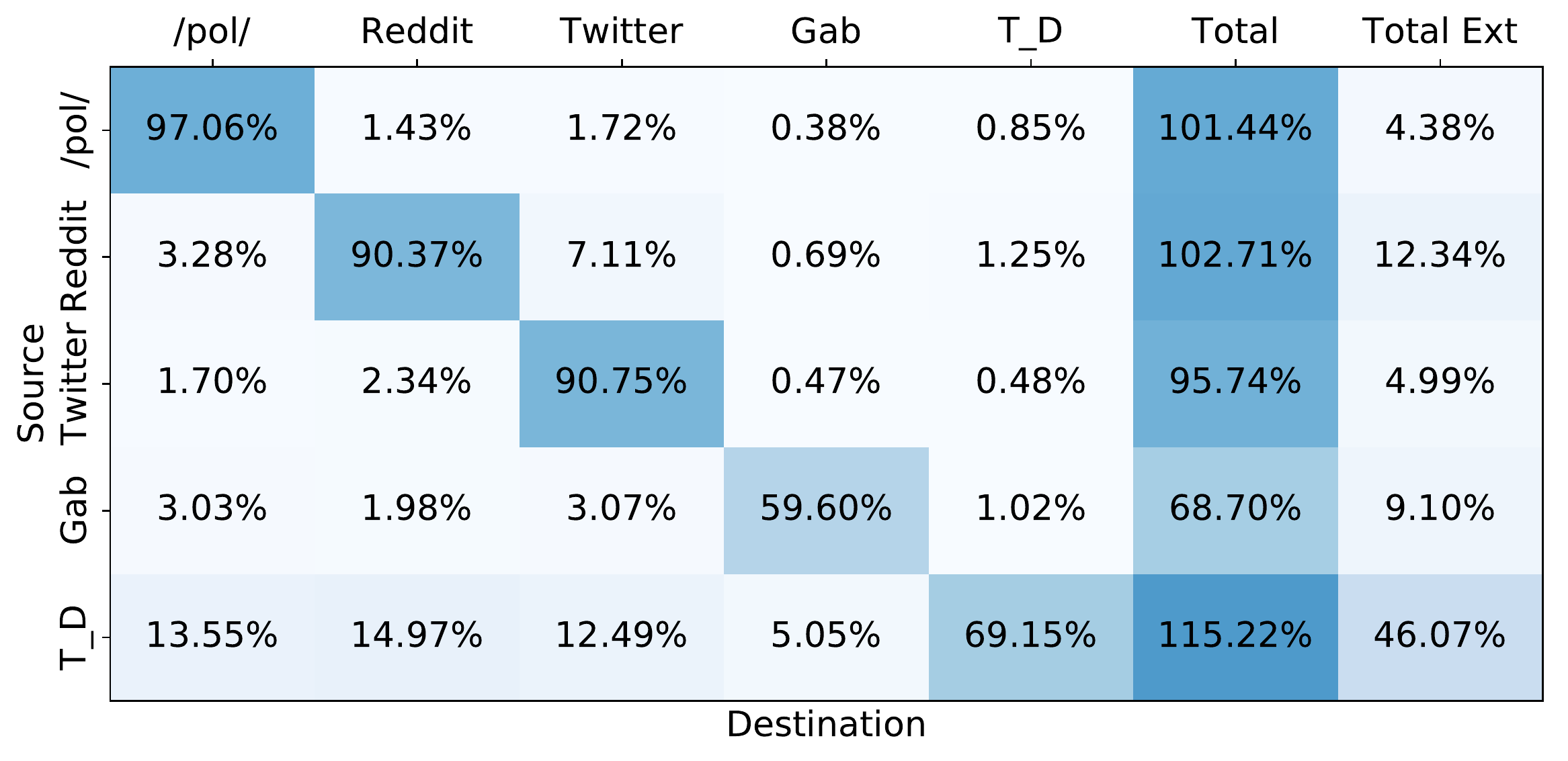}
\caption{Influence from source to destination community, normalized by the number of events in the \textit{source} community.  Columns for total influence and total external influence are shown.}
\label{fig:hawkes_cause_norm}
\end{figure}

Next, we look at the normalized influence of all clusters together.
Figure~\ref{fig:hawkes_cause_norm} shows the influence that a source community has on a destination community, normalized by the total number of memes posted on the \emph{source} community.
The values can be understood as an indication of how much influence a community has, relative to the frequency of memes posted.
For example, the influence Reddit has on Twitter is equal to 5.71\% of the total events on Reddit.
If the sum of values for a source is less than 100\%, it implies that many of the posts on the source community were caused by other communities, or that posts on the source community do not cause many posts on other communities.

There are several interesting things to note in Figure~\ref{fig:hawkes_cause_norm}.
First, \td has by far the greatest influence for the number of memes posted on it.
This is particularly apparent when looking at just external influence, where \td has more than 4 times as much influence than the rest of Reddit, the closest other community.
Memes from this community spread very well to all of the other communities.
While \dspol has a large total influence on the other communities (as seen in Figure~\ref{fig:hawkes_cause}), when normalized by its size, it has the smallest external influence: just 4.03\%.
Most of the memes posted on \dspol do not spread to other communities.
Both Gab and Twitter have a total normalized influence of less than 100\%; much less in Gab's case, although it has higher external influence.

\begin{figure}[t!]
\centering
\includegraphics[width=\columnwidth]{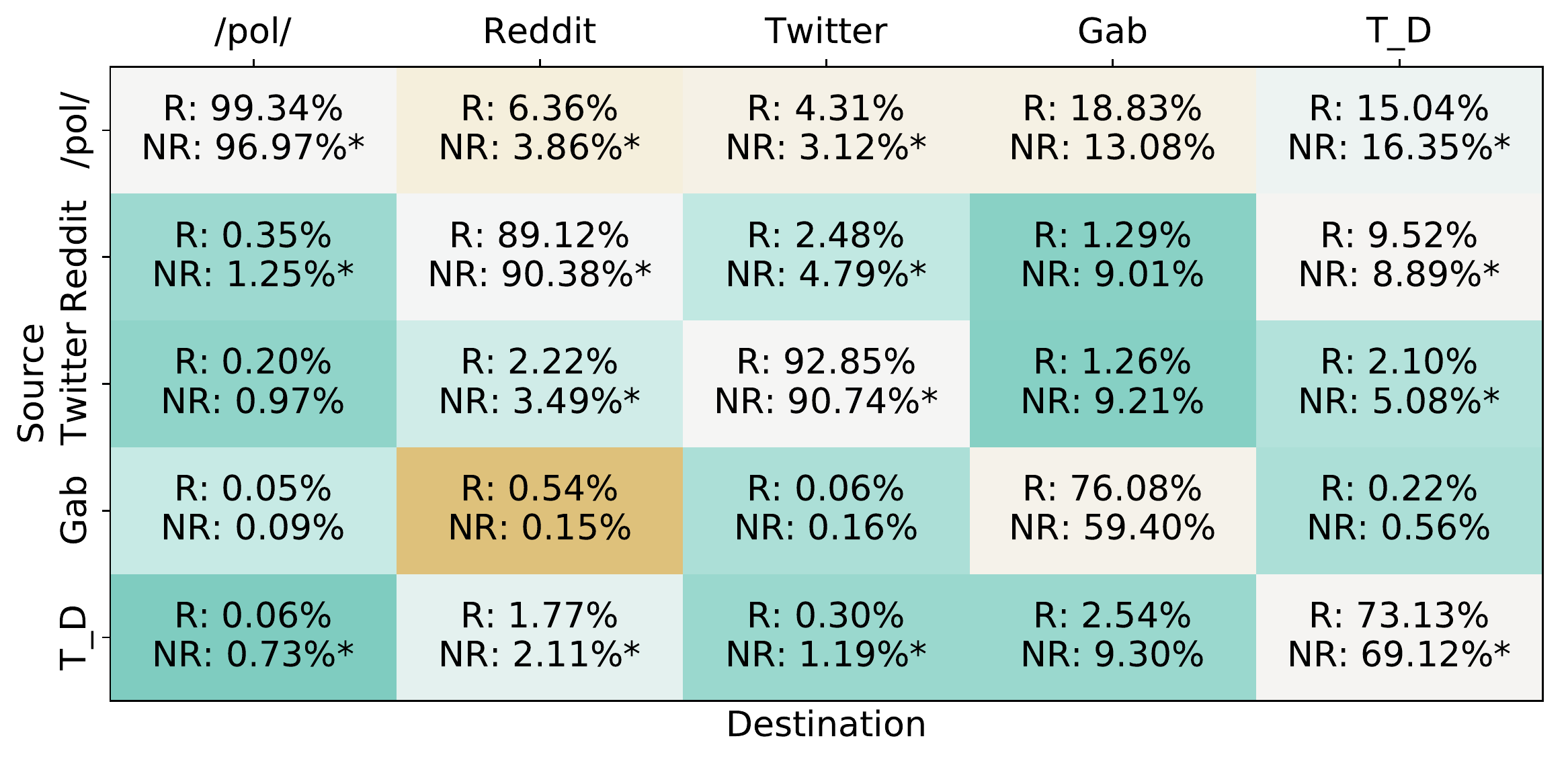}
\caption{Percent of the destination community's racist (R) and non-racist (NR) meme postings caused by the source community.  Colors indicate the percent difference between racist and non-racist.}
\label{fig:hawkes_racist_from}
\end{figure}

\begin{figure}[t!]
\centering
\includegraphics[width=\columnwidth]{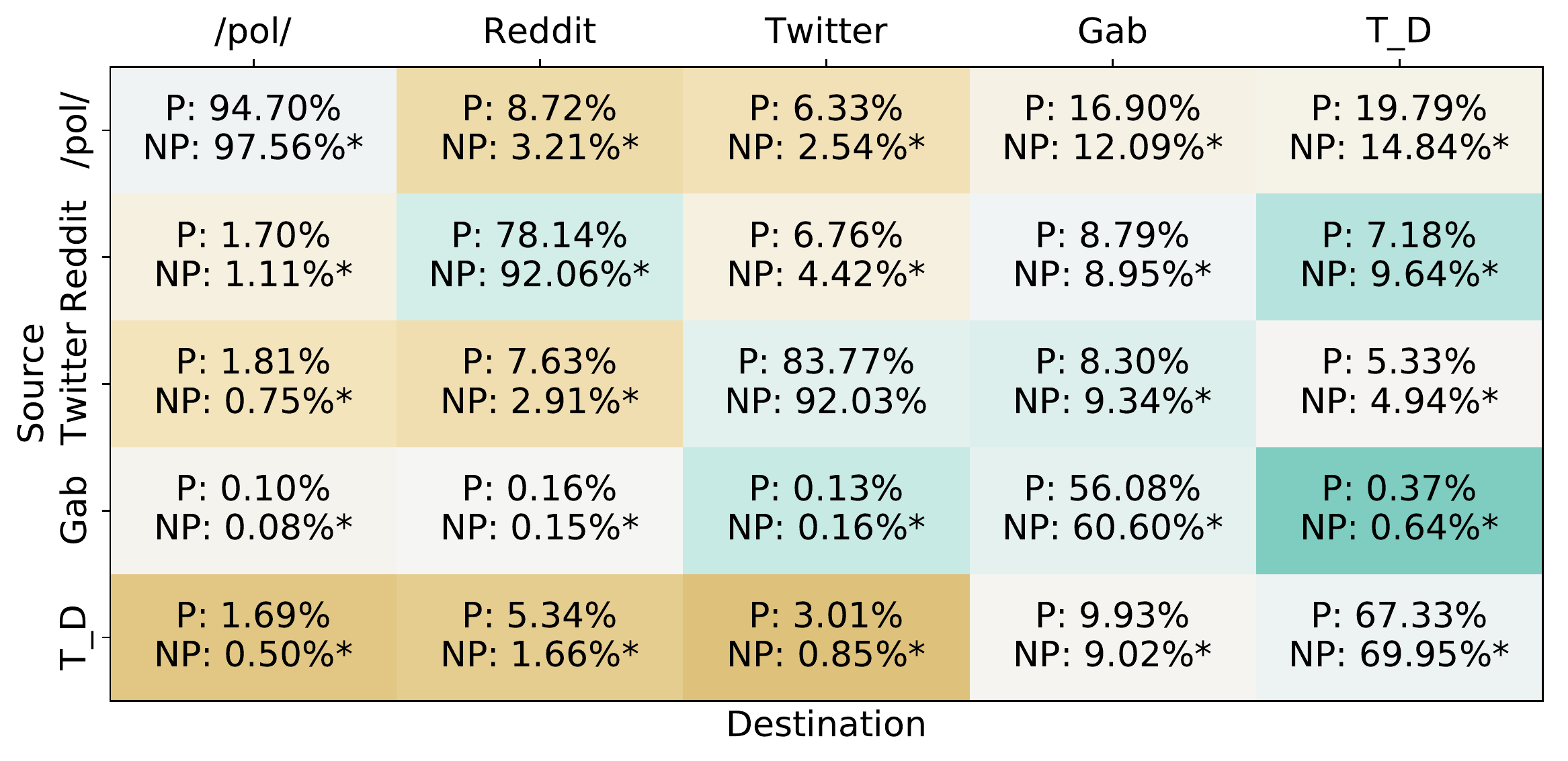}
\caption{Percent of the destination community's political (P) and non-political (NP) meme postings caused by the source community.  Colors indicate the percent difference between political and non-political.}
\label{fig:hawkes_political_from}
\end{figure}

Using the clusters identified as either racist or non-racist (see the end of Section~\ref{sec:meme_popularity}),
we compare how the communities influence the spread of these two types of content.
Figure~\ref{fig:hawkes_racist_from} shows the percentage of both the destination community's racist and non-racist meme posts caused by the source community.
We perform two-sample Kolmogorov-Smirnov tests to compare the distributions of influence from the racist and non-racist clusters; cells with statistically significant differences between influence of racist/non-racist memes (with $p {<} 0.01$) are reported with a * in the figure.
\dspol has the most \emph{total} influence for both racist and non-racist memes, with the notable exception of Twitter, where Reddit has the most the influence.
Interestingly, while the percentage of racist meme posts caused by \dspol is greater than non-racist for Reddit, Twitter, and Gab, this is \emph{not} the case for \td.
The only other cases where influence is greater for racist memes are Reddit to \td and Gab to Reddit.

When looking at political vs non political memes (Figure~\ref{fig:hawkes_political_from}), we see a somewhat different story.
Here, \dspol influences \td more in terms of political memes.
Further, we see differences in the \emph{percent} increase and decrease of influence between the two figures (as indicated by the cell colors).
For example, Twitter has a relatively larger difference in its influence on \dspol and Reddit for political and non-political memes than for racist and non-racist memes, but a smaller difference in its influence on Gab and \td.
This exposes how different communities have varying levels of influence depending on the type of memes they post.
\begin{figure}[t!]
\centering
\includegraphics[width=\columnwidth]{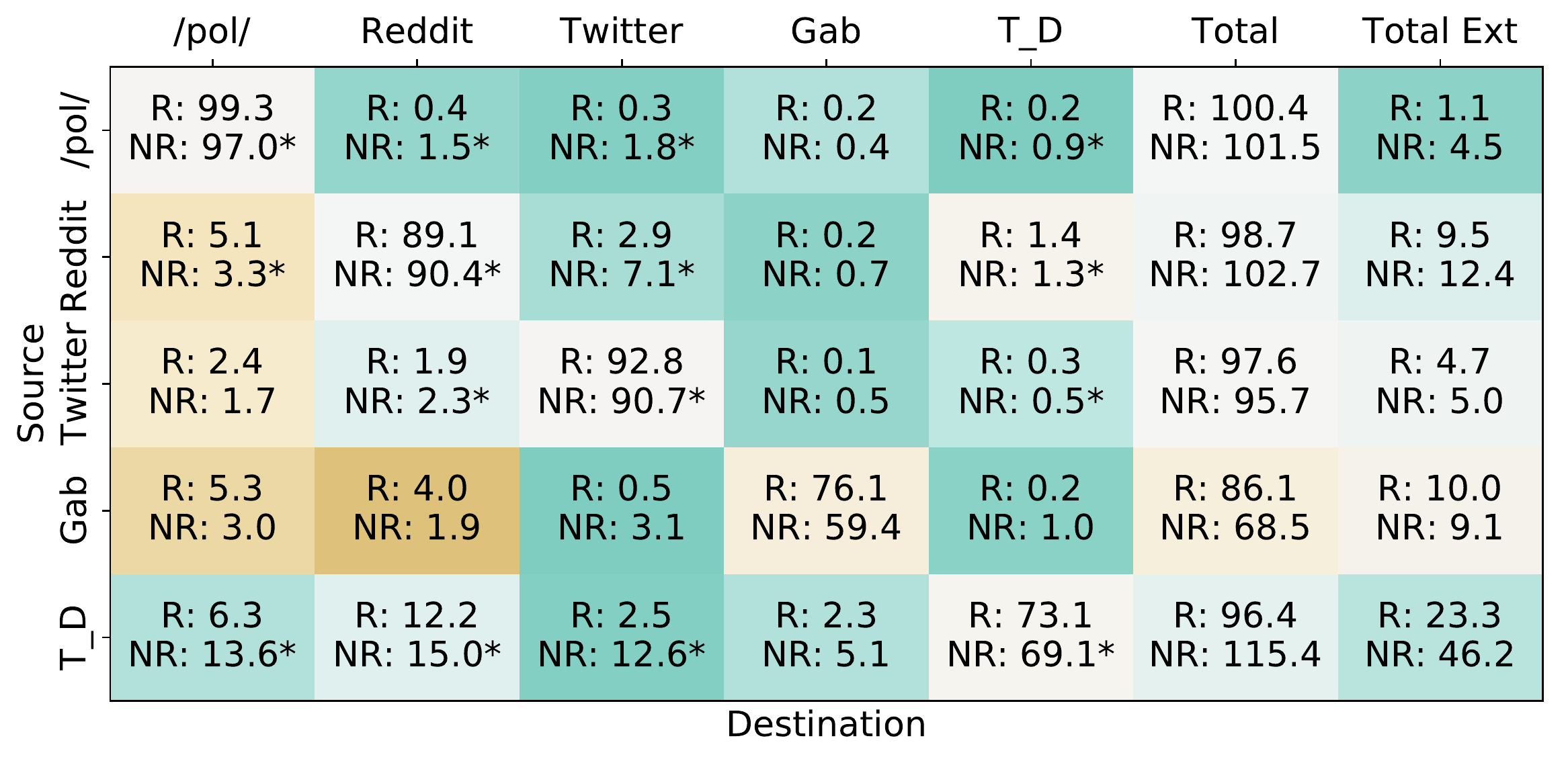}
\caption{Influence from source to destination community of racist and non-racist meme postings, normalized by the number of events in the \textit{source} community.}
\label{fig:hawkes_racist_norm}
\end{figure}

\begin{figure}[t!]
\centering
\includegraphics[width=\columnwidth]{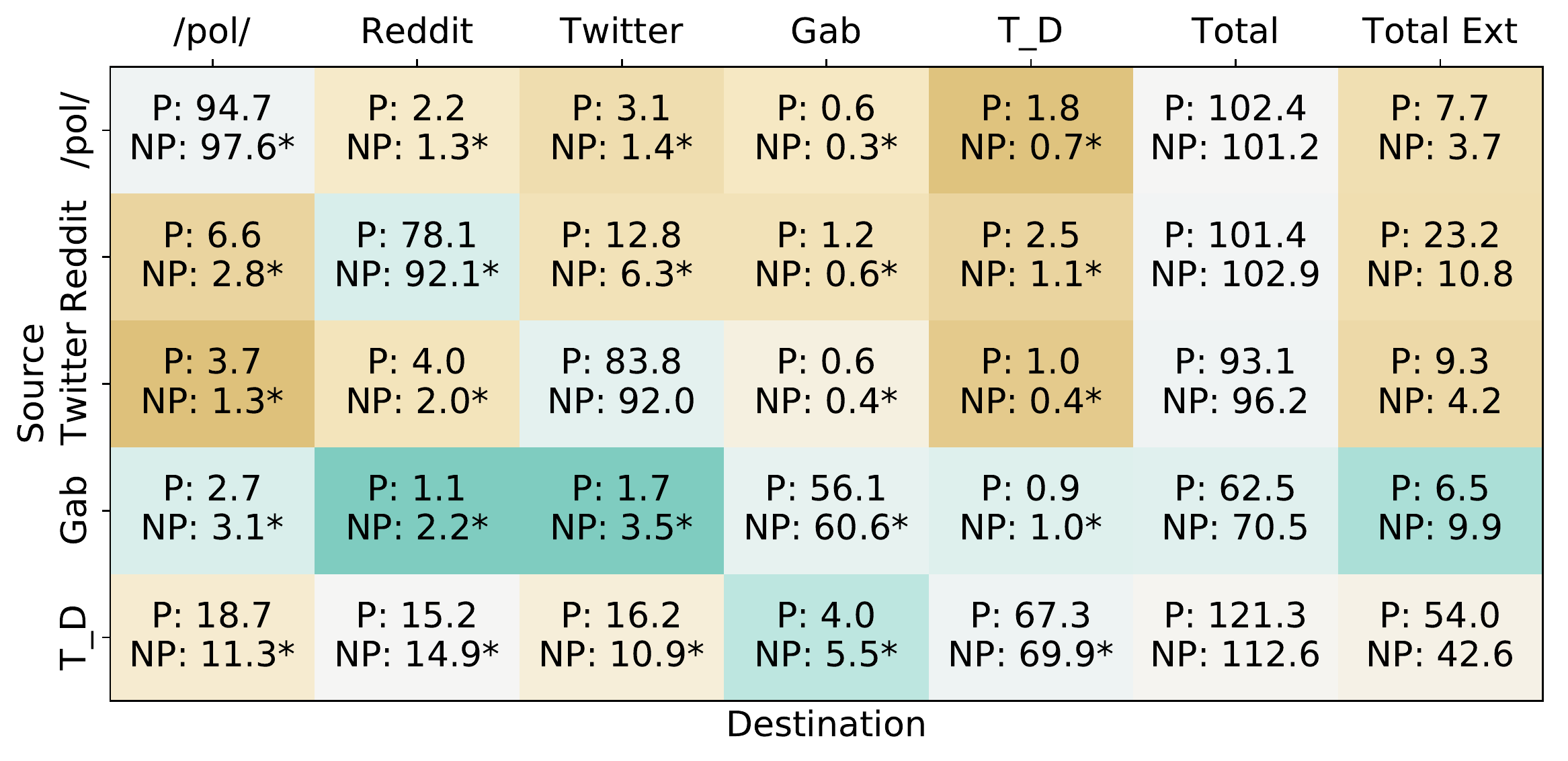}
\caption{Influence from source to destination community of political and non-political meme postings, normalized by the number of events in the \textit{source} community.}
\label{fig:hawkes_political_norm}
\end{figure}

While examining the raw influence provides insights into the meme ecosystem, it obscures notable differences in the meme posting behavior of the different communities.
To explore this, we look at the \emph{normalized} influence in Figure~\ref{fig:hawkes_racist_norm} (racist/non-racist memes) and Figure~\ref{fig:hawkes_political_norm} (political/non-political memes).
As mentioned previously, normalization reveals how \emph{efficient} the communities are in disseminating memes to other communities by revealing the \emph{per meme} influence of meme posts.
First, we note that the percent change in influence for the dissemination of racist/non-racist memes is quite a bit larger than that for political/non-political memes (again, indicated by the coloring of the cells).
More interestingly, both figures show that, contrary to the \emph{total} influence, \dspol is the \emph{least} influential when taking into account the number of memes posted.
While this might seem surprising, it actually yields a subtle, yet crucial aspect of \dspol's role in the meme ecosystem:
\dspol (and 4chan in general) acts as an evolutionary microcosm for memes.
The constant production of new content~\cite{hine2017kek} results in a ``survival of the fittest''~\cite{muhinterview} scenario.
A staggering number of memes are posted on \dspol, but only the {\em best} actually make it out to other communities.
To the best of our knowledge, this is the first result quantifying this analogy to evolutionary pressure.

\descr{Take-Aways.}
There are several take-aways from our measurement of influence.
We show that \dspol is, generally speaking, the most influential disseminator of memes in terms of raw influence.
In particular, it is more influential in spreading \emph{racist} memes than non-racist one, and this difference is deeper than in any other community.
There is one notable exception: \dspol is more influential in terms of \emph{non-racist} memes on \td.
Relatedly, \dspol has generally more influence in terms of spreading political memes than other communities.
When looking at the normalized influence, however, we surface a more interesting result: \dspol is the \emph{least} efficient in terms of influence while \td is the \emph{most} efficient.
This provides new insight into the meme ecosystem: there are clearly evolutionary effects.
Many meme postings do not result in further dissemination, and one of the key components to ensuring they are disseminated is ensuring that new ``offspring'' are continuously produced.
\dspol's ``famed'' meme magic, i.e., the propensity to produce and heavily push memes, is thus the most likely explanation for \dspol's influence on the Web in general.

\section{Related Work}
\label{sec:related_work}

We now review prior work studying the detection, evolution, and propagation of memes, their popularity, as well as various case studies. %
For each work, we also report whether they consider text, images, and/or videos.

\descr{Detection and Propagation of Memes.}
Leskovec et al.~\cite{leskovec2009memetracking} perform large-scale tracking of text-based memes, focusing on news outlets and blogs.
Ferrara et al.~\cite{ferrara2013clustering} detect text memes using an unsupervised framework based on clustering techniques.
Dang et al.~\cite{dang2015avf} study memes on Reddit by clustering submissions and using a set of similarity scores based on Google Tri-grams. 
Ratkiewicz et al.~\cite{ratkiewicz2010detecting} introduce Truthy, a framework supporting the analysis of the diffusion of text-based, politics-related memes on Twitter. Babaei et al.~\cite{babaei2016efficiency} study Twitter users' preferences, with respect to information sources, including how they acquire memes.
Romero et al.~\cite{romero2011differences} study meme propagation on Twitter via hashtags, finding differences according to the topic and that politically-related hashtags, mainly about controversial topics, are persistent on the platform. 
Dubey et al.~\cite{dubey2018memesequencer} extract a rich semantic embedding corresponding to the template used for the creation of meme images, using deep learning and optical character recognition techniques.
They demonstrate the efficacy of their approach on a variety of tasks ranging from image clustering to virality prediction on datasets obtained from Reddit as well as scraped data from sites for generating meme images like \url{memegenerator.net} and \url{quickmeme.com}.
By contrast, we focus on the detection and propagation of image-based memes without limiting our scope to image macros.
To this end, we reduce the dimensionality of raw images using perceptual hashing, and use clustering techniques to identify groups of memes. 
We also detect and study the propagation of memes across multiple Web communities, using publicly available data from memes annotation sites (i.e., KYM).

\descr{Popularity of Memes.}
Weng et al.~\cite{weng2012competition,weng2014predicting} study the popularity of memes spreading as hashtags on Twitter.
They model virality using an agent-based approach, taking into account that users have a limited capacity in receiving/viewing memes on Twitter.
They study the features that make memes popular, finding that those based on network community structures are strong indicators of popularity. %
Tsur and Rappoport~\cite{tsur2015dont} predict popularity of text-based memes on Twitter using linguistic characteristics as well as cognitive and domain features.
Ienco et al.~\cite{Ienco2010thememe} study memes propagating via text, images, audio, and video on the Yahoo! Meme platform (a platform discontinued in 2012), aiming to predict virality and select memes to be shown to users after login.
Coscia~\cite{coscia2013competition} 
studies meme images and distinguishes the traits that make them more likely to be popular on the Web, presenting a case study on the Quickmeme generator site.

Whereas, we also study the popularity of memes, however, unlike previous work, we rely on a multi-platform approach, encompassing data from \dspol, Reddit, Twitter, and Gab,
and show that the popularity of memes depends on the Web community and its ideology.
For instance, \dspol is well-known for its anti-semitic ideology and in fact the ``Happy Merchant'' meme~\cite{happy_merchant_meme} is the 3rd most popular meme on \dspol.

\descr{Evolution of Memes.}
Adamic et al.~\cite{adamic2016information} study the evolution of text-based memes on Facebook, showing that it can be modeled by the Yule process.
They find that memes significantly evolve and new variants appear as they propagate, and that specific communities within the network diffuse specific variants of a meme.
Bauckhage~\cite{bauckhage2011insights} study the temporal dynamics of 150 memes using data from Google Insights as well as social bookmarking services like Digg, showing that different communities exhibit different interests/behaviors for different memes, and that epidemiology models can be used to predict the evolution and popularity of memes.
Simmons et al.~\cite{simmons2011memes} focus on detecting and studying the evolution of memes that are propagated via quoted text, finding that mutations of text are surprisingly frequent.

By contrast, we study the temporal aspect of memes using Hawkes processes.
This statistical framework allows us to assess the causality of the posting of memes on various Web communities, thus modeling their evolution and their influence across multiple communities.

\descr{Generating Memes.} Oliveira et al.~\cite{oliveira2016one} study the challenges of creating memes, both for humans and automated bots.
They build and test an automated meme creator bot, which combines a headline with an image macro and posts the resulting meme on Twitter:
while generated posts are easily recognizable as memes, the bot fails to produce {\em humorous} memes -- a task that is challenging for humans too.
Wang and Wen~\cite{wang2015ican} study memes from sites like \url{memegenerator.net} that have images embedded with text, focusing on the correlations between the text and the characteristics of the image, ultimately proposing a non-paranormal approach for generating text from an image macro.

\descr{Case Studies.}
Heath et al.~\cite{heath2001emotional} present a case study of how people perceive memes with a focus on urban legends, finding that they are more willing to share memes that evoke stronger disgust.
Shifman and Thelwall~\cite{shifman2009assessing} study the diffusion, evolution, and translation of single memes (i.e., specific jokes) by relying on a mixed-methods approach based on clustering and search engine queries on the Web.
Shifman~\cite{shifman2012anatomy} analyzes 30 video memes on YouTube, both qualitatively and quantitatively, finding that meme videos have several common features like humor, simplicity, and repetitiveness.
Xie et al.~\cite{xie2011tracking} also focus on YouTube memes, performing a large-scale keyword-based search for videos related to the Iranian election in 2009, extracting frequently used images and video segments. 
They show that most of the videos are not original, thus, meme-related techniques can be exploited to deduplicate content and capture the content diffusion on the Web.
Finally, Dewan et al.~\cite{dewan2017towards} study the sentiment and content of images that are disseminated during crisis events like the 2015 Paris terror attacks.
They analyze 57K images related to the attacks,
finding instances of misinformation and conspiracy theories.

We also present a case study focusing on image memes of Pepe the Frog (see the discussion about Figure~\ref{fig:casestudy_frogs_dendrogram}).
This showcases both the overlap and the diversity of certain memes, as well as how memes can be influenced by real-world events, with new variants being generated.
For instance, after the UK Brexit referendum in 2016, memes with Pepe the Frog started to be used in the Brexit context (see Appendix~\ref{sec:appendix_interesting_images}).
This also demonstrates how our processing pipeline can effectively uncover interesting overlaps and characteristics {\em across} memes.

\descr{Fringe Communities.} Previous work has also shed light on fringe Web communities like 4chan, Gab, and sub-communities within Reddit.
Bernstein et al.~\cite{bernstein20114chan} study the ephemerality and anonymity features of the 4chan community using data from the Random board (\dsb).
Hine et al.~\cite{hine2017kek} focus on \dspol, analyzing 8M posts and detecting a high volume of hate speech as well as the phenomenon of ``raids,'' i.e., coordinated attacks aimed at disrupting other services.
Zannettou et al.~\cite{zannettou2018what} analyze 22M posts from 336K users on Gab, finding that hate speech occurs twice as much as in Twitter, but twice less than \dspol.
They also highlight a strong presence of alt-rights users previously banned from mainstream social networks.
Snyder et al.~\cite{snyder2017fifteen} measure doxing on 4chan and 8chan,
while Chandrasekharan et al.~\cite{chandrasekharan2017bag} introduce a computational approach to detect abusive content also looking at 4chan and Reddit.

Finally, Hawkes processes have also been used to quantify influence of fringe Web communities like \dspol and \td to mainstream ones like Twitter in the context of misinformation~\cite{zannettou2017web}.
We follow a similar approach here, but use an improved method of determining the influence of the different communities.

\section{Discussion \& Conclusion} \label{sec:conclusion}

In this paper, we presented a large-scale measurement study of the meme ecosystem.
We introduced a novel image processing pipeline and ran it over 160M images collected from four Web communities (4chan's \dspol, Reddit, Twitter, and Gab).
We clustered images from fringe communities (\dspol, Gab, and Reddit's \td) based on perceptual hashing and a custom distance metric, annotated the clusters using data gathered from Know Your Meme, and analyzed them along a variety of axes.
We then associated images from all the communities to the clusters to characterize them through the lens of memes and the influence they have on each other.

Our analysis highlights that the meme ecosystem is quite complex, with intricate relationships between different memes and their variants.
We found important differences between the memes posted on different communities (e.g., Reddit and Twitter tend to post ``fun'' memes, while Gab and \dspol racist or political ones).
When measuring the influence of each community toward disseminating memes to other Web communities, we found that \dspol has the largest overall influence for racist and political memes, however, \dspol was the least \emph{efficient}, i.e., in terms of influence w.r.t.~the total number of memes posted,
while \td is very successful in pushing memes to both fringe and mainstream Web communities.

Our work constitutes the first attempt to provide a multi-platform measurement of the meme ecosystem, with a focus on fringe and potentially dangerous communities.
Considering the increasing relevance of digital information on world events, our study provides a building block for future cultural anthropology work, as well as for building systems to protect against the dissemination of harmful ideologies.
Moreover, our pipeline can already be used by social network providers to assist the identification of hateful content; for instance, Facebook is taking steps to ban Pepe the Frog used in the context of hate~\cite{fb-pepe}, and our methodology can help them automatically identify hateful variants.
Finally, our pipeline can be used for tracking the propagation of images from any context or other language spheres, provided an appropriate annotation dataset.

\descr{Performance.}
We also measured the time that it takes to associate images posted on Web communities to memes.
All other steps in our system are one-time batch tasks, only executed if the annotations dataset is updated.
To ease presentation, we only report the time to compare all the 74M images from Twitter (the largest dataset) against the medoids of all 12K annotated clusters:
it took about 12 days on our infrastructure, equipped with two NVIDIA Titan Xp GPUs.
This corresponds to 14ms per image, or 73 images per second.
Note that, if new GPUs are added to our infrastructure, the workload would be divided equally across all GPUs.

\descr{Future work.} In future work, we plan to include memes in video format, thus extending to other communities (e.g., YouTube).
We also plan to study the {\em content} of the posts that contain memes, incorporating OCR techniques to capture associated text-based features that memes usually contain, and improving on KYM annotations via crowdsourced labeling.
While shedding light on the Internet meme ecosystem, our findings yield a number of future directions exploring, e.g., where memes are first created, understanding components of a meme that might increase/decrease its chance of dissemination, gaining a better understanding of the various families of memes, how they influence public opinion, and so on.

\descr{Acknowledgments.} We thank the anonymous reviewers and our shepherd Christo Wilson for their insightful feedback. This project has received funding from the European Union's Horizon 2020 Research and Innovation program under the Marie Sk\l{}odowska-Curie ENCASE project (Grant Agreement No. 691025).
We also gratefully acknowledge the support of the NVIDIA Corporation, for the donation of the two Titan Xp GPUs used for our experiments.
Finally, we want to thank /pol/'s anonymous users for their ``creativity'' and ``invaluable'' input to this paper (in meme form; see Appendix~\ref{sec:appendix_interesting_images}).

{\small
\bibliographystyle{abbrv}
%\bibliography{refs}

}

\appendix

\section{Clustering Parameter Selection} \label{sec:appendix_clustering}
Our implementation uses the DBSCAN algorithm with a clustering threshold equal to 8.
To select this threshold, we perform the clustering step while varying the distances.
Table~\ref{tbl:clustering_statistics_evaluation} shows the number of clusters and the percentage of images that are regarded as noise by the clustering algorithm for varying distances.
We observe that, for distances 2-4, we have a substantially larger percentage of noise, while with distance 10 we have the least percentage of noise.
With distances between 6 and 8 we observe that we get a larger number of clusters than the other distances, while the noise percentages are 73\% and 63\%, respectively.

\begin{figure}[t]
\centering
\includegraphics[width=0.7\columnwidth]{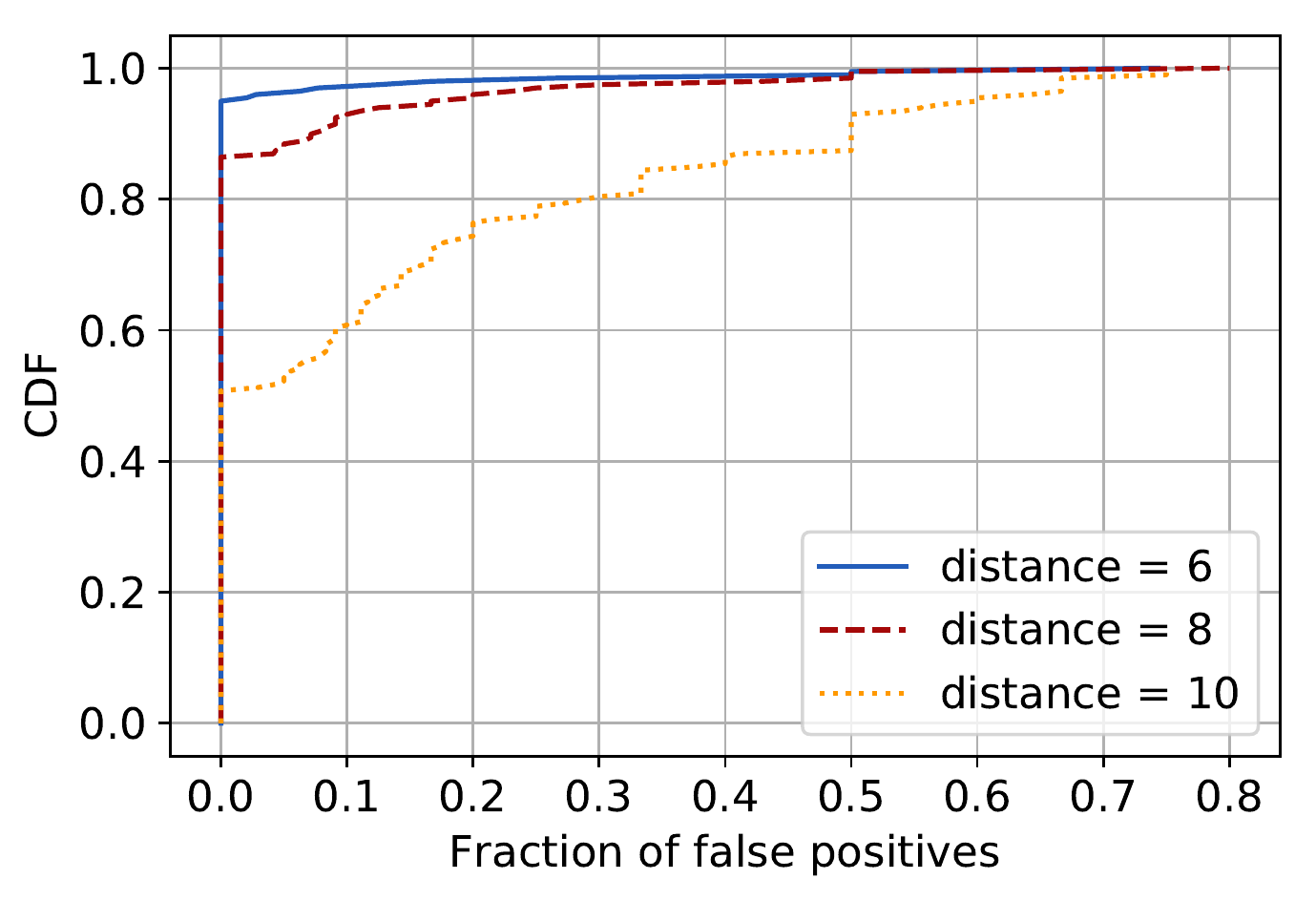}
\caption{Fraction of false positives in clusters with varying clustering distance. }
\label{fig:clustering_evaluation}
\end{figure}

To further evaluate the clustering performance for varying distances, we randomly select 200 clusters and manually calculate the number of images that are false positives within each cluster. Figure~\ref{fig:clustering_evaluation} shows the CDF of the false positive fraction in the random sample of clusters for distances 6, 8, and 10 (we disregard distances 2-4 due to the high percentage of noise).
Distance 10 yields a high number of false positives, while distances 6-8 the overall false positives are below 3\%.
Therefore, we investigate the impact of these false positives in the overall dataset, looking at all posts that contain false and true positives in the random sample of 200 clusters, using distance 8.
We find that the false positives have little impact as they occur substantially fewer times than true positives: the percentage of true positives over the set of false positives and true positives is 99.4\%.
Thus, due to the larger number of clusters, the acceptable false positive performance, and the smaller percentage of noise (when compared to distances 2-6), we elect to use as a threshold the perceptual distance that is equal to 8.

\begin{table}[t]
\centering
\small
\begin{tabular}{@{}rrr@{}}
\toprule
\multicolumn{1}{l}{\textbf{Distance}} & \multicolumn{1}{c}{\textbf{\#Clusters}} & \multicolumn{1}{c}{\textbf{\%Noise}} \\ \midrule
2 & 30,327 & 82.9\% \\
4 & 34,146 & 78.5\% \\
6 & 37,292 & 73.0\% \\
8 & 38,851 & 62.8\% \\
10 & 30,737 & 27.8\% \\ \bottomrule
\end{tabular}%
\caption{Number of clusters and percentage of noise for varying clustering distances.}
\label{tbl:clustering_statistics_evaluation}
\end{table}

\begin{figure*}[t]
\includegraphics[width=0.95\textwidth]{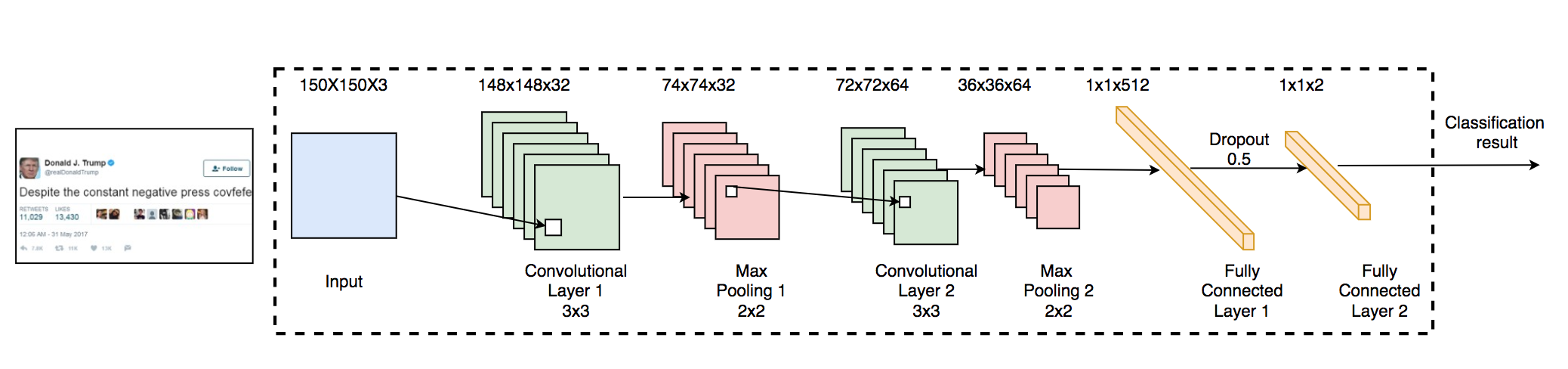}
\vspace{-0.2cm}
\caption{Architecture of the deep learning model for detecting screenshots from Twitter, /pol/, Reddit, Instagram, and Facebook.}
\label{fig:ml_architecture}
\vspace{-0.2cm}
\end{figure*}

\section{KYM and Clustering Annotation Evaluation} \label{sec:appendix_clustering_annotation_evaluation}

While KYM might not be a household name, the site is seemingly the largest curated collection of memes on the Web, i.e., KYM is as close to an ``authority'' on memes as there is.
That said, crowdsourcing \emph{is} an aspect of how KYM works, and thus there might be questions as to how ``legitimate'' some of the content is.
To this end, we set out to measure the quality of KYM by sampling a number of pages and manually examining them.
This is clearly a subjective task, and a fully specified definition of what makes a valid meme is approximately as difficult as defining ``art.''
Nevertheless, the authors of this paper have, for better or worse, collectively spent thousands of hours immersed in the communities we explore; thus, while we are not confident in providing a strict definition of a meme, we are in claiming that we know a meme when we see it.

Using the same randomly selected 200 clusters as mentioned in Appendix~\ref{sec:appendix_clustering}, we visited each KYM page the cluster was tagged with and noted whether or not it properly documented what we consider an ``actual'' meme.
The 200 clusters were mapped to 162 unique KYM pages, and of these 162 pages, 3 (1.85\%) we decided were ``bad.''
This is mainly due to the lack of completeness and relatively high number of random images in the gallery (see ~\cite{maxvidya_meme,xy_meme} for some examples of ``bad'' KYM entries).

Next, we set out to determine whether the label (i.e., KYM page) assigned to each of our randomly sampled clusters was appropriate.
Using three annotators, for each cluster we examined the KYM page, the medoid of the cluster, and the images in the cluster itself and noted whether the label does in fact apply to the cluster.
Here, again, there is a great degree of subjectivity.
To reign some of the subjectivity in, we used the following guidelines:
\begin{compactenum}
  \item If the exact image(s) in the cluster appear in the KYM gallery, then the label is correct.
  \item For images that do not appear in the KYM gallery, if the label is \emph{appropriate}, then it is a correct labeling.
\end{compactenum}

There are some important caveats with these guidelines.
First, KYM galleries are crowdsourced, and while curated to some extent, the possibility for what amounts to random images in a gallery \emph{does} exist; however, based on our assessment of KYM page validity, this occurs with low probability.
Second, we considered a label correct if it was \emph{appropriate}, even if it was not necessarily the \emph{best} possible label.
For example, as our results show, many memes are related, and many images mix and match pieces of various memes.
While it is definitely true that there might be better labels that exist for a given cluster, this straightforward and comprehensible labeling process is sufficient for our purposes.
We leave a more in-depth study of the subjective nature of memes for future work.
Finally, it is important to note that memes are a \emph{cultural} phenomenon, and thus the potential for cultural bias in our annotation is possible.
Note that our annotators were born in three different countries (USA, Italy, and Cyprus), only one is a native English speaker, and two have spent substantial time in the US.

After annotating clusters, we compute the Fleis agreement score ($\kappa$).
With our cluster samples, we achieve $\kappa {=} 0.67$, which is considered ``substantial'' agreement.
Finally, for each cluster we obtain the majority agreement of all annotators to assess the accuracy of our annotation process; we find that 89\% of the clusters had a legitimate annotation to a specific KYM entry.

\section{Screenshot Classifier} \label{sec:appendix_classifier}
We now provide details on our screenshot classifier mentioned in Step 4 in Figure~\ref{fig:pipeline}).

\descr{Dataset}. Table~\ref{tbl:curated_dataset} summarizes the dataset used for training the classifier.
It includes 28.8K images that depict posts from Twitter, 4chan, Reddit, Facebook, and Instagram, which we collect from public sources.
First, we download images from specific subreddits that only allow screenshots from a particular community.
For example, the 4chan subreddit require all submissions to be of a screenshot of a 4chan thread.
Next, we use the Pinterest platform to download specific boards that contain mostly screenshots from the communities we study.
Also, we search and obtain image datasets that are publicly available on Web archiving services like the Wayback Machine.
We then manually filter out images that were misplaced.
Finally, we include 10K random images posted on \dspol (i.e., a subset of the 4.3M images collected for our measurements).

\begin{table}[t]
\centering
\setlength{\tabcolsep}{5pt}
\resizebox{\columnwidth}{!}{%
\begin{tabular}{@{}lllllll@{}}
\toprule
\textbf{Platform}  & Twitter                    & 4chan                      & Reddit                    & Facebook               & Instagram                  & Other                      \\ \midrule
\textbf{\# images} & \multicolumn{1}{r}{14,602} & \multicolumn{1}{r}{10,127} & \multicolumn{1}{r}{2,181} & \multicolumn{1}{r}{1,414} & \multicolumn{1}{r}{497} &\multicolumn{1}{r}{10,630} \\ \bottomrule
\end{tabular}%
}
\caption{Curated dataset used to train the screenshot classifier.}
\label{tbl:curated_dataset}
\vspace{-0.2cm}
\end{table}

\descr{Classifier.} To detect screenshots that contain images from one of the social networks included in our dataset, we use Convolutional Neural Networks.
Figure~\ref{fig:ml_architecture} provides an overview of our classifier's architecture.
It includes two Convolutional Neural Networks, each followed by a max-pooling layer.
The output of these layers is fed to a fully-connected dense layer comprising 512 units.
Finally, we have another fully-connected layer with two units, which outputs the probability that a particular image is a screenshot from one of the five social networks and the probability that an image is a random one.
To avoid overfitting on the two last fully-connected layers, we apply Dropout with $d=0.5$~\cite{srivastava2014dropout}.
This means that, while training, 50\% of the units are randomly omitted from updating their parameters. %

\begin{figure}[t]
\centering
\includegraphics[width=0.725\columnwidth]{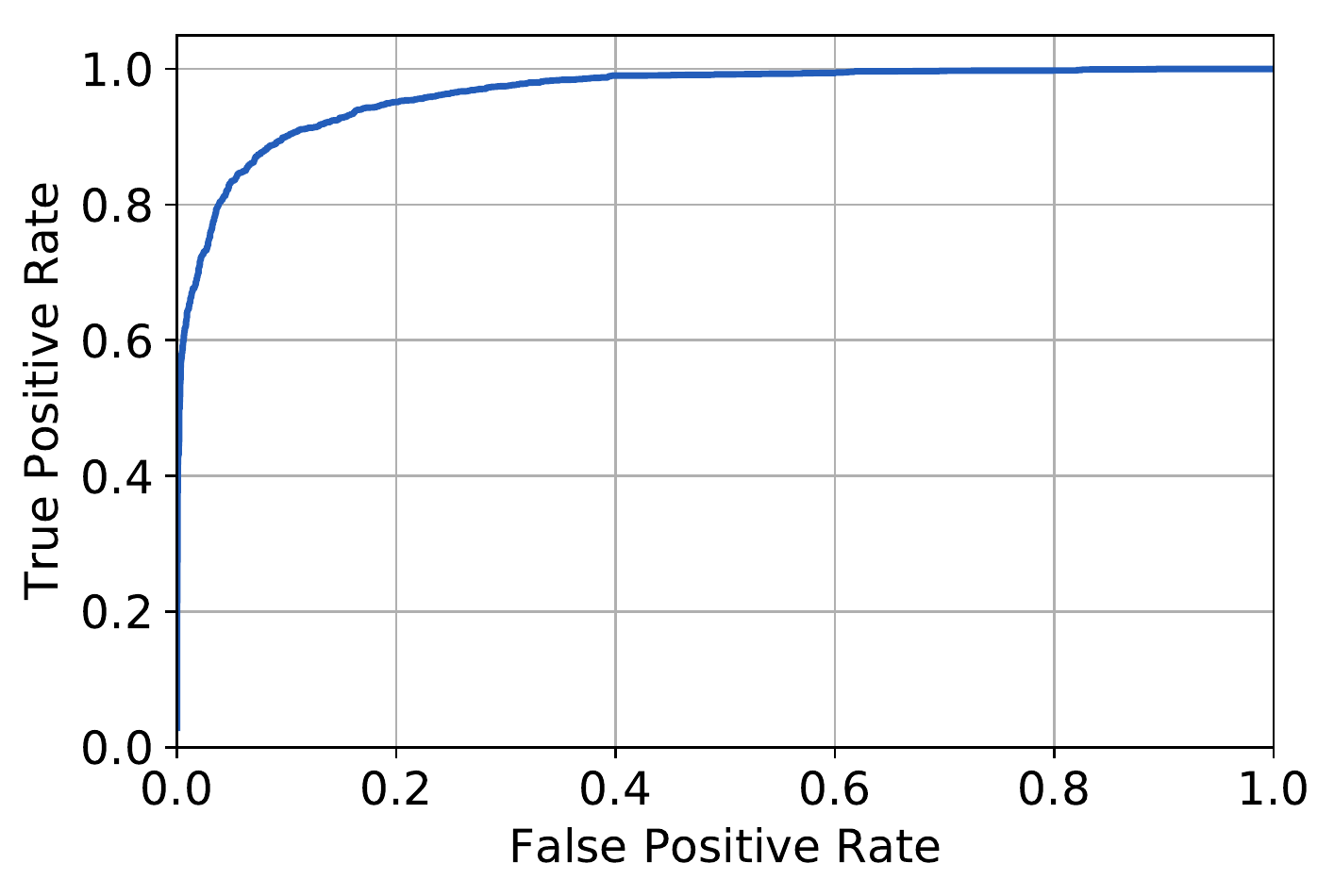}
\caption{ROC curve of the screenshot classifier.}
\label{fig:model_performance}
\vspace{-0.2cm}
\end{figure}

\descr{Experimental Evaluation.} Our implementation uses Keras~\cite{chollet2015keras} with TensorFlow as the backend~\cite{abadi2016tensorflow}.
To train our model, we randomly select 80\% of the images and evaluate based on the rest 20\% out-of-sample dataset.
Figure~\ref{fig:model_performance} shows the ROC curve of the model.
We observe that the devised classifier exhibits acceptable performance with an Area Under the Curve (AUC) of 0.96.
We also evaluate our model in terms of accuracy, precision, recall, and F1-score, which amount to 91.3\%, 94.3\%, 93.5\%, and 93.9\%, respectively.

\begin{figure}[t]
\centering
\includegraphics[width=0.6\columnwidth]{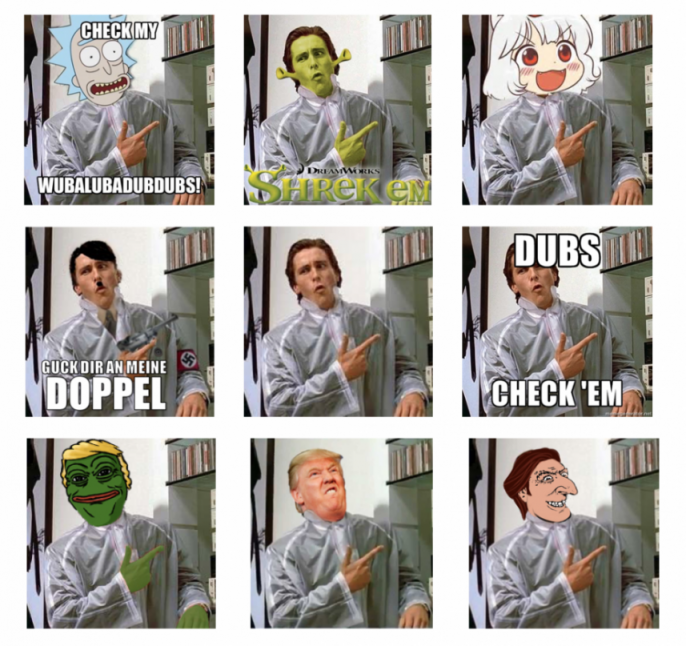}
\caption{Images that are part of the Dubs Guy/Check Em Meme.}
\label{fig:check_em_example}
\end{figure}
\begin{figure}[t]
\centering
\includegraphics[width=0.6\columnwidth]{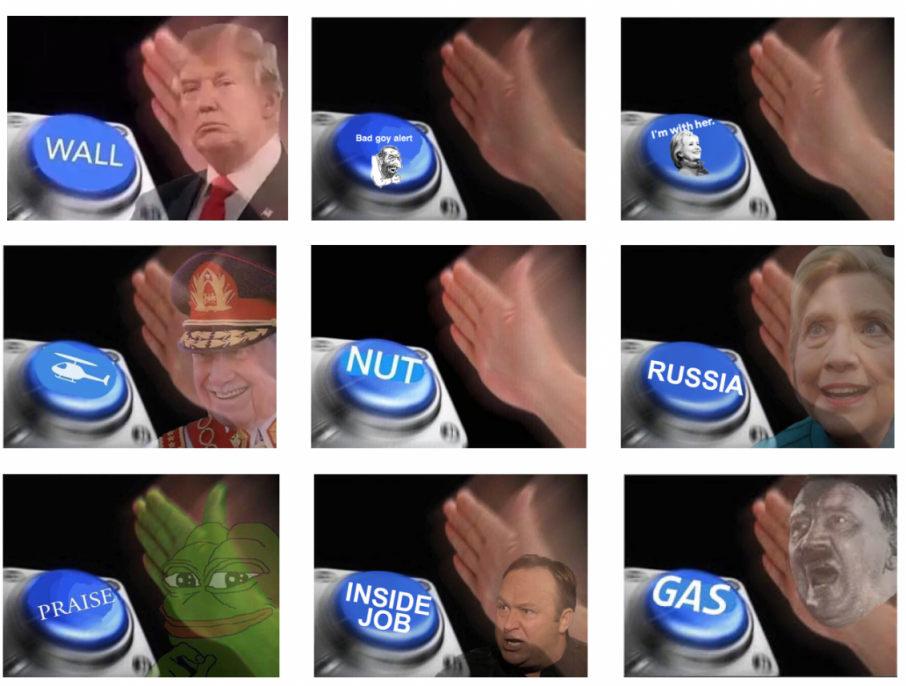}
\caption{Images that are part of the Nut Button Meme.}
\label{fig:nut_button}
\end{figure}
\begin{figure}[t]
\centering
\includegraphics[width=0.6\columnwidth]{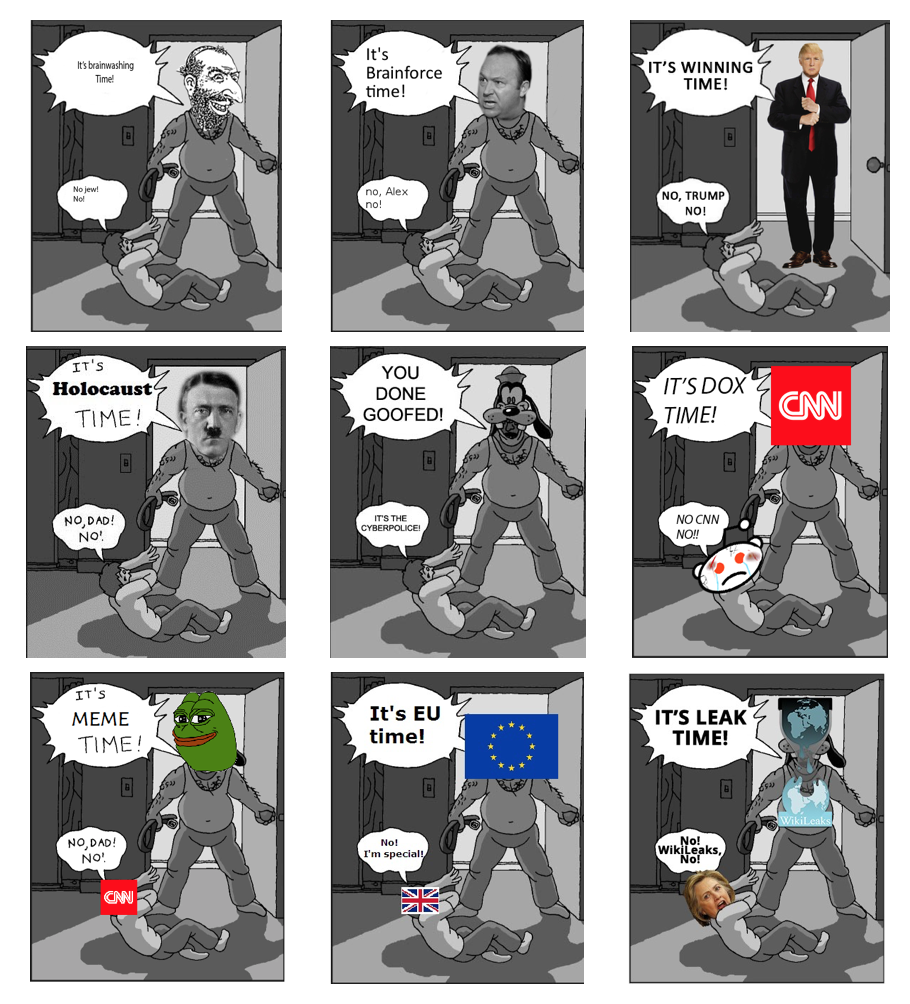}
\caption{Images that are part of the Goofy's Time Meme.}
\label{fig:goofys}
\end{figure}

\section{Clusters examples} \label{sec:appendix_clusters}
As anticipated in Section~\ref{sec:analysis:clusters}, we also present some examples of clusters showcasing how the proposed pipeline can effectively detect and group images that belong to the same meme.

Specifically, Figure~\ref{fig:check_em_example} shows a subset of the images from the Dubs Guy/Check Em meme~\cite{dubs_guy_meme}, Figure~\ref{fig:nut_button} a subset of images that belong to the Nut Button meme~\cite{nut_button_meme}, while Figure~\ref{fig:goofys} -- to the Goofy's Time meme~\cite{goofy_meme}.
Note that all these images are obtained from /pol/ clusters.

In all clusters, we observe similar variations, i.e., variations of Donald Trump, Adolf Hitler, The Happy Merchant, and Pepe the Frog appear in all examples.
Once again, this emphasizes the overlap that exists among memes.

\section{Interesting Images} \label{sec:appendix_interesting_images}
Finally, we report some ``interesting'' examples of images from our frogs case study (see Section~\ref{subsection:hierarchy}), as well as an example of an image for enhancing/penalizing the public image of specific politicians (as discussed in Section~\ref{sec:meme_popularity}).

\begin{figure}[t]
\centering
\includegraphics[width=0.625\columnwidth]{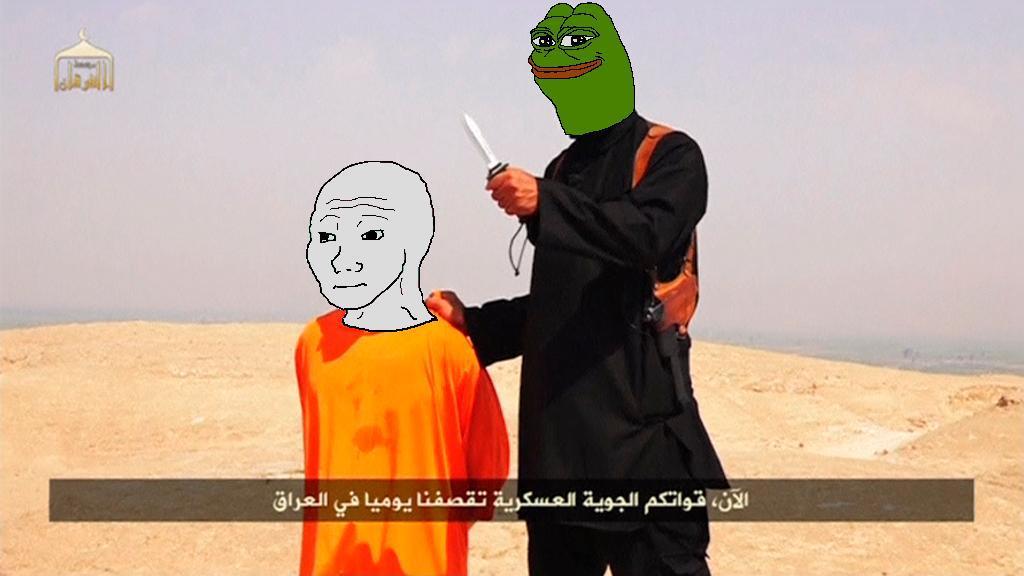}
\caption{Image that exists in the clusters that are connected with frogs and Isis Daesh.}
\label{fig:isis_pepe}
\end{figure}

\begin{figure}[t]
\centering
\includegraphics[width=0.625\columnwidth]{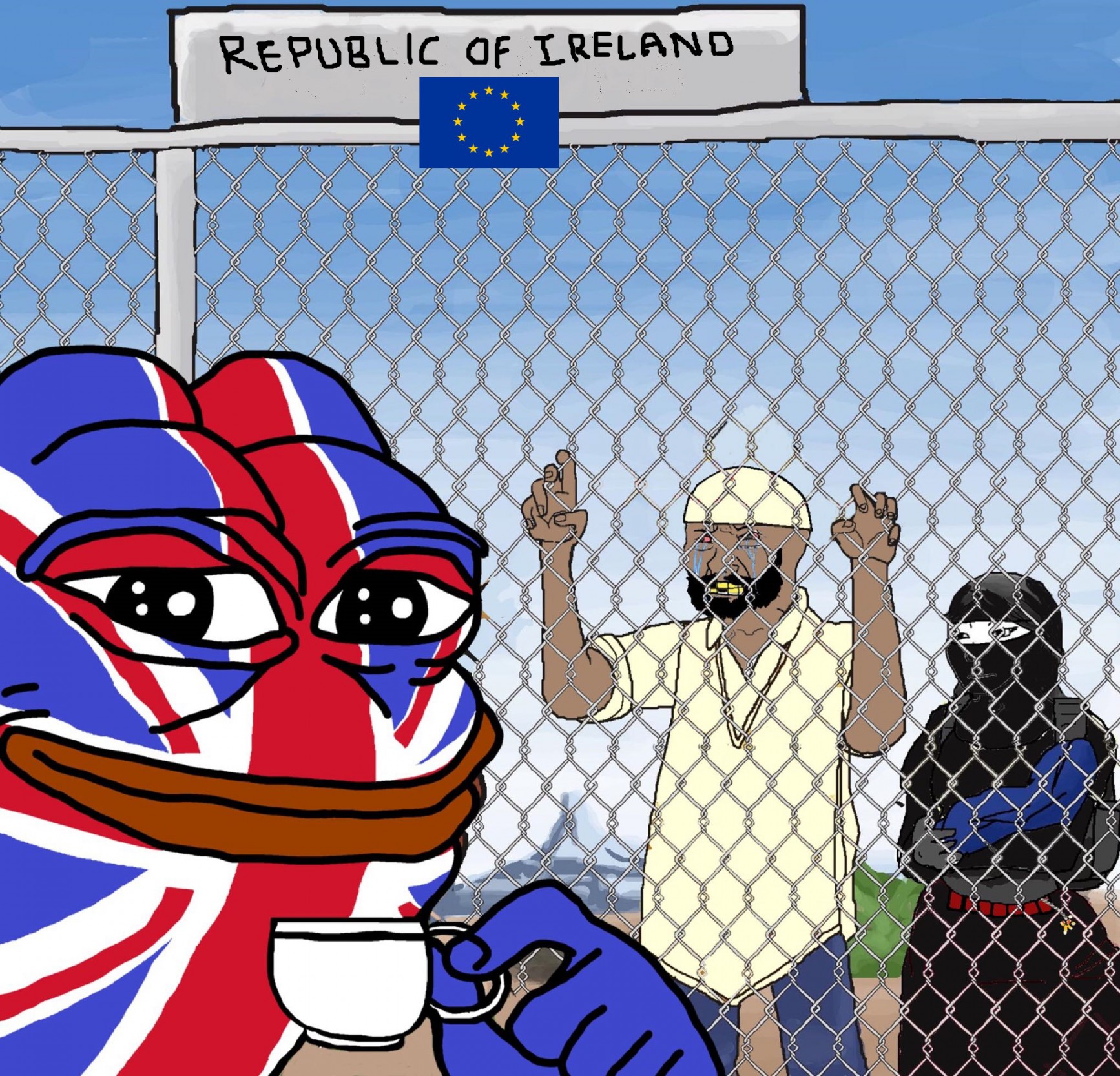}
\caption{Image that exists in the clusters that are connected with frogs and Brexit.}
\label{fig:pepe_brexit}
\vspace{-0.2cm}
\end{figure}

\begin{figure}[t]
\centering
\includegraphics[width=0.6\columnwidth]{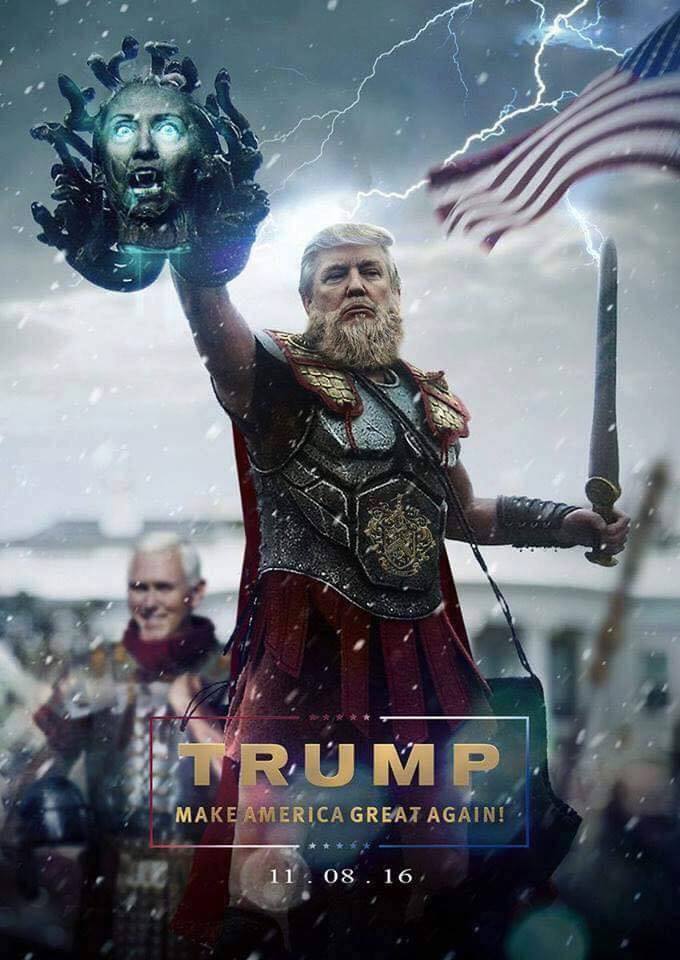}
\caption{Meme that is used for enhancing/penalizing the public image of specific politicians. Hillary Clinton is represented as Medusa, a monster, while Donald Trump is presented as Perseus (the hero who beheaded Medusa).}
\label{fig:trump_clinton_medusa}
\end{figure}

Specifically, Figure~\ref{fig:isis_pepe} shows an image connecting the Smug Frog~\cite{smug_frog_meme} and the ISIS memes~\cite{isis_meme}.
Also, Figure~\ref{fig:pepe_brexit} shows an image connecting the Smug Frog and the Brexit meme~\cite{brexit_meme}.
Finally, Figure~\ref{fig:trump_clinton_medusa} shows a graphic image found in \dspol that aims to attack the image of Hillary Clinton, while boosting that of Donald Trump.
(The image depicts Hillary Clinton as a monster, Medusa, while Donald Trump is presented as Perseus, the hero who beheaded Medusa.)

\begin{figure}[t!]
\centering
\includegraphics[width=0.45\columnwidth]{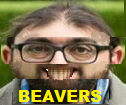}
\caption{A meme of the 6th author of this paper created by \dspol.}
\label{fig:beaver}
\end{figure}

\begin{figure}[t!]
\centering
\includegraphics[width=0.55\columnwidth]{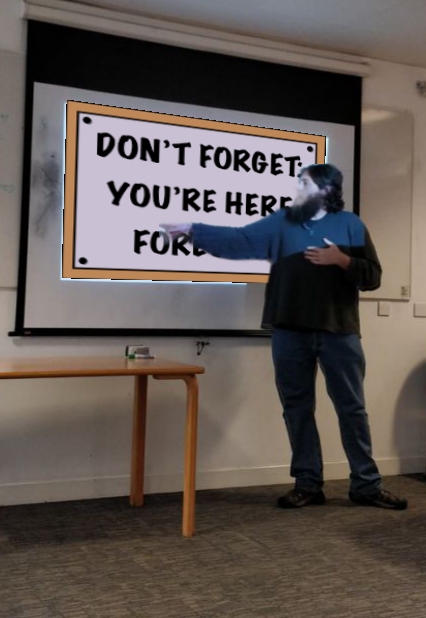}
\caption{A meme of the 3rd author of this paper created by \dspol.}
\label{fig:simpsons}
\end{figure}

Furthermore, some of the communities we study in the current work have taken an interest in our previous work.
\dspol has taken a particular interest, and as additional evidence to the community's meme creating ``ability,'' Figures~\ref{fig:beaver} and~\ref{fig:simpsons} are two sample memes created in response to our work.
Figure~\ref{fig:beaver} is a manipulated photo of the 6th author, originally taken as part of an interview for Nature News.
\dspol seized upon this opportunity and, during the course of one discussion thread\footnote{\url{http://archive.4plebs.org/pol/thread/129243152/}}, decided that he ``definitely masturbates to beavers,'' creating Figure~\ref{fig:beaver} to prove the point.
Figure~\ref{fig:simpsons}, which appeared in a recent \dspol thread\footnote{\url{http://archive.4plebs.org/pol/thread/172617415/}}, is an edited version of a photo, related to a Simpsons derived meme, of the 3rd author giving a talk on our previous work.

\begin{figure}[t!]
\centering
\includegraphics[width=0.55\columnwidth]{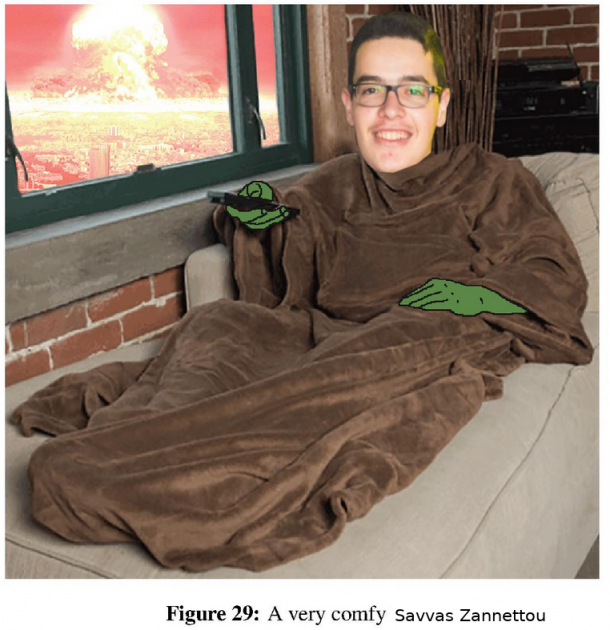}
\caption{A meme of the 1st author of this paper created by \dspol via modification of a Figure from the ``Rare Pepe'' appendix in~\cite{4chanarxiv}. This new variant, ``a very comfy Savvas Zannettou,'' is a modification of the ``comfy pepe'' meme.}
\label{fig:comfy_savvas}
\end{figure}

\begin{figure}[t!]
\centering
\includegraphics[width=0.8\columnwidth]{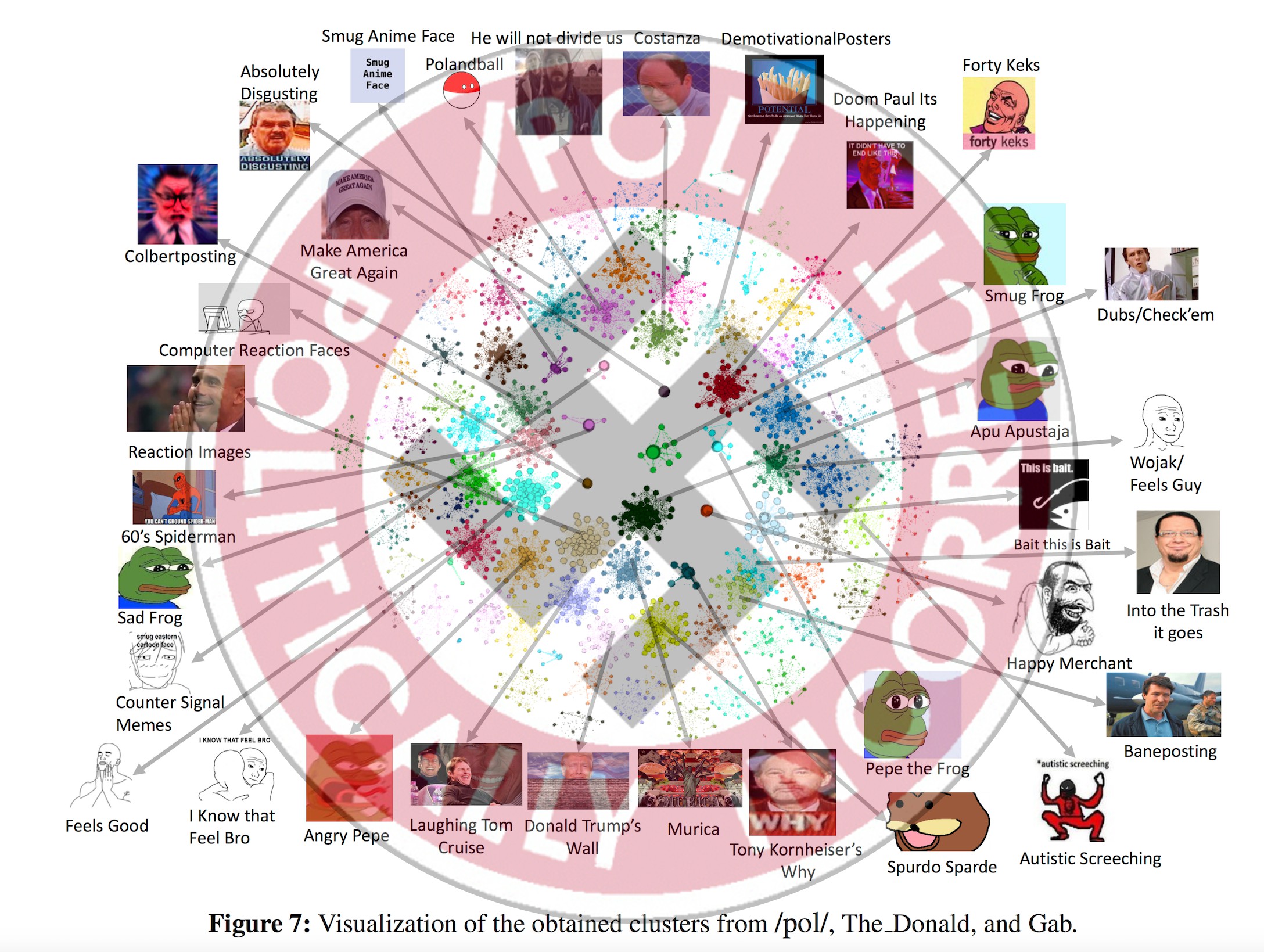}
\caption{A meme created by /pol/ users combining Figure~\ref{fig:clusters_graph} with \dspol's logo.}
\label{fig:fig7_pol}
\end{figure}

\begin{figure}[t!]
\centering
\includegraphics[width=0.8\columnwidth]{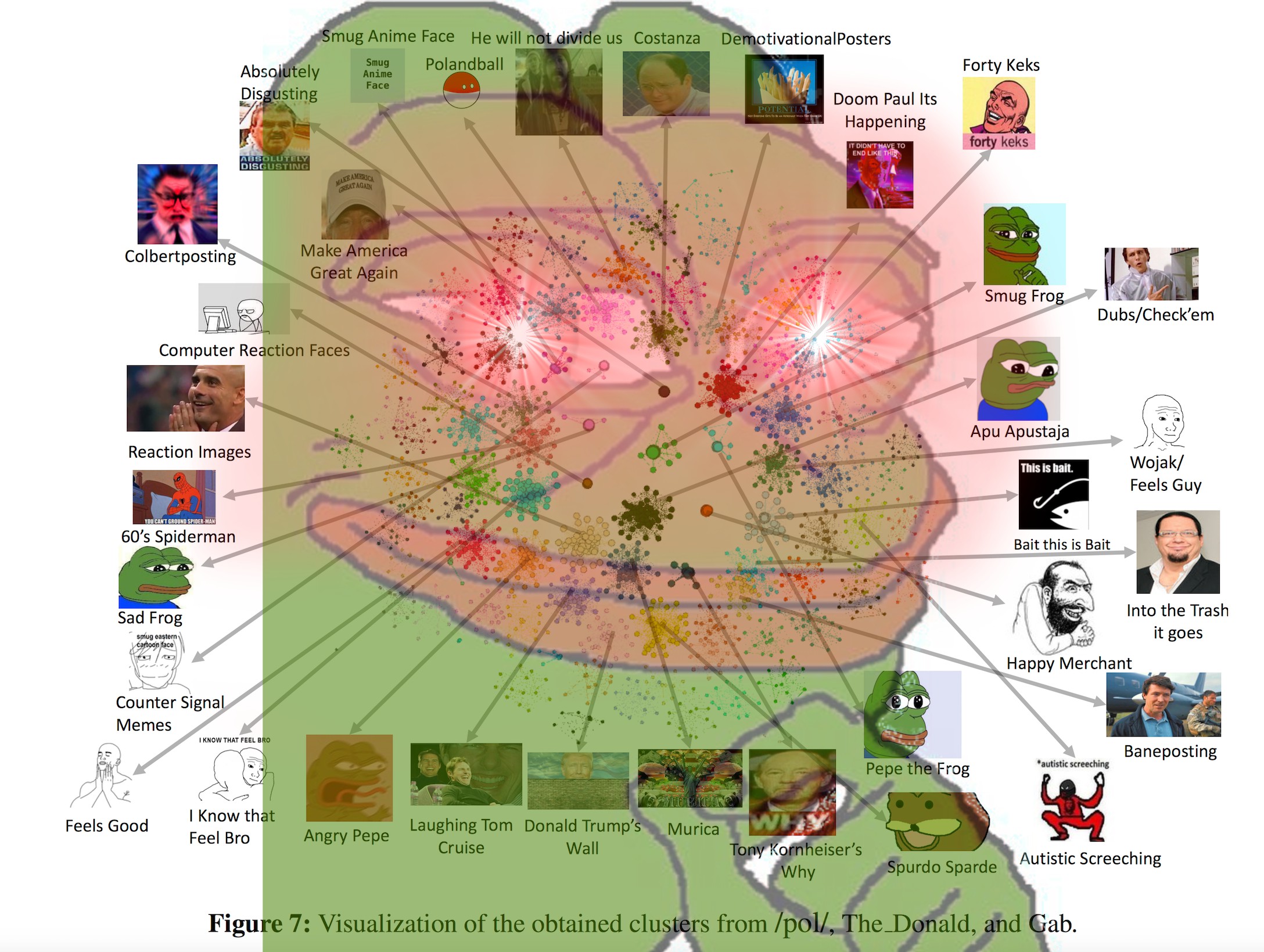}
\caption{A meme created by \dspol combining Figure~\ref{fig:clusters_graph} with the Smug Frog Meme.}
\label{fig:fig7_pepe}
\end{figure}

\begin{figure}[t!]
\centering
\includegraphics[width=0.8\columnwidth]{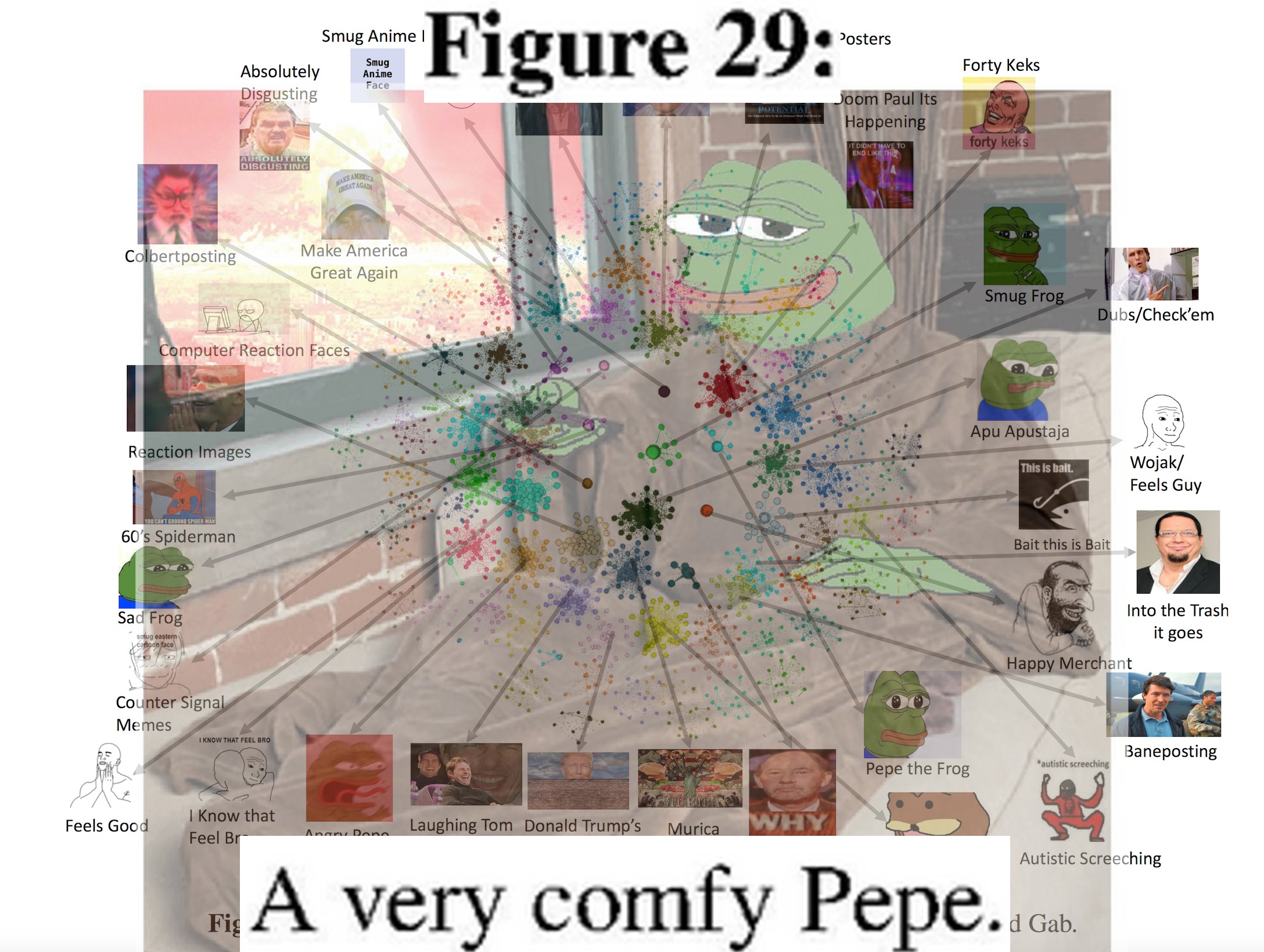}
\caption{A meme created by \dspol users combining Figure~\ref{fig:clusters_graph} with the ``comfy pepe'' meme.}
\label{fig:fig7_comfypepe}
\end{figure}

\begin{figure}[t!]
\centering
\includegraphics[width=0.8\columnwidth]{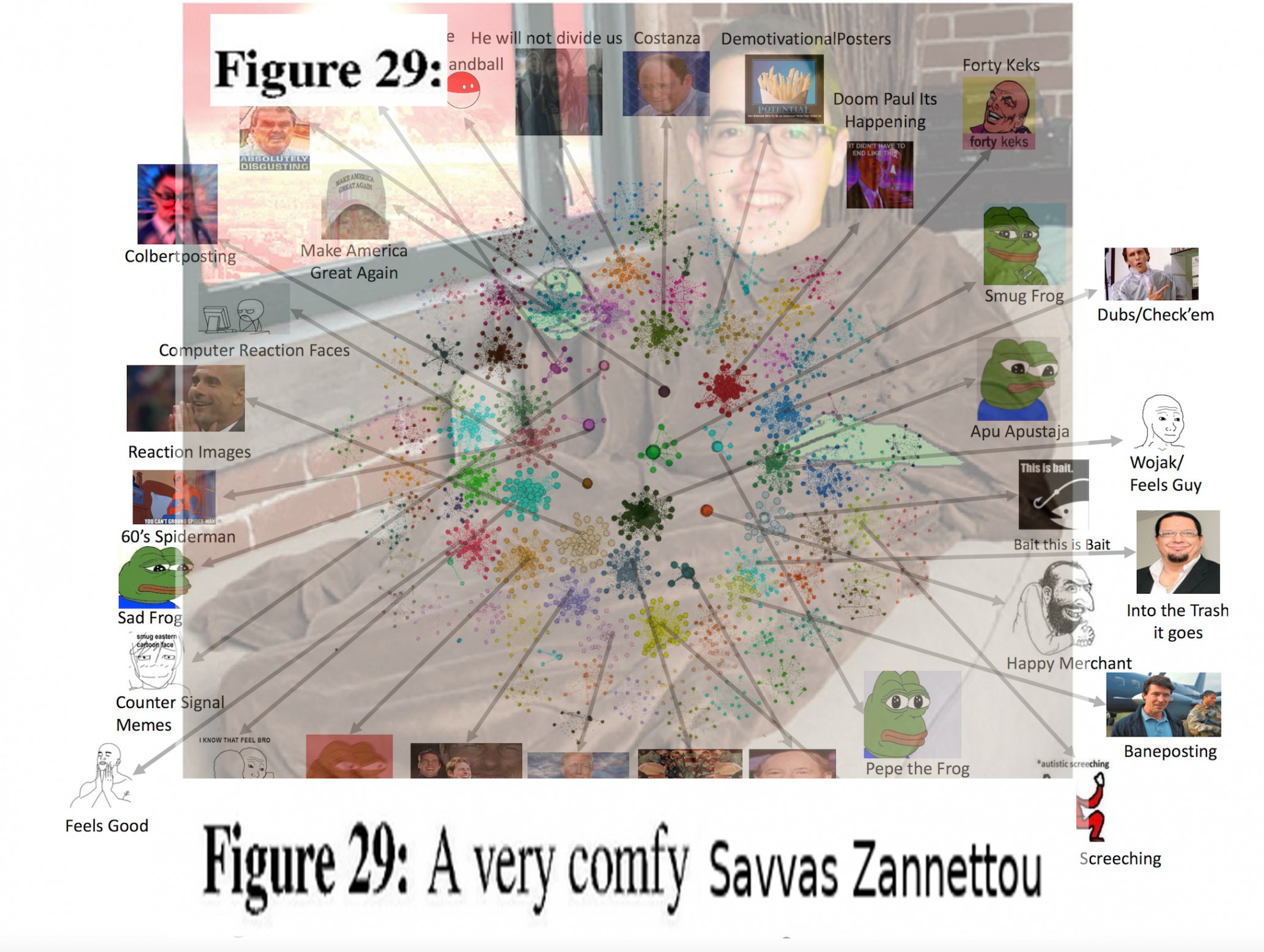}
\caption{A meme created by \dspol users combining Figure~\ref{fig:clusters_graph} with the ``very comfy Savvas Zannettou'' variant from Fig~\ref{fig:comfy_savvas}.}
\label{fig:fig7_savvas}
\end{figure}

In addition, after the release of the initial version of this paper, \dspol quickly picked up and its users started generating memes.
For example, they photoshopped the face of the first author on top of a ``comfy pepe'' meme, the particular variant of which is actually a screenshot from our previous work~\cite{4chanarxiv}, hence creating the new ``very comfy Savvas Zannettou'' variant in Figure~\ref{fig:comfy_savvas}.\footnote{\url{http://archive.4plebs.org/pol/thread/175115412}}
At the same time, Figure~\ref{fig:clusters_graph} from the current work became a meme itself, as \dspol users treated it as a base upon which to build further memes.
Figure~\ref{fig:fig7_pol} shows the graph with \dspol's ``logo,'' Figure~\ref{fig:fig7_pepe} the graph combined with a variation of the Smug Frog Meme, Figure~\ref{fig:fig7_comfypepe} the same gra[h with the ``comfy pepe'' meme, while the newly created ``very comfy Savvas Zannettou'' appears as a variant too (see Figure~\ref{fig:fig7_savvas}).

To conclude, our work also attracted the interest of the press, and journalists have created their own memes about our paper.
For instance, in a Quartz article\footnote{\burl{https://goo.gl/KubZjX}}, an Expanding Brain meme was created and shared showing the use of memes by people, while in an IFLSCIENCE article\footnote{\url{https://goo.gl/YKVjw2}} a Distracted Boyfriend Meme was shared showing our particular value to the scientific community: we have, in large part, quantified some fundamental aspects of memes on the Internet, leaving the rest of the community to focus on other areas of curiosity.

\begin{figure}[t!]
\centering
\includegraphics[width=0.8\columnwidth]{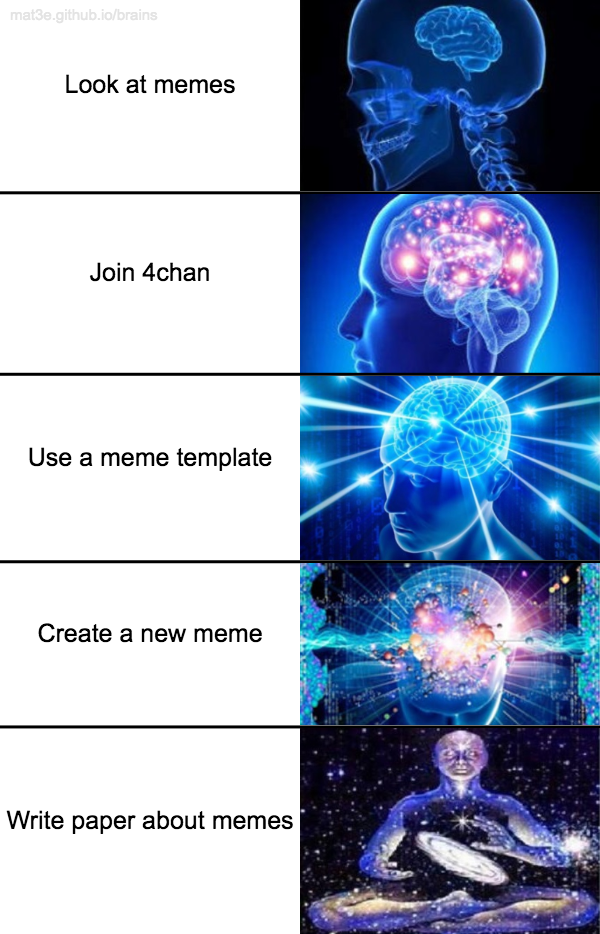}
\caption{A variant of the ``expanding brain'' meme created by QZ in an article about this paper. The meme illustrates a progression of ``intelligence,'' starting from people that just look at memes and culminating in those that write a paper about memes (i.e., the authors of this paper).}
\label{fig:expanding_brain}
\end{figure}

\begin{figure}[t!]
\centering
\includegraphics[width=0.8\columnwidth]{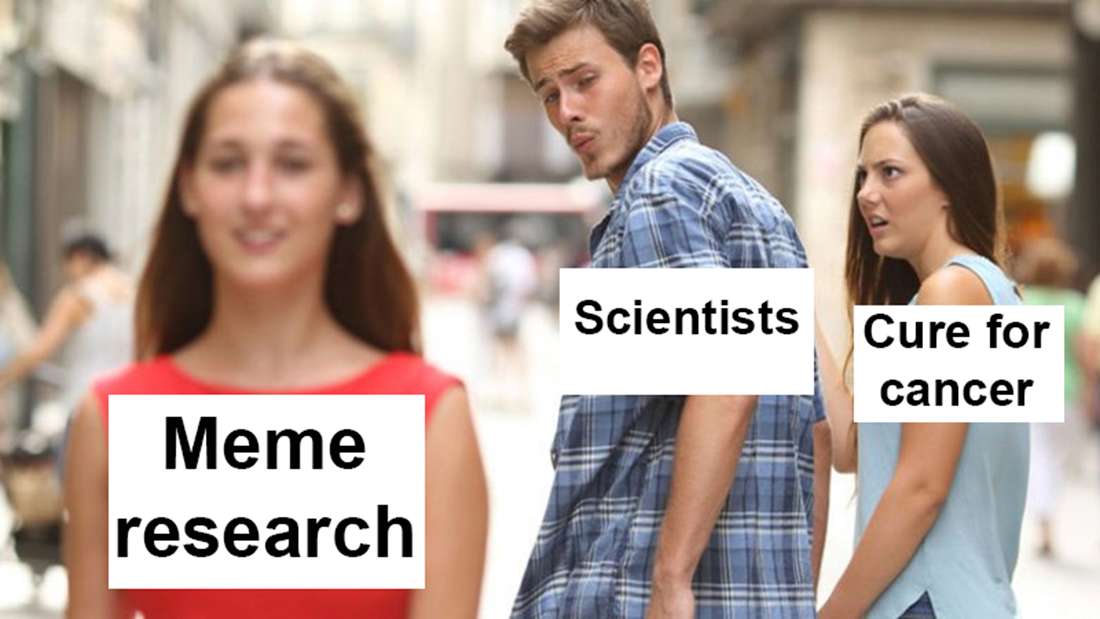}
\caption{A meme created by IFLSCIENCE in an article featuring this paper.}
\label{fig:distracted_boyfriend}
\end{figure}

\end{document}